%% file: thesis_final.tex
\documentclass{dmathesis}
\usepackage{graphicx, eepic}
\input{format}

\newcounter{multieqs}

\newcommand{\be}{\begin{equation}}
\newcommand{\ee}{\end{equation}}
\newcommand{\eq}[1]{(\ref{#1})}
\newcommand{\bit}{\begin{itemize}}  \newcommand{\eit}{\end{itemize}}

\newcommand{\bra}[1]{\langle #1|}
\newcommand{\ket}[1]{|#1 \rangle}
\newcommand{\ipr}[2]{\langle #1 | #2 \rangle}

\newcommand{\bm}[1]{\mbox{\boldmath $#1$}}
\newcommand{\rf}[1]{(\ref{#1})}

\def\bd{\begin{document}}
\def\ed{\end{document}}
\def\nn{\nonumber}
\def\bea{\begin{eqnarray}}
\def\eea{\end{eqnarray}}
\let\bm=\bibitem

\def\la{\langle}
\def\ra{\rangle}

\def\npb#1#2#3{Nucl. Phys. {\bf{B#1}} #3 (#2)}
\def\plb#1#2#3{Phys. Lett. {\bf{#1B}} #3 (#2)}
\def\prl#1#2#3{Phys. Rev. Lett. {\bf{#1}} #3 (#2)}
\def\prd#1#2#3{Phys. Rev. {D \bf{#1}} #3 (#2)}
\def\cmp#1#2#3{Comm. Math. Phys. {\bf{#1}} #3 (#2)}
\def\cqg#1#2#3{Class. Quantum Grav. {\bf{#1}} #3 (#2)}
\def\nppsa#1#2#3{Nucl. Phys. B (Proc. Suppl.) {\bf{#1A}}#3 (#2)}
\def\ap#1#2#3{Ann. of Phys. {\bf{#1}} #3 (#2)}
\def\ijmp#1#2#3{Int. J. Mod. Phys. {\bf{A#1}} #3 (#2)}
\def\rmp#1#2#3{Rev. Mod. Phys. {\bf{#1}} #3 (#2)}
\def\mpla#1#2#3{Mod. Phys. Lett. {\bf A#1} #3 (#2)}
\def\jhep#1#2#3{J. High Energy Phys. {\bf #1} #3 (#2)}
\def\atmp#1#2#3{Adv. Theor. Math. Phys. {\bf #1} #3 (#2)}

\def\N{{\cal N}}
\def\sst{\scriptscriptstyle}
\def\thetabar{\bar\theta}
\def\Tr{{\rm Tr}}
\def\one{\mbox{1 \kern-.59em {\rm l}}}

\def\a{\alpha}      \def\da{{\dot\alpha}}  \def\dA{{\dot A}}
\def\b{\beta}       \def\db{{\dot\beta}}  
\def\g{\gamma}  \def\G{\Gamma}  \def\dc{{\dot\gamma}}  
\def\d{\delta}  \def\D{\Delta}  \def\ddt{\dot\delta}  
\def\e{\epsilon}        \def\ve{\varepsilon}  
\def\f{\phi}    \def\F{\Phi}    \def\vvf{\f}  
\def\h{\eta}  
\def\k{\kappa}  
\def\l{\lambda} \def\L{\Lambda}  
\def\m{\mu} \def\n{\nu}  
\def\o{\omega}  
\def\p{\pi} \def\P{\Pi}  
\def\r{\rho}  
\def\s{\sigma}  \def\S{\Sigma}  
\def\t{\tau}  
\def\th{\theta} \def\Th{\Theta} \def\vth{\vartheta}  
\def\X{\Xeta}  
\def\z{\zeta}  

\def\na{\nabla}  

\def\cA{{\cal A}} \def\cB{{\cal B}} \def\cC{{\cal C}}  
\def\cD{{\cal D}} \def\cE{{\cal E}} \def\cF{{\cal F}}  
\def\cG{{\cal G}} \def\cH{{\cal H}} \def\cI{{\cal I}}  
\def\cJ{{\cal J}} \def\cK{{\cal K}} \def\cL{{\cal L}}  
\def\cM{{\cal M}} \def\cN{{\cal N}} \def\cO{{\cal O}}  
\def\cP{{\cal P}} \def\cQ{{\cal Q}} \def\cR{{\cal R}}  
\def\cS{{\cal S}} \def\cT{{\cal T}} \def\cU{{\cal U}}  
\def\cV{{\cal V}} \def\cW{{\cal W}} \def\cX{{\cal X}}  
\def\cY{{\cal Y}} \def\cZ{{\cal Z}}

\def\ua{\underline{\alpha}}  
\def\uc{\underline{\phantom{\alpha}}\!\!\!\gamma}  
\def\um{\underline{\mu}}  
\def\ud{\underline\delta}  
\def\ue{\underline\epsilon}  
\def\una{\underline a}\def\unA{\underline A}  
\def\unb{\underline b}\def\unB{\underline B}  
\def\unc{\underline c}\def\unC{\underline C}  
\def\und{\underline d}\def\unD{\underline D}  
\def\une{\underline e}\def\unE{\underline E}  
\def\unf{\underline{\phantom{e}}\!\!\!\! f}\def\unF{\underline F}  
\def\unm{\underline m}\def\unM{\underline M}  
\def\unn{\underline n}\def\unN{\underline N}  
\def\unp{\underline{\phantom{a}}\!\!\! p}\def\unP{\underline P}  
\def\unq{\underline{\phantom{a}}\!\!\! q}  
\def\unQ{\underline{\phantom{A}}\!\!\!\! Q}  
\def\unH{\underline{H}}  
  
\def\As {{A \hspace{-6.4pt} \slash}\;}  
\def\bs {{b \hspace{-6.4pt} \slash}\;}  
\def\Ds {{D \hspace{-6.4pt} \slash}\;}
\def\Gts {{\Gt \hspace{-6.4pt} \slash}\;}
\def\ds {{\del \hspace{-6.4pt} \slash}\;}  
\def\ss {{\s \hspace{-6.4pt} \slash}\;}  
\def\ks {{ k \hspace{-6.4pt} \slash}\;}  
\def\ps {{p \hspace{-6.4pt} \slash}\;}   
\def\xs {{x \hspace{-6.4pt} \slash}\;}  
\def\pas {{{p_1} \hspace{-6.4pt} \slash}\;}  
\def\pbs {{{p_2} \hspace{-6.4pt} \slash}\;}   
\def\cFs {{{\cal F} \hspace{-6.4pt} \slash}\;}

\def\Ah{{\hat{A}}}  
\def\Dh{{\hat{D}}}
\def\Gh{{\hat{G}}}
\def\Fh{{\hat{F}}}
\def\Ih{{\hat{I}}} 
\def\Jh{{\hat{J}}} 
\def\Kh{{\hat{K}}}
\def\Lh{{\hat{L}}} 
\def\Ph{{\hat{P}}}
\def\Rh{{\hat{R}}}
\def\Vh{{\hat{V}}} 
\def\Xh{{\hat{X}}}
 
\def\ah{{\hat{\a}}}
\def\bh{{\hat{\b}}}
\def\gh{{\hat{\g}}}
\def\dh{{\hat{\d}}}
\def\hh{\hat{h}}
\def\uh{\hat{u}}  
\def\xh{\hat{x}}  
\def\yh{\hat{y}}  
\def\ph{\hat{p}}  
\def\xih{\hat{\xi}}  
\def\chih{\hat{\chi}}  
\def\Psih{\hat{\Psi}}    
\def\cVh{\hat{\cV}} 

\def\psit{\tilde{\psi}}  
\def\Psit{\tilde{\Psi}}   
\def\Psibt{\tilde{\bar{Psi}}}  

\def\st{\tilde{\sigma}}  

\def\delt{\tilde{\delta}}
\def\Phit{\tilde{\Phi}}   
\def\Phitb{\overline{\tilde{Phi}}}  
\def\tht{\tilde{\th}}  
\def\lt{\tilde{\l}}
\def\chit{\tilde{\chi}}   
\def\phit{\tilde{\phi}} 

\def\At{\tilde{A}}
\def\Bt{\tilde{B}}
\def\Ct{\tilde{C}}
\def\Dt{\tilde{D}}
\def\Et{\tilde{E}}
\def\Ft{\tilde{F}}
\def\Gt{\tilde{G}}
\def\Ht{\tilde{H}}
\def\It{\tilde{I}}
\def\Jt{\tilde{J}}
\def\Qt{\tilde{Q}}  
\def\Rt{\tilde{R}}  
\def\Mt{\tilde{M }}  
\def\Nt{\tilde{N}}   
\def\St{\tilde{S}}
\def\Vt{\tilde{V}}
\def\Xt{\tilde{X}} 
\def\at{\tilde{a}}
\def\ct{\tilde{c}}
\def\dt{\tilde{d}}
\def\htt{\tilde{h}} 
\def\ft{\tilde{f}}
\def\gt{\tilde{g}}
\def\pt{\tilde{p}}  
\def\qt{\tilde{q}}  
\def\vt{\tilde{v}}  
\def\nt{\tilde{n}}  
\def\ut{\tilde{u}}  
\def\wt{\tilde{w}}  
\def\zt{\tilde{z}} 
\def\xt{\tilde{x}} 
\def\yt{\tilde{y}} 
\def\Psit{\tilde{\Psi}}
\def\vphit{\tilde{\varphi}}  
\def\Lt{\tilde{\L}}

\def\eb{\bar{\epsilon}} 
\def\delb{\bar{\partial}}  
\def\thb{\bar{\theta}}
\def\mub{\bar{\mu}}
\def\lamb{\bar{\l}}
\def\psib{\bar{\psi}}
\def\sb{\bar{\sigma}}
\def\xib{\bar{\xi}}
\def\chib{\bar{\chi}}

\def\Psib{\bar{\Psi}}
\def\Phib{\bar{\Phi}}
\def\Lamb{\bar{\Lambda}}
\def\Sb{{\overline \Sigma}}
\def\cb{\bar{c}}
\def\hb{\bar{h}}
\def\qb{\bar{q}}
\def\wb{\bar{w}}
\def\ub{\bar{u}}
\def\zb{{\bar{z}}}
\def\Hb{\bar{H}}
\def\Qb{{\bar Q}}
\def\Omegab{\overline{\Omega}}
\def\ob{\overline{\omega}}

\def\Ab{{\overline A}} \def\Bb{{\overline B}} \def\Cb{{\overline C}}  
\def\Db{{\overline D}} \def\Eb{{\overline E}} \def\Fb{{\overline F}}  
\def\Gb{{\overline G}} 
\def\Ib{{\overline I}}  
\def\Jb{{\overline J}} \def\Kb{{\overline K}} \def\Lb{{\overline L}}  
\def\Mb{{\overline M}} \def\Nb{{\overline N}} \def\Ob{{\overline O}}  
\def\Pb{{\overline P}}  \def\Rb{{\overline R}}  
 \def\Tb{{\overline T}} \def\Ub{{\overline U}}  
\def\Vb{{\overline V}} \def\Wb{{\overline W}} \def\Xb{{\overline X}}  
\def\Yb{{\overline Y}} \def\Zb{{\bar{Z}}}  

\def\fb{{\overline f}}
\def\gb{{\overline g}}
\def\mb{{\overline m}}
\def\lb{{\overline l}}
\def\yb{{\overline y}}
  
\def\ldel{{\overleftarrow{\del}}}
\def\rdel{{\overrightarrow{\del}}}
\def\ldeldel{{\overleftarrow{\del^2}}}
\def\rdeldel{{\overrightarrow{\del^2}}}
\def\ldelb{{\overleftarrow{\bar{\del}}}}
\def\rdelb{{\overrightarrow{\bar{\del}}}}

\def\ba{{\bf a}} 
\def\bk{{\bf k}}  
\def\bl{{\bf l}}  
\def\bp{{\bf p}}  
\def\bq{{\bf q}}  
\def\br{{\bf r}}
\def\bt{{\bf t}}
\def\bu{{\bf u}}
\def\bv{{\bf v}}
\def\bx{{\bf x}}  
\def\by{{\bf y}}  
\def\bR{{\bf R}}  
\def\bV{{\bf V}}

\def\bone{{\bf 1}}  

\def\va{{\vec a}}
\def\vk{{\vec k}}
\def\vp{{\vec p}}
\def\vq{{\vec q}}
\def\vx{{\vec x}}
\def\vy{{\vec y}}
\def\vu{{\vec u}}
\def\vv{{\vec v}}

\def\vs{{\vec \sigma}}
\def\vtau{{\vec \tau}}

\newcommand{\ov}[1]{\overrightarrow{#1}}

\def\frA{\mathfrak{A}}
\def\frB{\mathfrak{B}}
\def\frC{\mathfrak{C}}
\def\frD{\mathfrak{D}}
\def\frE{\mathfrak{E}}
\def\frF{\mathfrak{F}}
\def\frG{\mathfrak{G}}
\def\frH{\mathfrak{H}}
\def\frM{\mathfrak{M}}
\def\frN{\mathfrak{N}}
\def\frR{\mathfrak{R}}
\def\frW{\mathfrak{W}}

\def\fra{\mathfrak{a}}
\def\frb{\mathfrak{b}}
\def\frf{\mathfrak{f}}
\def\frg{\mathfrak{g}}
\def\frh{\mathfrak{h}}
\def\frl{\mathfrak{l}}
\def\frs{\mathfrak{s}}
\def\fri{\mathfrak{i}}
\def\frj{\mathfrak{j}}

\def\ma{\mathfrak{a}}
\def\mg{\mathfrak{g}}
\def\mh{\mathfrak{h}}
\def\mR{\mathfrak{R}}
\def\mN{\mathfrak{N}}

\def\d{\delta}\def\D{\Delta}\def\ddt{\dot\delta}  
  
\def\pa{\partial} \def\del{\partial}  
\def\xx{\times}  
\def\uno{\mbox{1 \kern-.59em {\rm l}}}    
  
\def\trp{^{\top}}  
\def\inv{^{-1}}  
\def\dag{{^{\dagger}}}  
\def\pr{^{\prime}}  
  
\def\rar{\rightarrow}  
\def\lar{\leftarrow}  
\def\lrar{\leftrightarrow}  
  
\newcommand{\0}{\,\!}      
\def\one{1\!\!1\,\,}  
\def\im{\imath}  
\def\jm{\jmath}  
  
\newcommand{\tr}{\mbox{tr}}  
\newcommand{\slsh}[1]{/ \!\!\!\! #1}  
  
\def\vac{|0\rangle}  
\def\lvac{\langle 0|}  
  
\def\hlf{\frac{1}{2}}  
\def\ove#1{\frac{1}{#1}}  

\def\Box{\square}  
\def\CC {\mathbb{C}}
\def\FF {\mathbb{F}}
\def\RR{\mathbb{R}}
\def\NN{\mathbb{N}}  
\def\ZZ{\mathbb{Z}}  
\def\bb#1{{\bf #1}}  
\def\bcomment#1{}  
\def\bfhat#1{{\bf \hat{#1}}}  
\def\VEV#1{\left\langle #1\right\rangle}  

\newcommand{\ex}[1]{{\rm e}^{#1}} \def\ii{{\rm i}}  

\newcommand{\lrbrk}[1]{\left(#1\right)}
\newcommand{\sfrac}[2]{{\textstyle\frac{#1}{#2}}}
 
\def\stw{{\sqrt{2}}}

\def\rf {{\rm f}}
\def\ri {{\rm i}}
\def\rj {{\rm j}}
\def\rk {{\rm k}}
\def\rl {{\rm l}}
\def\rs {{\scriptscriptstyle \rm S}}
\def\rt {{\scriptscriptstyle \rm T}}

\def\rQ {{\scriptscriptstyle \rm \cQ}}
\def\rR {{\scriptscriptstyle \rm \cR}}

\def\cQb{{\cal \Qb}}
\def\cRb{{\cal \Rb}}
\def\cWb{{\cal \Wb}}

\def\fd {{\rm N}}
\def\afd {{\overline{\rm N}}}

\def \II {I\hspace{-.1em}I\hspace{.1em}}
\def \IIA {\mbox{\II A\hspace{.2em}}}
\def \IIB {\mbox{\II B\hspace{.2em}}}
\def \gs {g^s}
\def \ls {\lambda^s}

\def \I {{\cal I}}
\def \qs {q\hspace{-.53em}/\hspace{.15em}}
\def \ks {k\hspace{-.53em}/\hspace{.15em}}
\def \YM {{\mbox{\tiny YM}}}
\def \gym {g_{\YM}}

\def \Lc {\L_c}
\def\IR{\relax{\rm I\kern-.18em R}}
\def \id {{\bf 1}}

\def\cci{\ell}
\def\ccj{\ell'}

\def \thbb{\overline{\th\th}}
\newcommand \ol{\overline}
\def \lamb{\bar{\lambda}}
\def \vphi{\varphi}
\def \lambh{\hat{\bar{\lambda}}}
\def \lh{\hat{\lambda}}
\def \dd{\ddagger}

\def \Xd{\dot{X}}

\begin{document}

\input{frontpage}

\pagenumbering{arabic}
\setcounter{page}{1}

\chapter{Introduction}
In this chapter we will discuss the emergence of M-theory from string theory, in particular we will discuss recent developments on the objects known as M2-branes and M5-branes.

\section{What is M-Theory?}
In this section we will discuss briefly the motivations for studying String Theory and M-Theory and what is known in the various theories that build up to M-Theory.
\subsection{Strings, D-branes and M-Theory}

This thesis explores and develops recent ideas in high energy physics known as String Theory and M-Theory, but let us first discuss the motivations for obtaining such theories in the first place. The Standard Model of Particle Physics gives a very accurate description of three of the four fundamental forces in nature namely the Strong, Weak, and Electromagnetic forces. It is a quantum field theory, known as a gauge theory, with gauge group $SU(3)\times SU(2)\times U(1)$ and at the time of writing this thesis is one of the most celebrated successes of human achievement as it describes three of the four forces and their interactions with great measurable accuracy. The fourth force, gravity, is somewhat unreconcilable with the Standard Model. But what we do have is General Relativity which provides us with a classical description of gravity at large scales. 

Many people wish to seek out a theory of `everything', i.e. a grand unified theory of nature which not only quantises gravity, but unifies all the four forces of nature and describes all of their interactions. String Theory and consequently M-Theory provides us with a candidate for such a description of our universe. One of the interesting features of  string theory is that it has a critical dimension in which the theory is mathematically consistent, this is $9+1$ spacetime dimensions for string theory and $10+1$ spacetime dimensions for M-theory. This is a feature which is found neither in the Standard Model nor General Relativity.

So what is String Theory? That question would take too long to describe here so we refer the reader to \cite{Pol1, Pol2} for a comprehensive review, so let us explain schematically what the theory is and the emergence of M-theory. String theory was originally formulated to describe strong interactions of QCD, however it turned out by examining the spectra of the theory that it in fact gave a massless spin 2 graviton as well as vectors and scalars. Originally just the bosonic sector of the theory was found and the critical dimension was $D=26$.  The theory was later supersymmetrised with fermions and this gave a superstring theory in $D=10$. There are two different types of string; one is the open string which comes with a boundary condition at the two end points of the string, the other is a closed string with no boundary. The types of boundary condition one can have for the open string can be either Neumann or Dirichlet. The latter of the two is an interesting condition and will lead to objects known as branes which we will expand on later.

When constructing string theory, one has various choices to make in its formulation, see \cite{Pol2} for a full description of the various Ramond and Neveu-Schwarz periodicity conditions in the Ramond-Neveu-Schwarz (RNS) formalism as well as mixed left and right movers in the 26 and 10 dimensional theories. It turns out that there are five unique string theories in $D=10$, these are Type \I, Type \IIA, Type \IIB, Heterotic $SO(32)$ and Heterotic $E_8\times E_8$. In this thesis we will primarily be concerned with the Type \IIA and Type \IIB theories. These are related by a symmetry known as T-duality, this is where Type \IIA theory on a circle of radius $R$ is equivalent to Type \IIB theory on a circle of radius $\tilde{R}$ with $R={\a'}/{\tilde{R}}$. We will make this more concrete later in Chapter 5.

Another advantage of string theory is the simplicity of its interactions. There is precisely one diagram per order of the string coupling $g_s$, this is in contrast to the Standard Model where one has $s,t,u$ channel diagrams per order of the coupling.

The string tension is given by 
\be
T_{\rm String} = \frac{1}{2\pi l_s^2},
\ee
where $l_s$ is the string scale or string length.
When considering the boundary terms of an open string, it was found that momentum is allowed to flow from the string to an object called a D-brane, these are higher dimensional analogs of strings, such as membranes etc. These so-called D-branes turn out to be fundamental objects in string theory and will be discussed at length in this thesis. The fact that open strings end on D-branes is due to the boundary conditions involved, so for a $(p+1)$-dimensional object called a D$p$-brane we have Neumann boundary conditions along the brane in the $\s^0,...,\s^p$ directions. The directions $\s^{p+1},..., \s^9$ have Dirichlet boundary conditions.

A D$p$-brane has a tension in a similar way as a string but the mass dimension of the tension depends on the type of D$p$-brane, namely
\be
T_p = \frac{1}{g_s}\frac{2\pi}{{(2\pi l_s)}^{p+1}}.
\ee
These D$p$-branes are stable BPS objects and have an action called the DBI action, they are allowed to interact and so we can add an interaction term $S_{\rm int}$ to the D-brane action
\be
S_{p} = S_{DBI} +S_{\rm int}.
\ee
The DBI action is given by
\be
S_{DBI} = -T_p\int d^{p+1}\s \sqrt{-\det(G_{\m\n} + 2\pi \a' F_{\m\n} +B_{\m\n})},
\ee
here $G_{\m\n}$ is the pullback of the spacetime metric onto the worldvolume of the brane, similarly with $F_{\m\n}$ and $B_{\m\n}$. The field strength $F_{\m\n}$ is that of the abelian gauge field $A_\m$ that lives on the single brane.\footnote{When considering a stack of branes, this becomes non-abelian.} For a gauge invariant theory, we introduce the Kalb-Ramond or NS-NS 2-form field $B_{\m\n}$ which is anti-symmetric. In chapter 5, we will discuss how this field is connected to noncommutative geometry. We can expand this DBI action to first order, this gives the low energy effective action of the worldvolume theory in some large tension limit which decouples us from gravity and makes the theory weakly coupled for fixed $l_s$. The action has an abelian $U(1)$ gauge symmetry by the open string ending on the single D-brane. The more interesting case is that of a stack of $N$ coincident such D-branes where the $U(1)^N$ factor is promoted to $U(N)$ in the unbroken phase of the stack of branes, the $U(N)$ gauge symmetry associated with a stack of $N$ D$p$-branes is given by the low energy effective action which is called Super Yang-Mills (SYM) theory in $(p+1)$-dimensions. See Chapter 5 for a further discussion of these theories. 

A D$p$-brane naturally couples to a RR potential\footnote{For the origin of these terms we refer the reader to \cite{Pol1,Pol2}} which for a D0-brane would be a gauge field $A_\m$, for a D1-brane would be a two-form $A_{\m\n}$ etc. More explicitly this is given by
\be
S_{\rm pot} = \frac{T_p}{(p+1)!}\int d^{p+1}\s \ve^{\m_1...\m_{p+1}}A_{\m_1...\m_{p+1}},
\ee
this potential term can be added to the brane action $S_p = S_{DBI} +S_{\mathrm int} + S_{\mathrm pot}$ to give the full action for the D$p$-brane in question. These RR gauge potentials have a field strength in which they are gauge invariant under transformations $\d A_{p+1} = {\rm d} \L_p$, where $\L_p$ is a $p$-form. 

\begin{figure}[htb]
\centering
\includegraphics[width=1\textwidth]{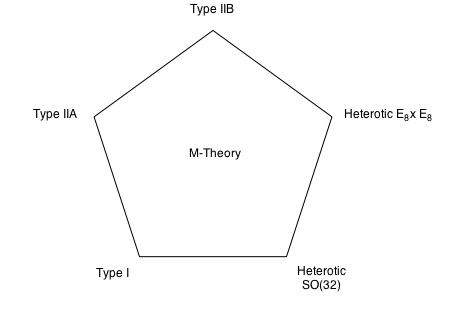}
\caption{Diagram of the web of dualities between the various superstring theories and M-theory.}
\label{m-theoryweb}
\end{figure}

In 1995 a new type of relationship was found in string theory by Witten \cite{wittenM}, this was built upon previous works on obtaining a UV completion to 11-dimensional supergravity as was obtained for the 10-dimensional superstring theories\cite{11d-1,11d-2,11d-3}. In Witten's proposal he considered the large string coupling limit of Type \IIA theory, $g_s\to \infty$, and found that this corresponded to a large extra dimension
\be
R_{11} = g_s l_s.
\ee
This lead to a mysterious 11-dimensional theory called M-Theory, it is an added mystery as to what the `M' stands for also. The existence of this extra dimension allows a unification of all five superstring theories via a web of dualities, see Figure~\ref{m-theoryweb}.

From eleven dimensional supergravity theories \cite{cjs}, we know that a three-form potential exists $C_{MNP}$, where $M,N,P=0,...,10$. So this lead to the discovery of the M2-brane which couples to this three-form electrically and the M5-brane  which couples magnetically. The focus of this thesis is on these two branes, the first half concentrates on recent developments in the M2-brane theory and the second half describes a new quantum geometry on the M5-brane theory.

\subsection{Outline}

The focus of this thesis is to obtain a better understanding of M2-branes, M5-branes and their interactions. In this Chapter we will review what is currently known in M-theory, specifically the multiple M2-brane theory and the recent advancements to obtain some understanding of the non-abelian M5-brane theory. For the M2-brane theory we begin by looking at the motivation behind a particular structure called a 3-bracket before reviewing the BLG theory which uses a Lie 3-bracket valued in a 3-algebra. The features of this theory are discussed as well as a Higgsing procedure to obtain a D2-brane theory. We then turn to the ABJM description of M2-branes. For the M5-brane, we will be discussing the issues with a worldvolume field theory action for a self-dual 3-form field strength of a tensor field $B_2$ in six dimensions and some attempts to get around this at the cost of full six dimensional Lorentz invariance. We then briefly discuss a duality between M5-branes on $S^1$ and 5D SYM theory.

This thesis is then comprised of two parts, we begin with an overview of Part I. In Chapter 2 we review the main results of the paper \cite{LR}, the authors constructed the closed $\cN=8$ M2-brane theory coupled to a flux term. The analysis was then repeated for the $\cN=6$ ABJM theory. In Chapter 3 we extend these results to the open M2-branes picture and consider the various possible boundary conditions corresponding to different M-theory objects. The results are then obtained for the ABJM theory also. Finally, in Chapter 4 we review the Lorentzian 3-algebra in three dimensions and its description of maximally supersymmetric D2-brane theory. This allows us to apply the reduction to the flux terms of the M2-brane theory to obtain the flux modified D2-brane theory. Once this is obtained, the D2-D4-brane system is then considered where we find a new fuzzy funnel solution.

In Part II, we begin by reviewing some basic concepts of matrix models and noncommutative geometry in Chapter 5. We discuss the motivations for generalising such a noncommutative geometry to 3-dimensions. In Chapter 6, we present a proposal for a new type of quantum geometry called the quantum Nambu geometry (QNG) before we describe its origins in the matrix model of D1-Strings in a RR 3-form flux. We demonstrate that there is a large flux double scaling limit which admits the QNG as a solution. We then construct large flux matrix models for Type \IIA, Type \IIB and M-theory. The D4-brane matrix model is then obtained as a result of the D1-Strings expanding over the QNG and then this is generalised to the M5-brane theory by the recent proposal of M5-branes on $S^1$ being equivalent to 5D SYM. A key feature of the QNG is that the 3-form field strength of the M5-branes is constructed from 1-forms instead of the 2-form $B_2$. In Chapter 7, we construct representations of the QNG. The first example is for finite $N$ representations where the Nambu bracket is just reduced to a statement about Lie algebras, these were constructed by Nambu. In the large $N$ limit we find two examples of an infinite dimensional representation of the QNG. The first is a generalisation of the Heisenberg algebra, i.e. we have `raising and lowering' operators which are constructed out of Hermitian operators. The second representation is where the operators can be complex but are unitarily related. In both cases the representations have three degrees of freedom.

\section{M2-branes}
In the previous section we introduced the M2-brane, the worldvolume theory for M2-branes was found recently by Bagger and Lambert \cite{BL1, BL2, BL3} and independently by Gustavsson \cite{Gut}. For a general review of the recent developments in the subject of membranes see \cite{lambrev1,lambrev2}, for a review on M-theory before this see \cite{berman}. The Bagger-Lambert-Gustavsson model (BLG) admits maximal $\cN=8$ supersymmetry but it has been shown \cite{papa1,gaunt} that the theory in fact only describes a pair of M2-branes in a certain orbifold. One year later Aharony, Bergman, Jafferis and Maldacena (ABJM) wrote down a theory of $N$ M2-branes but the supersymmetry was reduced to $\cN=6$ \cite{ABJM}. The entropy scales like $N^{3/2}$ for $N$ M2-branes, this is quite different to the usual $N^2$ scaling we are used to from D-branes \cite{kleb-entropy}. See Chapter 8 for further discussions on this. We now explain how the BLG model was constructed and will look at some applications.
\subsection{BLG Theory}

The motivation behind the BLG theory was to obtain the gauge symmetry and supersymmetry of multiple M2-branes with maximal $\cN=8$ supersymmetry. The key to writing down the gauge theory of multiple M2-branes relies on the use of a 3-bracket structure. The idea of a 3-bracket came from a BPS equation proposed by Basu and Harvey \cite{BH}, the Basu-Harvey equation is an M-theory BPS equation for multiple coincident M2-branes ending on a single M5-brane.

\subsubsection{Basu-Harvey Equation}
To understand what the Basu-Harvey equation describes we first go to the string theory analogue known as a Nahm equation, see \cite{nahm2} for a review of it within string theory. This is analogous with the M-theory Basu-Harvey equation as we can perform a reduction (via dimensional reduction and a T-duality) to obtain the Nahm equation which describes multiple coincident D1-strings ending ending on a D3-brane
\be
\label{nahm}
\frac{dX^i}{ds} = \frac{i}{2}\ve^{ijk}[X^j,X^k],
\ee
where $i,j,k=2,3,4$ are the transverse indices to the D1-strings along the worldvolume of the D3-branes. Here $s = x^1$ is the distance along the spatial coordinate of the D1-strings and so can be thought of as the distance between the D1's and the D3-brane in the fuzzy funnel setup which will become clear shortly. The Nahm equation is used in the study of monopoles, the D1-strings can be thought of as monopoles on the D3-brane. To see this let us consider the solution
\be
\label{nahmsol}
X^i = \frac{1}{2s}\t^i,
\ee
where $\t^i$ are the Lie algebra generators of SU(2) satisfying 
\be
[\t^i,\t^j] = 2i \ve^{ijk}\t^k
\ee
and we take $\t^i$ to be in the $N$-dimensional irreducible representation such that its quadratic Casimir is given by
\be
C= \sum_{i=2,3,4} \t^i\t^i = N^2-1.
\ee
Now we can find the radius of the fuzzy funnel solution
\be
R = \sqrt{\sum_i (X^i)^2} = \frac{\sqrt{N^2-1}}{2s},
\ee
as we can see in order for the radius to blow up we need to have a very small $s$. 

\begin{figure}[htb]
\centering
\includegraphics[width=0.8\textwidth]{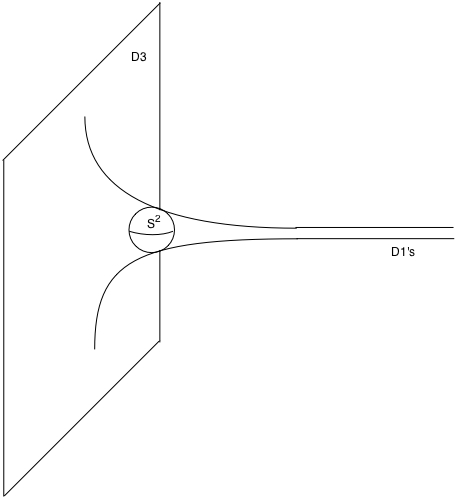}
\caption{Multiple coincident D1-strings blowing up into an Abelian D3-brane at spatial infinity via a fuzzy $S^2$.}
\end{figure}

So to summarise, what we have for the D1-D3 system is a set of D1-strings blowing up into a D3-brane at infinity by a fuzzy funnel which can be thought of as fuzzy spheres $S^2$ giving a round sphere in the limit $s\rightarrow 0$.

\begin{figure}[htb]
\centering
\includegraphics[width=0.8\textwidth]{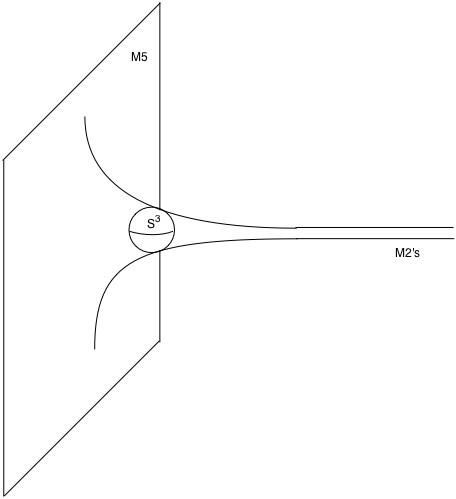}
\caption{Multiple coincident M2-branes blowing up into an Abelian M5-brane at spatial infinity via a fuzzy $S^3$.}
\label{m2m5fig}
\end{figure}

We will now look at the M-theory generalisation of the Nahm equation and its interpretation. The key to the generalisation is the use of a 3-bracket instead of a Lie-algebra valued commutator due to the enhancement of the fuzzy $S^2$ to a fuzzy $S^3$ for a system with relative dimension 3. This became concrete when considering the BLG theory of multiple coincident M2-branes, the construction by Basu and Harvey was originally thought to be quite ad-hoc and it was not clear where the origin of the 3-bracket structure came from. The Basu-Harvey equation describes a system of multiple coincident M2-branes ending on an Abelian M5-brane as in Figure \ref{m2m5fig}. We shall see in this section that the idea of a 3-bracket was key to providing the means to write down the BLG theory and therefore to provide us with an origin for the Basu-Harvey equation. It reads
\be
\label{bhorig}
\frac{dX^i}{ds} + \frac{\l}{4!}\ve^{ijkl}[G_5,X^j,X^k,X^l]=0,
\ee
where $\l = M^3/8\pi\sqrt{2N}$ is a constant, $s =x^2$ and $i,j,k,l= 3,4,5,6.$ 
The matrix $G_5$ is determined from the representation of the $spin(4) = SU(2)\times SU(2)$\footnote{The spin group $Spin(n)$ is the double covering of the Special Orthogonal group $SO(n)$ and the $Spin(n)$ group and $spin(n)$ algebra share the same dimension as their Special Orthogonal counterparts.} algebra of the fuzzy $S^3$ and satisfies $G_5^2 =1$. The bracket in \eq{bhorig} is multilinear and antisymmetric, in particular it is trilinear in the scalar fields $X^i$. The M2-branes have been shown to be equivalent to a higher dimensional analogue of the D1-string monopole interpretation of the D1-D3 system known as the self-dual string solitons on the M5-brane worldvolume \cite{sdstring, CS2}. 

\subsubsection{Gauge theories and 3-algebras}
Now we understand where the 3-bracket originated in the literature, we can study the BLG theory and why it was so revolutionary to use such a structure. The M2-brane is the strong coupling limit of the D2-brane from \IIA theory, the Lagrangian for such a D2-brane is given by a three dimensional $\cN=8$ maximally supersymmetric Yang-Mills theory with an $SO(7)$ R-symmetry and a $U(N)$ gauge symmetry. For the M2-brane theory we expect a maximally supersymmetric theory in three dimensions with an $SO(8)$ R-symmetry. In \cite{BL1,BL2,BL3}, the authors proposed that one could use a new type of algebra for the gauge symmetry for the multiple M2-brane theory, these are named 3-algebras denoted by $\cA$ with a Lie 3-bracket.
%

Now let us discuss the gauge symmetry of M2-branes using the 3-algebra, but before we do this it is helpful to recall the ordinary Yang-Mills gauge symmetry from D-branes and then build up to the concept of a 3-algebra, the global transformation is given by
\be
\d X = [\a,X]
\ee
where $\a, X \in SU(N)$. A derivation on the commutator gives the Jacobi identity, which can then be written in terms of structure constants $f^{ab}{}_c$ as defined by $X = X_a T^a$ and $[T^a,T^b] = f^{ab}{}_cT^c.$ For the M2-branes, we find that the supersymmetric closure\footnote{We shall make this clear when we examine the supersymmetry transformations for the BLG theory, the motivation here is purely for historic reasons.} on the fields gives a local symmetry proportional to the 3-bracket \cite{BL1}. The global version of this symmetry is given by
\be
\label{gsym1}
\d X = [\a,\b,X]
\ee
where $\a,\b \in \cA$. Imposing the derivation property on the 3-bracket then gives us
\be
\d [X,Y,Z] = [\d X, Y, Z] + [X, \d Y, Z] + [X, Y, \d Z],
\ee
this is known as the `fundamental identity' in the literature and is a generalisation of the Jacobi identity for 3-brackets. More explicitly the fundamental identity reads
\be
\label{fund}
[\a,\b,[X,Y,Z]]= [[\a,\b, X], Y, Z] + [X, [\a,\b, Y], Z] + [X, Y, [\a,\b, Z]],
\ee
where we will now try to find an analogous relation to the Jacobi identity for the fundamental identity in terms of a structure constant.

We can expand the fields $X$ in terms of a Hermitian basis $T^a$ of $\cA$ as $X = X_a T^a$, where $a=1,..., \dim \cA$. Now we can introduce the equivalent Lie algebra relation for a 3-bracket known as the Lie 3-algebra
\be
[T^a, T^b, T^c] = f^{abc}{}_d T^d,
\ee
also we introduce the natural metric from the trace $\Tr$:
\be
h^{ab} = \Tr(T^a, T^b),
\ee
here we assume that the metric is Euclidean\footnote{The signature could be taken to be  Lorentzian, this will be explored further in detail in Chapter 4. It turns out that the negative modes correspond to a ghost contribution to the action which is shown to un-couple from the D2-brane theory.} in signature and positive definite $h^{ab}=\d^{ab}$. This metric can be used to raise and lower gauge indices. Now that we have a trace, we can construct a property similar to the invariance of the trace for YM gauge theories for a 3-algebra as
\be
\label{3fund}
\Tr([T^a,T^b,T^c], T^d) = -\Tr(T^a,[T^b,T^c,T^d]),
\ee 
or equivalently
\be
f^{abcd} = -f^{dbca},
\ee
note that the 3-algebra analogue of the invariance of the trace property has an important sign.
Using the invariance of the trace \eq{3fund} and that the 3-bracket is anti-symmetric gives us 
\be
f^{abcd} = f^{[abcd]},
\ee
thus the fundamental identity \eq{fund} can be written as 
\be
\label{fi}
f^{efg}{}_d f^{abc}{}_g = f^{efa}{}_g f^{bcg}{}_d + f^{efb}{}_g f^{cag}{}_d + f^{efc}{}_g f^{abg}{}_d.
\ee

We now wish to study the gauge symmetries of the BLG theory, to begin with we note that the global symmetry \eq{gsym1} can be written in the form
\be
\d X_d = f^{abc}{}_d\a_a\b_b X_c,
\ee
this can be generalised to the global transformation
\be
\label{gsym2}
\d X_d = f^{abc}{}_d \L_{ab} X_c.
\ee

We can now introduce a covariant derivative and gauge the symmetry \eq{gsym2}, the covariant derivative $D_\m$ is defined such that $\d (D_\m X) = D_\m(\d X) + (\d D_\m) X$. So then by promoting the global symmetry \eq{gsym2} to a local gauge symmetry, we define
\be
\d X_d = f^{abc}{}_d \L_{ab} X_c = \Lt^c{}_d X_c.
\ee 
So the covariant derivative is given explicitly by
\be
\label{cderiv}
(D_\m X)_a = \del_\m X_a -\At_\m{}^b{}_aX_b,
\ee
where $\At_\m{}^b{}_a \equiv f^{cdb}{}_a A_{\m cd}$ is the BLG non-Abelian gauge field living in the generalisation of the adjoint representation for a 3-algebra $\cA$ where
\bea
\d \At_\m{}^b{}_a &=& \del_\m \Lt{}^b{}_a - \Lt{}^b{}_c \At_\m{}^c{}_a + \At_\m{}^b{}_c \Lt^c{}_a \nn\\
&\equiv& D_\m \Lt^b{}_a.
\eea
This transforms the same way as a Lie algebra valued gauge field would, so we now define the field strength of the gauge field by
\be
([D_\m, D_\n] X)_a = \Ft_{\m\n}{}^b{}_a X_b.
\ee
Using \eq{cderiv} we obtain
\be
\Ft_{\m\n}{}^b{}_a = \del_\n\At_\m{}^b{}_a - \del_\m \At_\n{}^b{}_a -\At_\m{}^b{}_c\At_\n{}^c{}_a + \At_\n{}^b{}_c\At_\m{}^c{}_a,
\ee
which satisfies the Bianchi identity 
\be
D_{[\m} \Ft_{\n\l]}{}^b{}_a =0.
\ee
We have seen that the 3-algebra valued fields have similar gauge transformations as their Lie algebra counterparts, however the gauge field $\At_\m{}^b{}_a$ has two algebraic indices.

\subsubsection{Supersymmetry}
We now wish to supersymmetrise\footnote{Originally in \cite{BL1} the authors constructed supersymmetry transformations of only the scalars and fermions for the ungauged theory. This lead to the concept of using a 3-bracket based on a nonassociative 3-algebra being the analogy of a commutator for the uplift from a D2-brane theory to the M2-brane. } the gauged theory and construct an action; it should have superconformal invariance in three dimensions with 16 supersymmetries and an $SO(8)$ R-symmetry on the eight transverse scalar fields. The natural supersymmetry transformations to use would be an uplift of the D2-brane supersymmetry transformations:
\bea
\label{blsusy}
\d X^I_a &=& i\eb \G^I \Psi_a\nn \\
\d \Psi_a &=& D_\m X^I_a \G^\m\G^I \e -\frac{1}{3!}X_b^IX_c^JX_d^K f^{bcd}{}_a \G^{IJK}\e\nn \\
\d \At_\m{}^b{}_a &=& i\eb\G_\m\G^IX^I_c\Psi_d f^{cdb}{}_a.
\eea
Here $\m = 0,1,2$ and $I=3,...,10$ and $\Psi$ is a 16 component Majorana spinor that satisfies
\be
\G^{012}\Psi = -\Psi,
\ee
also the supersymmetry parameter $\e$ satisfies
\be
\G^{012}\e=\e.
\ee

We can make this algebra close on-shell after imposing certain equations of motion. These arise by considering the off-shell closure relations which are of the form of translations and gauge transformations plus equations of motion. The scalars close on-shell:
\be
\label{scalarclose}
[\d_1,\d_2]X^I_a = v^\m D_\m X^I_a +\Lt^b{}_a X^I_b,
\ee
where
\bea
\label{translation}
v^\m &=&-2i\eb_2\G^\m\e_1 \\ 
\label{lambda}\Lt^b{}_a &=& -i\eb_2\G_{JK}\e_1 X^J_c X^K_d f^{cdb}{}_a. 
\eea
For the fermions we use the Fierz identity in \eq{fierzblg} to obtain the off-shell closure:
\bea
[\d_1,\d_2]\Psi_a =&& v^\m D_\m\Psi_a +\Lt^b{}_a \Psi_b \nn\\
&&+i\eb_2\G_\n\e_1\G^\n\left(\G^\m D_\m\Psi_a +\frac{1}{2}\G^{IJ}X^I_cX^J_d\Psi_b f^{cdb}{}_a\right) \nn\\
&& +\frac{i}{4}\eb_2\G^{KL}\e_1\G^{KL}\left(\G^\m D_\m\Psi_a +\frac{1}{2}\G^{IJ}X^I_cX^J_d\Psi_b f^{cdb}{}_a\right).
\eea
We must impose the equations of motion
\be
\label{feom}
\G^\m D_\m\Psi_a +\frac{1}{2}\G^{IJ}X^I_cX^J_d\Psi_b f^{cdb}{}_a=0,
\ee
then the fermions close on-shell:
\be
[\d_1,\d_2]\Psi_a = v^\m D_\m\Psi_a +\Lt^b{}_a \Psi_b
\ee
We are left with the gauge field $\At_\m$. After using the fundamental identity to eliminate the $\G^{(5)}$ term, we obtain the off-shell closure:
\bea
[\d_1,\d_2]\At_\m{}^b{}_a = &&2i\eb_2\G^\n\e_1\ve_{\m\n\l}(X^I_c D^\l X^I_d +\frac{i}{2}\Psib_c\G^\l\Psi_d)f^{cdb}{}_a \nn\\
&&-2i\eb_2\G^{IJ}\e_1 X^I_c D_\m X^I_d f^{cdb}{}_a.
\eea
This can be made to close on-shell by imposing the equation of motion
\be
\label{gaugeeom}
\Ft_{\m\n}{}^a{}_b +\ve_{\m\n\l}(X^I_c D^\l X^I_d +\frac{i}{2}\Psib_c\G^\l\Psi_d)f^{cdb}{}_a =0,
\ee
then we have the closure:
\be
\label{gtransA}
[\d_1,\d_2]\At_\m{}^b{}_a = v^\n\Ft_{\m\n}{}^b{}_a + D_\m \Lt^b{}_a.
\ee
The gauge field is non-dynamical as it has no propagating degrees of freedom.

Finally we derive the bosonic equation of motion, this can be found by taking the supersymmetric variation of the fermionic equation of motion \eq{feom} and imposing \eq{gaugeeom}:
\be
\label{seom}
D^2 X^I_a -\frac{i}{2}\Psib_c\G^{IJ}X^J_d \Psi_b f^{cdb}{}_a - V'(X^I_a),
\ee
where $V(X^I_a)$ is a sextic potential in the scalar fields $X^I$
\be
V = \frac{1}{2\cdot 3!}\Tr([X^I,X^J,X^K],[X^I,X^J,X^K]).
\ee

Now that we have verified that the superalgebra closes on-shell and that the equations of motion respect the supersymmetries \eq{blsusy}, we can write down an action which has scalar and fermionic kinetic terms and an interaction term for these fields. Adding in a Chern-Simons term will allow us to have a non-propagating gauge field\footnote{In 2004 it was shown in \cite{JS} that supersymmetric Chern-Simons theories with non-propagating degrees of freedom can be constructed with $\cN=1,2$ supersymmetry but not $\cN=8$. The author conjectured that there was no Lagrangian description for such an $\cN=8$ theory and we shall see that this is no longer the case.}, the bosonic potential will be given by $V$. Understanding where the $\Tr$ products are, we have the BLG action:
\bea
\label{blgaction}
S = \int \mathrm{d}^3x\, \Big(-\frac{1}{2}D_\m X^{aI}D^\m X^I_a + \frac{i}{2}\Psib^a\G^\m D_\m \Psi_a+\frac{i}{4}\Psib_b\G^{IJ}X^I_cX^J_d\Psi_a f^{abcd} \nn\\
\hphantom{\int \mathrm{d}^3x\, \Big(-}+\frac{1}{2}\ve^{\m\n\l}(f^{abcd}A_{\m ab}\del_\n A_{\l cd} +\frac{2}{3}f^{cda}{}_g f^{efgb}A_{\m ab}A_{\n cd}A_{\l ef}) -V\Big).
\eea
The equations of motion from the above action are precisely equations \eq{feom}, \eq{gaugeeom} and \eq{seom} and the action is invariant under the supersymmetry transformations \eq{blsusy} up to a total derivative. The action contains no free parameters as expected in a theory of M2-branes, the structure constants $f^{abc}{}_d$ can be rescaled but they must remain quantised due to the presence of the Chern-Simons term. The gauge fields $A_\m$ that appear in the Chern-Simons term are not the physical fields $\At_\m$, they come as part of a twisted Chern-Simons term which remains invariant under shifts of $A_\m$ that leave $\At_\m{}^b{}_a=A_{\m cd}f^{cdb}{}_a$ invariant under the supersymmetry transformations \eq{blsusy}.

\subsubsection{Quantisation and Vacuum Moduli Space}

It was shown in \cite{papa1,gaunt} that there is a solution\footnote{It was also shown that this solution is in fact unique.} to the fundamental identity with four generators $T^a,\, a=1,...,4$ and the nonassociative algebra is denoted by $\cA_4$ in the literature. The gauge algebra is given by $SO(4)$ and we have the invariant four-tensor:
\be
f^{abcd} \propto \ve^{abcd},
\ee
where generators are normalised as $\Tr(T^a,T^b) \propto \d^{ab}$.
This is a rather restrictive solution as we would like the gauge theory to describe $N$ M2-branes, what we have is an $SO(4) = SU(2) \times SU(2)$ gauge algebra and when the fields are $SU(2)\times SU(2)$ valued it is called a bifundamental gauge theory. It turns out that the BLG model describes no more than two M2-branes \cite{papa1,gaunt} when we consider the Euclidean metric on the 3-algebra, this is a result of the fact that there is no way to recover the $U(N)$ gauge symmetry of D2-branes from the Lie 3-algebra of M2-branes for $N > 2$.

The quantisation of the structure constant $f^{abcd}$ will now be considered. Classically, the action structure constants can be rescaled and we can preserve the equations of motion. In a quantum theory the Chern-Simons term must be a quantised quantity, so we expect the $f^{abcd}$ to behave this way too. In \cite{csquant} a path integral quantisation of the Chern-Simons action was considered, it was shown that for a well-defined path integral we require the coefficient of the Chern-Simons term to be $k/4\p$ where $k\in \mathbb{Z}$. The quantity $k$ is known as the Chern-Simons level\footnote{The BLG theory is only valid for $k=1,2$ if we want an M-theory interpretation, this will be seen with a moduli space argument later.
} and is quantised for a compact gauge group. The solution for the structure constant is
\be
f^{abcd} = \frac{2\p}{k}\ve^{abcd},
\ee
where $a=1,...,4$ and $k\in \mathbb{Z}$.

The vacuum moduli spaces for the BLG theory were first considered in \cite{BL3} and then in greater detail in the two overlapping papers \cite{mod1} and \cite{mod2}. The moduli are found by requiring that all the BLG fields $\del_\m X^I = \Psi = \At_\m =0$ except for the scalars $X^I$ vanish and
\be
[X^I, X^J, X^K]=0
\ee
to satisfy the equations of motion. The moduli space for $k=1$ was found to be 
\be
\cM_1 \equiv (\mathbb{R}^8/\mathbb{Z}_2) \times (\mathbb{R}^8/\mathbb{Z}_2),
\ee
while the $k=2$ moduli space is given by
\be
\cM_2 \equiv \frac{(\mathbb{R}^8/\mathbb{Z}_2) \times (\mathbb{R}^8/\mathbb{Z}_2)}{\mathbb{Z}_2}.
\ee

\subsubsection{Higgsing the Theory}

We now consider a reduction of the multiple\footnote{In the case of a single M2-brane one may apply an abelian dualisation on one of the eight scalar fields to obtain a gauge field.} M2-brane theory to give us a nonabelian D2-brane theory. There are two approaches that we shall consider in this thesis; the first is using a novel Higgs mechanism for the 3-algebra $\cA_4$ by Mukhi and Papageorgakis \cite{MP} and the second is by considering a Lorentzian 3-algebra \cite{Ho1,gomis3,v-lor,schwarz}, the latter case will be discussed in Chapter 4 where we will collate the literature on the subject and make it consistent. There has been work on more general 3-algebras for the $\cA$-type in \cite{MP}.

The first order D2-brane theory is a maximally supersymmetric Yang-Mills theory in $2+1$ dimensions, it should have seven scalar fields with an $SO(7)$ R-symmetry and a single gauge field contributing to the Yang-Mills piece of the action. The M2-brane theory is expected to be the conformally invariant IR fixed point of the D2-brane theory, this translates to taking the limit $g_{YM}\to \infty$.

A Yang-Mills coupling constant can be obtained by Higgsing a scalar field, say $X^{4(10)}$, along the M-theory direction. The 4 refers to the fourth gauge index of $\cA_4$, more generally we could denote this as $\phi \equiv \dim \cA$. Now the scalar fields have dimension $1/2$ while the M-theory radius $R$ has dimension $-1$, therefore the VEV must be
\be
\la X^{\phi(10)}\ra = \frac{R}{l_p^{3/2}}.
\ee
Under the compactification to \IIA theory we obtain
\be
\frac{R}{l_p^{3/2}} = \sqrt{\frac{g_s}{l_s}} \equiv g_{YM},
\ee
where we have the string coupling and length $g_s$ and $l_s$ respectively. The coupling constant for Yang-Mills theory in $2+1$ dimensions is dimensionful. Giving this field a VEV does not break any supersymmetry so we still have an $\cN=8$ supersymmetric gauge theory, we will now make some more definitions and then substitute into the BLG theory to find out whether we do indeed obtain a super Yang-Mills theory.

The next thing to consider is the gauge field; in the BLG theory it has two gauge indices $A_\m{}^b{}_a$, so we must reduce the gauge field in such a way that we will obtain a single gauge index. It can be done in the following way involving two gauge fields
\bea
A_\m^A &\equiv& A_\m^{A\phi} \\
B_\m^A &\equiv& \frac{1}{2}\ve^{A}{}_{BC}A_\m^{BC},
\eea
where $A,B \in \{1,2,3\}$ and $a,b = \{1,2,3,\phi\}$. It turns out that the $B_\m$ gauge field will obtain a mass term and be integrated out from the theory, so we define a covariant derivative in terms of the gauge field $A_\m^A$ only and its associated field strength
\bea
D_\m' X^{A(I)} &=& \del_\m X^{A(I)} -2\ve^A{}_{BC}A_\m{}^B X^{C(I)}, \\
F'_{\m\n}{}^A &=& \del_\m A_\n^A - \del_\n A_\m^A -2\ve^A{}_{BC}A_\m^B A_\n^C.
\eea
If we examine the bosonic part of the action, using the above substitutions we obtain
\be
\cL = -2g_{YM}^2 B_\m^A B^{\m}_A -2g_{YM} B_\m^A D'^\m X_A^{(10)} + 2\ve^{\m\n\l}B_\m^A F'_{\n\l A} +\text{higher order}.
\ee
The higher order terms are inversely proportional to the Yang-Mills coupling and so vanish in the IR fixed point limit. We can integrate $B_\m$ out by using the equation of motion
\be
B_\m^A = \frac{1}{2g_{YM}^2}\ve_\m{}^{\n\l} F'_{\n\l}{}^A - \frac{1}{2g_{YM}}D'_\m X^{A(10)},
\ee
then we see that the Lagrangian splits into a coupled $SU(2)$ action of SYM in 2+1 dimensions and a decoupled $U(1)$ abelian multiplet, the scalar $X^{(10)}$ is dualised into a gauge field for this multiplet and so we indeed have an SO(7) R-symmetry for the D2-brane theory. The Lagrangian for the Higgsed theory is given by
\be
\cL = \cL_{\text{decoupled}} + \cL_{\text{coupled}} \\
\ee
with
\be
\cL_{\text{decoupled}} = -\frac{1}{2} \del_\m X^{\phi(I)}\del^\m X^{(I)}_\phi +\frac{i}{2}\Psib^\phi \G^\m\del_\m \Psi_\phi.
\ee
The coupled action is allowed a field redefinition, namely a rescaling of $1/g_{YM}$, so we obtain
\be
\cL_{\text{coupled}} = \frac{1}{g_{YM}^2}\cL_0 + \frac{1}{g_{YM}^3}\cL_1 +\cO(g_{YM}^{-4}).
\ee
The action $\cL_0$ is the $\cN=8$ maximally supersymmetric Yang-Mills theory in $2+1$ dimensions given by
\bea
\cL_0 = &&-\frac{1}{4}F_{\m\n A}F^{\m\n A} -\frac{1}{2}D_\m X^{A(i)}D^\m X^{(i)}_A +\frac{1}{4}\big(\ve_{ABC}X^{A(i)}X^{B(j)}\big)\big(\ve_{DE}{}^C X^{D(i)}X^{E(j)}\big) \nn\\
&&+\frac{i}{2}\Psib^A \G^\m D_\m \Psi_A +\frac{i}{2}\ve_{ABC}\Psib^A \G^i X^{B(i)}\Psi^C.
\eea
We note that the action has higher order couplings, this will not be a feature of the Lorentzian D2-brane theory.

\subsubsection{Bifundamental gauge theory}
Before we discuss the ABJM theory, it will be useful to review \cite{ramsbif} where Van Raamsdonk recast the BLG theory gauge group in terms of $SO(4)= SU(2) \times SU(2)$. This allows us to write the field content in a bifundamental representation of $SU(2) \times SU(2)$ which is a special case of the ABJM theory where the gauge group is $U(N) \times U(N)$. 

Let us begin by noting that fundamental fields in the $\cA_4$ 3-algebra, for example\footnote{We can repeat this for the spinor.} $X^I$, are in a vector {\bf 4} of $SO(4)$.
\be
X^I = \begin{pmatrix}
x_1^I \\
x_2^I \\
x_3^I \\
x_4^I
\end{pmatrix}
\ee 
This can be decomposed into a $({\bf 2},{\bf 2})$ representation of $SU(2)\times SU(2)$, i.e. the field is now in the bifundamental representation of $SU(2)\times SU(2)$. The fields obey a reality condition
\be
\label{reality}
X_{\a\dot{\b}}^I = \ve_{\a\b}\ve_{\dot{\b}\dot{\a}}(X^I\dag)^{\dot{\a}\b}.
\ee
Written in terms of a Pauli basis\footnote{The Pauli matrices $\s_i$ are normalised such that $\Tr(\s_i\s_j)=2\d_{ij}$.} of $SU(2)$, the fields have the form
\be
X^I = \frac{1}{2}(x_4^I \id +ix_i^I \s^i) = \frac{1}{2}\begin{pmatrix}
x_4^I + ix_3^I & x_2^I +ix_1^I \\
-x_2^I +ix_1^I & x_4^I - ix_3^I
\end{pmatrix}.
\ee

For the gauge field $A_{\m ab}$, we must decompose it into the sum of its self-dual and anti-self-dual parts in order to get an adjoint gauge field for each gauge group $SU(2)$
\be
A_{\m ab} = -\frac{k}{2\p}(A_{\m ab}^+ +A_{\m ab}^-)
\ee
where
\be
A_{\m ab}^\pm = \pm\frac{1}{2}\ve_{ab}{}^{cd}A_{\m cd}^\pm.
\ee
By defining 
\bea
A_\m = A^+_{\m4i}\s_i, \\
\Ah_\m = A^-_{\m4i}\s_i,
\eea
we can write the bifundamental covariant derivative
\be
D_\m X^I = \del_\m X^I +iA_\m X^I -iX^I \Ah_\m.
\ee
The BLG Lagrangian in terms of bifundamental matter and adjoint gauge fields is given by
\bea
\label{bifund}
\cL &=&\tr(-(D_\m X^I)\dag D^\m X^I +i\Psib\dag\G^\m D_\m \Psi) \nn\\
&&+\tr\Big(-\frac{4\p i}{3k}\Psib\dag \G^{IJ}(X^I X^J\dag \Psi +X^J\Psi\dag X^I + \Psi X^I\dag X^J) \nn\\
&&\hphantom{\tr\Big(-}-\frac{32\p^2}{3k^2}X^{[I}X^J\dag X^{K]}X^K\dag X^J X^I\dag  \Big)\nn\\
&&+\frac{k}{4\pi}\ve^{\m\n\l}\Tr(A_\m \del_\n A_\l + \frac{2i}{3}A_\m A_\n A_\l) -\frac{k}{4\pi}\ve^{\m\n\l}\tr(\Ah_\m \del_\n \Ah_\l + \frac{2i}{3}\Ah_\m \Ah_\n \Ah_\l). \nn\\ 
\eea
The above action is invariant under a new set of supersymmetry rules which are obtained by applying the decomposition rules above to the original BLG supersymmetry transformations \eq{blsusy}
\bea
\d X^I &=&i\eb\G^I\Psi, \\
\d \Psi &= &D_\m X^I \G^\m\G^I\e +\frac{4\p}{3}X^I X^J\dag X^K \G^{IJK}\e, \\
\d A_\m &=& \frac{2\p}{k}\eb\G_\m\G^I(X^I\Psi\dag -\Psi X^I\dag), \\
\d \Ah_\m &=& \frac{2\p}{k}\eb\G_\m\G^I(\Psi\dag X^I - X^I\dag\Psi).
\eea
The Chern-Simons terms in the Lagrangian \eq{bifund} have opposite signs, so we say that the gauge theory is of level $(k,-k)$. Note that the single twisted Chern-Simons term in the original BLG theory \eq{blgaction} decomposes into this nice form of two ordinary Chern-Simons terms with opposite level for the gauge group $SO(4)$.

\subsection{ABJM Theory}
In \cite{ABJM} the authors constructed a three dimensional superconformal Chern-Simons-Matter theory with $\cN=6$ supersymmetry\footnote{For Chern-Simons level $k=1,2$ and gauge group $SU(2)\times SU(2)$, the $\cN=6$ theory becomes enhanced to the full $\cN=8$ theory. This is because the $SU(2)$ R-symmetry and the global $SU(4)$ flavour symmetry can be combined to give an $SO(8)$ R-symmetry. 
For higher rank gauge groups $U(N)\times U(N)$ it was proposed in \cite{mono-op} that for Chern-Simons level $k=1,2$, the full $\cN=8$ theory can be obtained from the ABJM theory by introducing monopole operators. This is still unclear in the literature and so we will not discuss it here.}, the gauge group of the theory is given by a quiver gauge group $U(N)\times U(N)$ and is argued to describe the low energy action for $N$ M2-branes probing an orbifold with a $\mathbb{C}^4/\mathbb{Z}_k$ singularity. The supergravity background dual to the three dimensional CFT was shown to be $AdS_4 \times S^7/\mathbb{Z}_k$, note that this geometry is different to that found in eleven dimensional supergravity in the sense that the space has a $\mathbb{Z}_k$ orbifold structure. In the ABJM theory the parameters $N$ and $k\in \mathbb{Z}$ are free and do not suffer the same restrictions as the BLG theory, as such one is able to construct a 't Hooft coupling $\l \equiv N/k$ for a reduction to the Type \IIA theory on $AdS_4 \times \mathbb{CP}^3$. We will begin by constructing the BLG theory in superspace without any mention of a 3-bracket structure. Then we will show how we can modify the construction to generalise the gauge group. Finally we will discuss how to write the ABJM theory in terms of a 3-bracket, this 3-bracket structure is not exactly the same as in the BLG theory.

\subsubsection{Superspace formalism of BLG theory}

In the previous section we constructed a bifundamental formalism for the BLG theory with $SU(2)\times SU(2)$ gauge group, this will allow us to write the CFT in terms of $\cN=2$ superspace and then solve constraints to show that we do indeed get back the bifundamental Lagrangian as in \eq{bifund}. We do this because it is more natural to describe the ABJM theories in superspace, the component field actions can then be easily derived from these and in the case of BLG theory be shown to be exactly the same. 

In \cite{klebanov} the scalar fields from the BLG theory were put into an $SU(4)$ representation by setting
\be
Z^A = X^A +i X^{A+4},\quad A=1,...,4.
\ee
So only the subgroup $U(1)_R \times SU(4)$ of the $SO(8)$ R-symmetry is now manifest, the $SU(4)$ symmetry acts on the $A$ index. In the bifundamental theory we have scalars (and fermions) associated to the fundamental representation of each of the $SU(2)$'s associated with gauge symmetry, so we promote the scalars to chiral superfields $\cZ^A$ and anti-chiral superfields $\bar{\cZ}^A$ which transform under the fundamental and anti-fundamental representation respectively. Suppressing the $SU(4)$ indices for aesthetics, we have
\bea
\cZ &=& Z(y) + \sqrt{2}\th\z(y) +\th\th F(y), \\
\bar{\cZ} &=& \Zb(\yb) - \sqrt{2}\thb\z\dag(\yb) -\thbb \Fb\dag(\yb),
\eea
where we employ the (anti)-chiral coordinates as outlined in the Appendices. There are two conjugations we can perform on the (anti)-chrial superfields, (we write the example for the scalars but they hold for the fermions and auxiliary fields too), the first acts on the $SU(2)$ from the gauge group and the second is the $SU(4)$ from the R-symmetry
\bea
Z^{\ddag A} &\equiv& X\dag{}^A +iX\dag{}^{A+4}, \\
\Zb_A &\equiv& X^A -iX^{A+4}.
\eea
For the case of gauge group $SU(2)\times SU(2)$, we are able to identify two unique conjugations due to the reality condition \eq{reality}, in general this is not possible and so the only conjugation one can write down is given by the hermitian conjugate $Z\dag =\Zb^\ddag$. Hence only for $SU(2)\times SU(2)$ can we invert back to the original real scalars 
\be
X^A = \frac{1}{2}(Z^A + \Zb_A),
\ee
and 
\be
X^{A+4} = \frac{1}{2i}(Z^A - \Zb_A).
\ee

We now need to examine the vector supermultiplet in three dimensions as this will contribute to the Chern-Simons action and the matter piece too. There are two vector supermultiplets $\cV,\cVh$ due to the quiver gauge group $SU(2)\times SU(2)$, with a vector $A_\m$ and $\Ah_\m$ respectively at each node. Let us write the vector supermultiplet in the Wess-Zumino gauge
\bea
\label{vsm}
\cV = 2i\th\thb\t +2\th\g^\m\thb A_\m +\sqrt{2}i\th\th\ol{\th\l} - \sqrt{2}i\thbb\th\l +\th\th \thbb D,
\eea
similarly with hatted components for $\cVh$. Note that $\l, \lamb, \t, D$ are auxiliary fields and will be integrated out later. 
The Chern-Simons-matter action can then be constructed by the following terms
\be
\label{Kaction}
S_{\mathrm{BL}} = S_{\mathrm{CS}} + S_{\mathrm{mat}} + S_{\mathrm{pot}},
\ee
where
\bea
S_{\mathrm{CS}} &=& -iK \int \mathrm{d}^3x\,\mathrm{d}^4\th\int_0^1\mathrm{d}t\, \Tr\cV\Db^\a\left(e^{t\cV}D_\a e^{-t\cV}\right) - \cVh\Db^\a\left(e^{t\cVh}D_\a e^{-t\cVh}\right), \\
S_{\mathrm{mat}} &=& -\int \mathrm{d}^3x\,\mathrm{d}^4\th\, \Tr \cZ_A e^{-\cV} \cZ^A e^{\cVh}, \\
S_{\mathrm{pot}} &=& L \int \mathrm{d}^3x\,\mathrm{d}^2\th\, W(\cZ) + L \int \mathrm{d}^3x\,\mathrm{d}^2\thb\, \Wb(\bar{\cZ}),
\eea
where the superpotential is given by
\bea
\label{blgsuperpot}
W(\cZ) &=& \frac{1}{4!}\ve_{ABCD}\Tr(\cZ^A\cZ^{\dd B}\cZ^C\cZ^{\dd D}), \\ 
\Wb(\bar{\cZ}) &=& \frac{1}{4!}\ve_{ABCD}\Tr(\bar{\cZ}^A\bar{\cZ}^{\dd B}\bar{\cZ}^C\bar{\cZ}^{\dd D}).
\eea
The original component field bifundamental action \eq{bifund} can be obtained by integrating out all the auxiliary fields $\t, \l, \lamb, D, \z, F, \z\dag, \Fb\dag$ and making the following relations
\be
\label{su4enh}
K = \frac{1}{L},\quad L= 4f = \frac{8\p}{k},\quad k \in \mathbb{Z}.
\ee
So the manifest $U(1)_R \times SU(4)$ R-symmetry of the superpotential becomes enhanced to the full $SO(8)$ R-symmetry that we want for the BLG theory.

\subsubsection{A $U(N) \times U(N)$ Gauge Theory}
It is not clear how to generalise the BLG theory to an arbitrary unitary gauge theory with gauge group $U(N)\times U(N)$ for example. The problem lies in the global $SU(4)$ invariance of the BLG superpotential \eq{blgsuperpot} only being gauge invariant for the gauge group $SU(2)\times SU(2)$. In 2008, Aharony, Bergman, Jafferis and Maldacena \cite{ABJM} proposed to give up the manifest $SU(4)$ global symmetry of the superpotential and created the following superpotentials
\bea
\label{abjmsuperpot}
W(\cZ,\cW) &=& \frac{1}{4!}\ve_{AC}\ve^{BD}\Tr(\cZ^A\cW_B \cZ^C \cW_D), \\ 
\Wb(\bar{\cZ},\bar{\cW}) &=& \frac{1}{4!}\ve^{AC}\ve_{BD}\Tr(\bar{\cZ}_A\bar{\cW}^B \bar{\cZ}_C \bar{\cW}^D),
\eea
with an $SU(2)\times SU(2)$ {\it global} symmetry\footnote{Not to be confused with the BLG gauge symmetry.} and $A,B,C,D = 1,2$.
Here the chiral superfields are given by 
\bea
\cZ &=& Z(y) + \sqrt{2}\th\z(y) +\th\th F(y), \\
\bar{\cZ} &=& \Zb(\yb) - \sqrt{2}\thb\z\dag(\yb) -\thbb \Fb\dag(\yb), \\
\cW &=& W(y) + \sqrt{2}\th\o(y) +\th\th G(y), \\
\bar{\cW} &=& \Wb(\yb) - \sqrt{2}\thb\o\dag(\yb) -\thbb \Gb\dag(\yb).
\eea
The fields have been organised into $SU(2)\times SU(2)$ multiplets as
\bea
Z^1 &=& X^1 +i X^5, \\
Z^2 &=& X^2 +i X^6, \\
W_1 &=& X^3\dag +i X^7\dag, \\
W_2 &=& X^4\dag +i X^8\dag.
\eea
The new potential is then given by
\be
S_{\mathrm{pot}} = L \int \mathrm{d}^3x\,\mathrm{d}^2\th\, W(\cZ,\cW) + L \int \mathrm{d}^3x\,\mathrm{d}^2\thb\, \Wb(\bar{\cZ},\bar{\cW}).
\ee
The superpotential \eq{abjmsuperpot} is also invariant under an additional Baryonic $U(1)$ symmetry given by
\bea
\cZ^A &\rightarrow& e^{i\a}\cZ^A,\\
\cW_B &\rightarrow& e^{-i\a}\cW_B,
\eea
this symmetry will be gauged just as in the D3-brane case and will contribute to the gauge symmetry, so it can be combined with the $SU(N)\times SU(N)$ gauge symmetry of the superpotential to give a $U(N)\times U(N)$ gauge theory.
The $SU(N)\times SU(N)$ factor of the gauge symmetry comes from the fact that the superpotential is no longer restricted by the $\ddag$ conjugation of the BLG theory.

We now turn to the R-symmetry which is given by $U(1)_R \times SU(2)\times SU(2)$ thus far, the ABJM theory is given in terms of a $SU(4) = SO(6)$ R-symmetry and we shall show this is indeed the symmetry we have. Under the global $SU(2)\times SU(2)$, the fields $\cZ,\cW$ transform as $({\bf 2},{\bf 1})$ and $({\bf 1},{\bf \bar{2}})$ respectively while for the gauge symmetry $U(N)\times U(N)$ they transform in the $({\bf N},{\bf \bar{N}})$ and $({\bf \bar{N}}, {\bf N})$ respectively. We add a matter part to the action which is similar to that of the BLG theory \eq{Kaction} but with an addition $\cW$ term which comes from the splitting of the chiral superfields
\be
S_{\mathrm{mat}} = -\int \mathrm{d}^3x\,\mathrm{d}^4\th\, \Tr \Big[ \bar{\cZ}_A e^{-\cV} \cZ^A e^{\cVh} +\bar{\cW}^A e^{-\cVh} \cW_A e^{\cV} \Big],
\ee
the Chern-Simons term is unaffected by the modification to the chiral superfields and so is as given in \eq{Kaction}. After integrating out all auxiliary fields and making the identifications \eq{su4enh}, we obtain the action
\bea
S = \int d^3x &\Big[&\frac{k}{4\pi}\ve^{\m\n\l} \Tr(A_\m \del_\n A_\l + \frac{2i}{3}A_\m A_\n A_\l -\Ah_\m \del_\n \Ah_\l - \frac{2i}{3}\Ah_\m \Ah_\n \Ah_\l) \nn\\
&&-\Tr(D_\m Z)\dag D^\m Z -\Tr(D_\m W)\dag D^\m W -i\Tr \z\dag \g^\m D_\m\z -i\Tr \o\dag \g^\m D_\m\o \nn\\
&&-V\Big].
\eea
Here $V=V_{\text bos} + V_{\text ferm}$ is the potential which is quite complicated due to the $SU(4)$ R-symmetry not having a canonical form. However, we can arrange the field content into multiplets of the $SU(4)_R$, namely the fundamental and anti-fundamental representations respectively
\bea
Y^A &=& \{Z^A, W\dag{}^A\}, \\
Y^\dagger_A &=& \{Z^\dagger_A, W_A\},
\eea
here the $Y^A$ runs from $A=1,...,4$. In a similar fashion, we organise the fermions into
\bea
\psi_A &=& \{\ve_{AB}\z^B e^{-i\pi/4}, -\ve_{AB}\o\dag{}^B e^{i\pi/4}\},\\
\psi^A\dag &=& \{-\ve^{AB}\z_B^\dagger e^{i\pi/4}, \ve^{AB}\o_B e^{-i\pi/4}\}.
\eea
Now we may write the bosonic and fermionic potentials as
\bea
V_{\text bos} = -\frac{4\pi}{3k^2}&\Tr(&Y^A Y_A^\dagger Y^B Y_B^\dagger Y^C Y_C^\dagger +Y_A^\dagger Y^A Y_B^\dagger Y^B Y_C^\dagger Y^C \nn\\
&&+4Y^A Y_B^\dagger Y^C Y_A^\dagger Y^B Y_C^\dagger -6Y^A Y_B^\dagger Y^B Y_A^\dagger Y^C Y_C^\dagger)
\eea
and
\bea
V_{\text ferm} = \frac{2\pi i}{k} &\Tr(&Y_A^\dagger Y^A \psi^B\dag \psi_B -Y^A Y_A^\dagger \psi_B\psi^B\dag +2Y^A Y_B^\dagger \psi_A \psi^B\dag -2Y_A^\dagger Y^B \psi^A\dag \psi_B\nn\\
&&-\ve^{ABCD}Y_A^\dagger \psi_B Y_C^\dagger \psi_D +\ve_{ABCD}Y^A \psi^B\dag Y^C \psi^D\dag).
\eea
So we see that the R-symmetry group has been enhanced from $U(1)_R\times SU(2)\times SU(2)$ to $SU(4)_R$, hence we have an $\cN=6$ supersymmetric theory of multiple M2-branes with 12 supercharges.

The moduli space of the ABJM theory can now be analysed, we do this for the $U(1)\times U(1)$ abelian case first for simplicity and then generalise to the non-abelian $U(N)\times U(N)$ case. For the abelian theory the potential and interaction terms vanish and the action reduces to a free field action for the scalars and fermions $Y^I$ and $\Psi$, with $A=1,...,4$, here the $Y^I$ are organised into $Y^I \equiv (Z^1, Z^2, W_1, W_2)$ and similarly for the fermions. Naively the moduli space is given by $\mathbb{C}^4$ but we need to be careful as we have a Chern-Simons term which provides us with a slightly non-trivial moduli space\footnote{For an abelian Yang-Mills gauge theory with gauge transformations $A\rightarrow A-\mathrm{d}\L$, the gauge field can be gauged to zero. Hence the moduli space would be given by $\mathbb{C}^4$.}. Under the gauge transformations $A\rightarrow A-\mathrm{d}\L,\, \Ah\rightarrow \Ah-\mathrm{d}\hat{\L}$ we can gauge fix $A,\Ah$ to zero but we still have to consider the large gauge transformations due to the $\L, \hat{\L}$ terms. We may use the generalised Stokes' theorem to obtain the boundary action for an abelian Chern-Simons term (assuming a boundary here);
\be
\label{ACS}
\d S_{ACS} = \frac{k}{2\pi}\int_{\text{bdry}} (\L \wedge F -\hat{\L} \wedge \Fh).
\ee
Recall that the field strength of a one-form over a 2-manifold satisfies 
\be
\frac{1}{2\pi}\int F \in \mathbb{Z},
\ee
we may use this in \eq{ACS} to find the form of $\L$. Before we do so, let us now explain why $k\in \ZZ$. 

The Chern-Simons action contains a level $k$ as previously mentioned, this level is an integer for compact gauge groups. Let us consider the simpler case of a single gauge field $A_\m$ in a gauge group $SU(N)$ and then generalise to the case of ABJM theory. The action is given by
\be
S = \frac{k}{4\pi}\int d^3 x\,\ve^{\m\n\l}\Tr\left(A_\m\del_\n A_\l +\frac{2}{3}A_\m A_\n A_\l \right),
\ee
this is a non-linear sigma model of the spacetime $S^3$ (which is a compactification of the Euclideanised spacetime $\RR^{1,2}$), as a map from $S^3 \to SU(N)$. The Chern-Simons action transforms under large gauge transformations \cite{cslevel} as
\be
S \to S +2\pi k w,
\ee
where $w\in \ZZ$ is the winding number of the map $S^3 \to SU(N)$. Since the winding number is an integer, for a well defined quantum theory we require $k\in \ZZ$ for the path integral
\be
Z= e^{iS}
\ee
to remain invariant under such gauge transformations. The Chern-Simons terms in the ABJM action are also quantised as sums of Chern-Simons terms are still gauge invariant for the same reasons described above. 

Gauge transformations of the Chern-Simons action must leave the path integral invariant and we know $k\in \ZZ$, so we require 
\be
\L,\hat{\L} = \frac{2\pi n}{k}, \quad n\in \mathbb{Z}.
\ee
The scalar fields then transform as $Y^I \rightarrow e^{i(\L -\hat{\L})}Y^I$ under the gauge transformations where the transformations act on the $Y^I$ to the left and right respectively; so the moduli space is not $\mathbb{C}^4$, but $\mathbb{C}^4/\mathbb{Z}_k$. This orbifold is understood as the scalar fields obeying the $\mathbb{Z}_k$ symmetry as
\be
Y^I \rightarrow e^{\frac{2\pi i}{k}}Y^I.
\ee
So we interpret this CFT as a supersymmetric sigma model on the orbifold $\mathbb{C}^4/\mathbb{Z}_k$.

For the non-abelian $U(N)\times U(N)$ moduli space it is clear that the field configuration that gives a vanishing potential is when all the matrices $Y^I$ are in a diagonal form, indeed this is the full moduli space of the theory as any off-diagonal component will be massive. Thus the gauge symmetry is broken to $U(1)^N\times U(1)^N \times S_N$, where $S_N$ is the permutation group of order $N$ that permutes the diagonal components of the $U(1)$'s. So the moduli space for $N$ M2-branes in the ABJM theory is given by
\be
\cM_{\text ABJM} = (\mathbb{C}^4/\mathbb{Z}_k)^N/S_N.
\ee

We now motivate why this theory of multiple M2-branes has $\cN=6$ supersymmetry. The above moduli space is the same as the moduli space of $N$ M2-branes with a $\mathbb{C}^4/\mathbb{Z}_k$ singularity. If we consider the original $SO(8)$ R-symmetry from the $\cN=8$ theory, then the ABJM theory has an $SU(4)\times U(1)$ subgroup which is preserved under the orbifold action with $\cN=6$ supersymmetry and 12 supercharges. To see this we consider the decomposition of the fermion; the original fermions were in the ${\bf 8}_c$ of $SO(8)$, so under the decomposition they become in the ${\bf 6}_0 +{\bf 1}_2 +{\bf 1}_{-2}$ representation\footnote{Here the subscripts on the representations indicate the Chern-Simons level.} of $SU(4)\times U(1)$. The orbifold projects out the singlets if $k >2$ but are kept if $k=1,2$. This means that the case $k=1,2$ should have $\cN=8$ supersymmetry with 16 supercharges while the arbitrary $k\in \mathbb{Z}$ theory should have $\cN=6$ supersymmetry with 12 supercharges.

\subsubsection{An $\cN=6$ Theory in the 3-bracket Formalism}
Now we review the 3-bracket formalism of the ABJM theory \cite{BL4} as it will be useful for Chapters 2 and 3.
The $\cN=6$ theory can be written in terms of a 3-algebra which is now complex and does not have a fully anti-symmetric 3-bracket as in the BLG theory. In \cite{BL4} the authors proposed a basis $T^a, a=1,...,N$ of a 3-algebra which is a complex vector space equipped with a triple product
\be
[T^a, T^b; \Tb^{\cb}\, ] = f^{ab \cb}{}_d T^d,
\ee
which is anti-symmetric in only the first two indices. The fundamental identity is given by 
\be
f^{ab\cb}{}_e f^{ef\gb}{}_d = f^{af\gb}{}_e f^{eb\cb}{}_d + f^{bf\gb}{}_e f^{ae\cb}{}_d - f_{\bar{e}}{}^{f\gb \cb} f^{ab\bar{e}}{}_d,
\ee
with a positive definite inner product
\be
h^{\bar{a} b} = \Tr(\Tb^{\bar{a}}, T^b).
\ee
The relaxation of the constraints on $f^{ab\cb}{}_d$ allows us to construct an action with 12 supercharges; i.e. $\cN=6$ supersymmetry, which has $SU(4)_R$ R-symmetry which agrees with the ABJM theory.

The scalar field content is given by four hermitian 3-algebra valued scalars $Z^A_a$, where  $A=1,...,4$ are the $SU(4)$ indices and their complex conjugates $\Zb_{A \bar{a}}$. The fermions are given by $\psi_{Aa}$ and their complex conjugates are $\psi^A_{\bar{a}}$. In terms of the representations of the R-symmetry group $SU(4)_R$, a raised index $A$ means the field is transforming in the ${\bf 4}$ of $SU(4)_R$, a lowered index means the field is in the ${\bf \bar{4}}$ of $SU(4)_R$. This action of complex conjugation is realised by raising/lowering the R-symmetry index $A$ and swapping the gauge indices for barred if unbarred and vice versa. The supersymmetry parameter $\e_{AB}$ transforms in the ${\bf 6}$ of $SU(4)_R$.

The original $SO(8)$ BLG scalar fields will be organised into a representation of $SU(4)_R$ in the following way
\bea
Z^1 &=& \frac{1}{\sqrt{2}}(X^3 +iX^4), \quad \quad Z^2 = \frac{1}{\sqrt{2}}(X^5 +iX^6), \nn\\
Z^3 &=& \frac{1}{\sqrt{2}}(X^7 -iX^9), \quad \quad Z^4 = \frac{1}{\sqrt{2}}(X^8 -iX^{10}).
\eea

The Lagrangian for the $\cN=6$ theory is given by
\bea
\cL &=& -\Tr( D^\m \bar{Z}_A ,D_\m Z^A) 
-i\Tr( \psib^A, \g^\m D_\m \psi_A )
-V +\cL_{\mathrm{CS}} \nn\\
&& -i\Tr (\psib^A,[\psi_A,Z^B;\bar{Z}_B])
+2i\Tr (\psib^A,[\psi_B,Z^B;\bar{Z}_A]) \nn\\
&& +\frac{i}{2}\,\ve_{ABCD}\Tr(\psib^A,[Z^C,Z^D;\psi^B]) 
-\frac{i}{2}\,\ve^{ABCD}\Tr(\bar{Z}_D,[\psib_A,\psi_B;\bar{Z}_C]), 
\eea
where
\be
\label{n6BLpot}
V = \frac{2}{3}\Tr(\Upsilon_B^{CD}\bar{\Upsilon}^B_{CD})
\ee
and
\be
\Upsilon_B^{CD} = [Z^C, Z^D; \Zb_B] -\frac{1}{2}\d^C_B[Z^E,Z^D;\Zb_E] + \frac{1}{2}\d^D_B[Z^E,Z^C;\Zb_E].
\ee
The $\g^\m$ are the 3-dimensional gamma matrices given by the usual Pauli matrices as in the Appendix. Also $\cL_{CS}$ is the twisted Chern-Simons term as in the BLG case but with the new structure constants of the complex 3-algebra.

The supersymmetry transformations of the $\cN=6$ theory are given by
\bea
\d Z^A_d &=& i \eb^{AB}\psi_{Bd}, \\
\d \At_\m{}^c{}_d &=&  -i\eb_{AB}\g_\m Z^A_a\psi^B_{\bar{b}} f^{ca \bar{b}}{}_d +i\eb^{AB}\g_\m\Zb_{A\bar{b}}\psi_{Ba} f^{ca\bar{b}}{}_d, \\
\d \psi_{Bd} &=& \g^\m D_\m Z^A\e_{AB} + f^{ab\cb}{}_d Z^C_a Z^A_b \Zb_{C\cb}\e_{AB} +f^{ab\cb}{}_d Z^C_a Z^D_b \Zb_{B\cb}\e_{CD}.
\eea
On examining the closure of the supersymmetry algebra we find that it can be made to close on-shell and the gauge invariance property of the metric implies that gauge transformations $\L^c{}_d$ are elements of $u(N)$. 

There are an infinite class of examples of the complex 3-algebras one could have, so let us examine the general case. Take $V_1, V_2$ to be complex vector spaces of dimension $N_1, N_2$ respectively, then we can construct a vector space $\cA$ of linear maps $A : V_1 \rightarrow V_2$. One may then construct the triple product on this space;
\be
\label{3bn6}
[A,B;\Cb] = \l(AC\dag B - B C\dag A)\quad\quad A,B,C \in \cA,
\ee
where $\l$ is a normalisation constant, this will be given by $\l = 2\pi/k$ for the Chern-Simons-Matter theories to ensure a well defined path integral. The $\dag$ is the transpose conjugate for matrices, so we can introduce the following product
\be
\Tr(\Ab, B) = \tr(A\dag B)
\ee
which is the usual trace form for matrices. 

We can take the complex vector spaces to be $V_1 = \mathbb{C}^{N_1}$ and $V_2 = \mathbb{C}^{N_2}$, then we can consider these spaces as the vector spaces of the fundamental representations of $U(N_1)$ and $U(N_2)$. As such, the matrices $A\in \cA$ such that $A: \mathbb{C}^{N_1} \rightarrow \mathbb{C}^{N_2}$ are in the bifundamental representation $({\bf{N}}_1, {\bf{\Nb}}_2)$. So we may write down the bifundamental matter Lagrangian
\bea
\label{n6abjmabj}
\cL &=& -\tr( D^\m Z_A^\dagger D_\m Z^A) 
-i\tr( \psib^A\dag \g^\m D_\m \psi_A )
-V +\cL_{\mathrm{CS}} \nn\\
&& -\frac{2\pi i}{k}\tr (\psib^A\dag \psi_A Z_B^\dagger Z^B -\psib^A\dag  Z^B Z_B^\dagger\psi_A) \nn\\
&& +\frac{4\pi i}{k}\tr (\psib^A\dag \psi_B Z_A^\dagger Z^B -\psib^A\dag Z^B  Z_A^\dagger \psi_B) \nn\\
&& +\frac{2\pi i}{k}\,\ve_{ABCD}\tr(\psib^A\dag Z^C \psi^B\dag Z^D)
 -\frac{2\pi i}{k}\,\ve^{ABCD}\tr(Z_D^\dagger \psib_A Z_C^\dagger \psi_B).
\eea
For the choice $N_1=N_2$ we obtain the ABJM theory \cite{ABJM} with gauge group $U(N)\times U(N)$, for the choice $N_1 \neq N_2$ with $N_1> N_2$ we obtain the ABJ theory \cite{ABJ} with gauge group $U(N_1)\times U(N_2)$.

The Lagrangian \eq{n6abjmabj} for the choice $N_1=N_2$, i.e. the ABJM model, is invariant under the following set of supersymmetry transformations
\bea
\d Z^A &=& i \eb^{AB}\psi_{B}, \label{6d1}\\
\d A_\m &=&  \frac{2\pi}{k}\left[Z^B\psib^A\g_\m\e_{AB} 
+ \e^{AB}\g_\m\psi_A\bar{Z}_B\right], \label{6d2}\\
\d \psi_{A} &=& \g^\m\e_{AB}D_\m Z^B +N_A, \label{6d3}
\eea
and their conjugates, where 
\be
N_A = \frac{2\pi}{k}\left[-\e_{AB}\left(Z^C\bar{Z}_CZ^B 
-Z^B\bar{Z}_CZ^C \right) +2\e_{CD}Z^c\bar{Z}_AZ^D\right].
\ee

\section{M5-branes}

\subsection{Overview of M5-brane theory}
The theory of multiple M5-branes has been an area of focus recently. The low energy theory of multiple M5-branes is given by a chiral tensor-gauge theory in six dimensions known as a $(2,0)$ superconformal field theory \cite{m51,m52,m53,m54}. The abelian theory has been known for some time \cite{howe1,howe2} and a non-Lorentz invariant action in 6D has been developed in \cite{schw,pst1,pst2,pst3,pst4}, we shall expand on the reference \cite{pst2} called the PST action next. We will then discuss some other recent developments in understanding the M5-brane theory better.

\subsection{Non-Lorentz Manifest Action in 6D}
The reason why the action cannot be written down in a manifestly Lorentz invariant way in 6D is due to the 3-form field strength $H_{\m\n\l}$ of the 2-form tensor-gauge field $B_{\m\n}$ being self-dual, so any attempt to write a tensor-gauge kinetic term is trivially zero
\be
\int_{\S_6}H_3 \wedge *H_3=0.
\ee

In \cite{schw,schw-2} a non-Lorentz invariant action was constructed for the abelian M5-brane by choosing a non-manifestly 6D Lorentz invariant action. In \cite{pst1,pst2,pst3,pst4} PST constructed an off-shell action for an abelian M5-brane in 6D which is coupled to an auxiliary scalar. It turns out that a certain gauge will allow us to write this theory in 6D but with only 5D manifest Lorentz invariance. 
The construction for a covariant action for a self-dual 3-form field strength  $H= dB$ is now derived, the action reads
\be \label{pst}
S_{PST} = \int d^6 \s  \left[\sqrt{-g} 
\frac{1 }{4(\del a)^2}\del_{\m} a
H^{*\m \n \l} H_{\n \l \r} \del^\r a + Q(\Ht)
\right].
\ee
Here the Greek indices $\m=0,1,\cdots,5 $ and
\be
\Ht_{\m\n} := \frac{1}{\sqrt{(\del a)^2}}H^*_{\m\n\l} \del^\l a.
\ee
The 3-form $H_{\m\n\l}$ satisfies the self-duality condition, where
\be
\label{sdeqn}
 H^{*\m\n\l} := \frac{1}{6 \sqrt{-g}} \e^{\m\n\l\r\a\b}H_{\r\a\b}
\ee
is the Hodge dual of $H_{\r\a\b}$. Our convention for the  Hodge duality operation is
$\e_{012345} =1 =-g \e^{012345}$. 
The action \eq{pst} is invariant under the following local transformations:
\bea  
 && \quad \d B_{\m\n} = \del_{[\m} \L_{\n]}, \quad \d a =0;\label{T1}  \nn\\
 &&\quad \d B_{\m\n} = \del_{[\m } a \; \vphi_{\n]}
\quad \d a =0; \label{T2}\nn\\
&& \quad \d B_{\m\n} 
= \frac{\vphi}{2 (\del a)^2} (H_{\m\n\r} \del^\r a - \cV_{\m\n}),
\quad \d a = \vphi, \label{T3}
\eea
where
\be
\cV^{\m\n} := -2 \sqrt{\frac{(\del a)^2}{-g}} \frac{\d Q}{\d \Ht_{\m\n}}.
\ee
The equation of motion of the 2-form potential $B_{\m\n}$ is
\be\label{pst-eom}
\e^{\m\n\l\r\a\b} \del_\l \left(
\frac{\del_\r a}{(\del a)^2} (H_{\a\b \g} \del^\g a - \cV_{\a\b})
\right) =0.
\ee
Using the local symmetry \eq{T2}, one can then show that it is equivalent to
the self-duality condition 
\be \label{sdeom}
H_{\m\n\l} \del^\l a - \cV_{\m\n} =0. 
\ee
The scalar field $a$ is introduced to allow six dimensional covariance
and is completely auxiliary due to the symmetry \eq{T3}. 
If we choose a gauge $a=x^5$ and consider the linearised equation of motion with
\be
Q= -\frac{1}{4} \Ht_{\m\n} \Ht^{\m\n} \sqrt{-g},
\ee 
then 
\be 
\cV_{\m\n} = \Ht_{\m\n} \sqrt{(\del a)^2}
\ee
and \eq{sdeom} becomes the standard self-duality condition
\be
H_{\m\n 5} = H^*_{\m\n 5}.
\ee
In this case the gauge-fixed PST action reads
\be \label{pst1}
S_{PST} = -\frac{1}{4}  \int d^6 \s\left(
 \frac{1}{6} \e_{abcde}H^{abc} H^{de5} + H^{* ab5} H^*_{ab5}  \sqrt{-g} 
\right),
\ee
where $a=0, \cdots, 4$ etc. See \cite{schw2} 
for a detailed discussion of the non-covariant and covariant PST
formulations of the M5-brane action.

\subsection{A proposal for M5-branes as 5D SYM}

We now discuss the recent proposal for a duality between M5-branes and 5D SYM, this allows us to describe the M5-branes on a circle as a SYM theory. This result is quite remarkable as all of the information of the M5-brane theory is conjectured to be contained within the 5D SYM theory. We will make use of this in Part II of the thesis when we discuss matrix models of D4-branes and their generalisations to M5-branes. We save discussion of other recent advances in the theory of M5-branes to the conclusions in Chapter 8.

It was proposed that M5-branes compactified along an $S^1$ are equivalent to D4-branes of type \IIA string theory where the Kaluza-Klein modes of the M5-brane along the $S^1$ are identified with instantons in the D4-brane theory \cite{d1,d2}.

We know that the low energy action of D4-branes at weak coupling is given by 5D SYM with a $U(N)$ gauge group. The M5-branes are the strong coupling limit of the D4-branes and so we expect the UV fixed point of 5D SYM to be given by the (2,0) 6D superconformal field theory. This is due to the duality between type \IIA string theory and M-theory. The proposal is that the (2,0) theory compactified on $S^1$ gives 5D SYM and that the Kaluza-Klein modes in the M5-brane theory match up precisely with the instantons in the SYM theory. 

The action for the 5D SYM theory is given by
\bea
S &=& -\frac{1}{g^2_{YM}}\int d^5x\;\tr\Big(\frac{1}{4}F_{\mu\nu} F^{\mu\nu} + \frac{1}{2}D_\mu X^I D^\mu X^I -\frac{i}{2}\bar\Psi\Gamma^\mu D_\mu \Psi \nn\\
&&\hskip3cm+ \frac{1}{2}\bar\Psi \Gamma^5\Gamma^I[X^I,\Psi]- \frac{1}{4}[X^I,X^J]^2\Big),
\eea
where
\be
D_\mu X^I = \partial_\mu X^I - i[A_\mu,X^I]
\ee
and
\be
F_{\mu\nu} = \partial_\mu A_\nu - \partial_\nu A_\mu - i [A_\mu,A_\nu].
\ee
The content of the non-abelian theory consists of a gauge field $A_\m$ with $\m=0,...,4$, five $SO(5)_R$ invariant scalar fields $X^I$ with $I=6,...,10$ and a fermion $\Psi$. We have left the 5 direction as the M-theory direction here to match the literature. The supersymmetry transformations of the 5D SYM theory are given by
\bea
\delta X^I  &=& i\bar\epsilon \Gamma^I\Psi \nn\\
\delta A_\mu &=& i\bar\epsilon\Gamma_\mu\Gamma_5\Psi \nn\\
\delta\Psi  &=&\frac{1}{2}F_{\mu\nu}\Gamma^{\mu\nu}\Gamma_{5}\epsilon + D_\mu X^I\Gamma^\mu\Gamma^I\epsilon   - \frac{i}{2}[X^I,X^J]\Gamma^{IJ}\Gamma^5\epsilon,
\eea
where the supersymmetry parameter $\e$ satisfies the projection condition
\be
\G_{012345}\e=\e
\ee
and similarly for the fermion
\be
\G_{012345}\Psi=-\Psi.
\ee
The claim is that the 5D SYM theory has all the information for the (2,0) theory already encoded within it, to be more precise the instantons of the SYM theory can be calculated and then matched with the states of the (2,0) theory. This can be done by computing the superalgebras $\{Q_\a,Q_\b\}$ of both theories and then performing a dimensional reduction, along $x^5$, of the (2,0) superalgebra and then find the matching quantities. Omitting the details, it was shown in \cite{rozali1,rozali2,d2}  
that one must identify the M5-brane momentum in the M-theory direction $P_5$ with the central extension $Z_5$ of the D4-brane theory which is the instanton number
\be
P_5 = Z_5 =\frac{k}{R_{5}}=-\frac{1}{8g^2_{YM}}  \int d^4x \; {\rm tr}( F_{ij}F_{kl}\varepsilon_{ijkl}), \quad k \in \ZZ.
\ee
Note that the KK momenta $P_5$ and the instanton are quantised, so we set
\be
R_5 = \frac{g_{YM}^2}{4\pi^2}.
\ee
Further analyses of the KK modes and the instantons were carried out in \cite{d1,d2} but we will not consider them here.

\part{Boundary Dynamics of M2-branes with Flux}
\chapter{Multiple M2-branes in a Flux}

In this chapter we review the work of Lambert and Richmond \cite{LR} which is a construction of closed multiple M2-branes in a certain flux background. We first look at the $\cN=8$ case as it is simpler than the $\cN=6$ theory.

\section{Closed $\cN=8$ M2-branes with Flux}

Recall that the bosonic part of the effective abelian M2-brane action is given by
\bea
\label{abelianm2}
S_{M2} = -T_{M2}\int \mathrm{d}^3\s\, \sqrt{-\det P(g_{mn})} + \frac{T_{M2}}{3!}\int \mathrm{d}^3\s\, P(C_{mnp}),
\eea
where we pull-back the eleven-dimensional metric $g_{mn}$ to the M2-brane worldvolume, $m,n = 0,...,10$. The tension of the M2-brane is given by 
\be
T_{M2}=\frac{1}{4\pi^2 l_p^3}.
\ee
The 3-form potential $C_{mnp}$ is the natural (electrical) coupling for the M2-brane. When we pull-back the metric $g_{mn}$ to the worldvolume of the M2-brane, i.e. 
\be
g_{\m\n} = \del_\m x^m\del_\n x^n g_{mn},
\ee
we can rescale the scalars so that they have mass dimension one half
\be
\frac{x^m}{\sqrt{T_{M2}}} = X^m.
\ee
Then the metric and 3-form are dimensionless. In a static gauge $\s^\m=x^\m, \m =0,1,2,$ we have 8 transverse scalars $X^I, I=3,...,8.$
\subsection{Non-Abelian M2-branes in a Certain Flux}
We may now generalise the action \eq{abelianm2} to the full non-abelian case by replacing the first term with the BLG model and the second term with the full Wess-Zumino term obtained by the Myers effect \cite{myers}. It is interesting to note that while the Myers terms are always linear in the RR gauge fields in the case of D-branes, here we must allow for higher order RR gauge fields since the NS-NS fields come from a dimensional reduction of the M-theory 3-form. For the reduction of an M2-brane to a D2-brane, the 3-form potential of M-theory $C_{mnp}$ decomposes into a 3-form RR gauge field $C_{ijk}$ and an NS-NS 2-form $B_{(10)ij}$, where the indices $i,j,k=0,...,9$ and $x^{10}$ is the direction of compactification for M-theory. The 3-form potential has a 4-form field strength given by $G_4 = \mathrm{d}C_3$, so under gauge transformations of $C_3$, the gauge invariant electromagnetic dual field strength is $G_7 = \mathrm{d}C_6 = \star G_4 -\frac{1}{2}C_3\wedge G_4$.

The general form of the non-abelian pull-back of the 3-form is given by
\bea
\label{napb}
S_C = &&\frac{1}{3!}\int \mathrm{d}^3x\, \ve^{\m\n\r}\Big (aT_{M2} C_{\m\n\r} + 3bC_{\m IJ}\Tr(D_\n X^I, D_\r X^J) \nn\\
&&+12cC_{\m\n IJKL} \Tr(D_\r X^I, [X^J,X^K,X^L]) \nn\\
&&+12dC_{[\m IJ}C_{\n KL]} \Tr(D_\r X^I, [X^J,X^K,X^L]) + \cO(T_{M2}^{-1}) \Big),
\eea
here we ignore terms of order $\cO(T_{M2}^{-1})$ as we later take a decoupling limit. The constants $a,b,c,d$ are real numbers. In the action above we only included terms with even number of scalars $X^I$ due to gauge invariance, this is due to the fact that the trace $\Tr$ is an inner-product on the 3-algebra and not a number; so quantities like $\ve^{\m\n\r}C_{\m\n I}\Tr(D_\r X^I)$ are not invariant. The BLG theory describes M2-branes in an $\RR^8/\ZZ_2$ orbifold, this means that scalars should always appear in pairs in the Lagrangian is consistent with the orbifold.

In the action \eq{napb} we have 4 unknown constants, we can identify some of them straight away. The first term is the usual coupling of an M2-brane to the 3-form, so for $N$ M2-branes we take $a=N$. The second term is similar to a $C_{\m IJ}$ contribution from \eq{abelianm2}, there the coefficient is one and so for the non-abelian case we will also take this to be $b=1$. It is useful to note that this term is a non-Lorentz invariant quantity that would contribute towards the kinetic term of the BLG theory. The constant $c$ shall be determined by considering a back-reaction of the flux. The last term can be set to zero due to it being symmetric under $I,J \leftrightarrow K,L$.

Integrating by parts allows us to guarantee gauge invariance of the 3-form under \\$\d C_3~=~\mathrm{d} \L_2$, this does not guarantee the gauge invariance of the action however as we shall see. Ignoring any boundary terms when integrating by parts gives us
\bea
\label{napbbp}
S_C = &&\frac{1}{3!}\int \mathrm{d}^3x\, \ve^{\m\n\r}\Big (NT_{M2} C_{\m\n\r} + \frac{3}{2}G_{\m \n IJ}\Tr(X^I, D_\r X^J) \nn\\
&&-\frac{3}{2}C_{\m IJ}\Tr(X^I, \Ft_{\n\r} X^J) -cG_{\m\n\r IJKL} \Tr(X^I, [X^J,X^K,X^L]) \Big ),
\eea
where $G_{\m\n IJ} = 2\del_{[\m}C_{\n ]IJ}$ and $G_{\m\n\r IJKL}=3\del_{[\m}C_{\n\r]IJKL}$. The term proportional to the worldvolume field strength $\Ft_2$ is not invariant under the gauge transformation $\d C_3=\mathrm{d} \L_2$, so we add in the same quantity with opposite sign to cancel this term from the action
\be
S_F = \frac{1}{4}\int \mathrm{d}^3x\, \ve^{\m\n\r} C_{\m IJ}\Tr(X^I, \Ft_{\n\r} X^J).
\ee

Now we look at the final term in \eq{napbbp}, it is proportional to $G_7$ which is not invariant under the gauge transformation $\d C_3=\mathrm{d} \L_2$. We therefore add in a term proportional to $C_3 \wedge G_4$ since $G_7 + \frac{1}{2}C_3 \wedge G_4$ is invariant under the gauge transformation $\d C_3=\mathrm{d} \L_2$,
\be
S_{CG} = -\frac{c}{12}\int \mathrm{d}^3x\, \ve^{\m\n\r}\Tr(X^I, [X^J, X^K, X^L])
(C_3\wedge G_4)_{\m\n\r IJKL}.
\ee

Since we wish to consider the field theory and no gravitational effects, we employ the decoupling limit $T_{M2}\rightarrow \infty$ as there are no other parameters in our theory to tune. The total flux action is given by
\be
S_{flux} = S_C + S_F + S_{CG},
\ee
but the $C_{\m\n\r}$ term is just a constant if Lorentz invariant so we can ignore this. If we only want terms that preserve 3-dimensional Lorentz invariance we must discard the $G_{\m\n IJ}$ term. So we take 
\be
S_{flux} = c \int \mathrm{d}^3x\, \Gt_{IJKL}\Tr(X^I, [X^J, X^K, X^L]),
\ee
where we define
\be
\label{specialflux}
\Gt_{IJKL} = -\frac{1}{3!}\ve^{\m\n\r}(G_7 +\frac{1}{2}C_3 \wedge G_4)_{\m\n\r IJKL}.
\ee
\subsection{Supersymmetry}
We now wish to supersymmetrise the theory so that we may consider the supersymmetric boundary conditions in the next chapter. 
In the limit $T_{M2}\rightarrow \infty$,
the Lagrangian for $N$ closed M2-branes in the specific background field given by \eq{specialflux} consists of a flux and 
a mass term modification to the 
BLG Lagrangian $\cL_{\cN=8}$:
\be \label{LR1}
\cL = \cL_{\cN=8} + \cL_{flux} + \cL_{mass},
\ee
where 
\bea 
\cL_{\cN=8} &=& -\frac{1}{2} \Tr(D^\m X^{I}, D_\m X^I) 
+ \frac{i}{2}\Tr( \Psib, \G^\m D_\m \Psi) 
+ \frac{i}{4} \Tr( \Psib, \G_{IJ} [X^I, X^J, \Psi]) \nn\\
&& -\frac{1}{12} \Tr \big( [X^I,X^J,X^K],  [X^I,X^J,X^K]\big) + \cL_{CS}, 
\label{L-BL}\\
\cL_{flux} &=& 
c \Gt_{IJKL} \Tr(X^I, [X^J,X^K,X^L]) , \label{L-flux} \\
\cL_{mass} &=& -\frac{1}{2} m^2 \delta_{IJ}\Tr(X^I,X^J) 
+ \b\Tr(\Psib\G^{IJKL}, \Psi)\tilde{G}_{IJKL}. \label{L-mass}
\eea 
The background gauge field has transverse components $G_{IJKL}$ 
turned on and  
$\Gt_{IJKL}$ is defined by
\be 
\Gt_{IJKL} 
= \frac{1}{4!}\e_{IJKLMNPQ} G^{MNPQ},
\ee
where $I,J,K,L =3,4,\cdots, 10$.

The supersymmetry transformations for the deformed theory are given by \\$\d=\d_0 + \d'$ where
$\d_0$ are the supersymmetry transformations of the original BLG theory,
\bea
\d_0 X^I_a &=& i \eb \G^I \Psi_a, \label{d1}\\
\d_0 \At_\m{}^b{}_a &=&  i \eb \G_\m \G_I X^I_c \Psi_d f^{cdb}{}_a, \label{d2}\\
\d_0 \Psi_a &=& D_\m X^I_a \G^\m \G^I \e 
- \frac{1}{6} X^I_b X^J_c X^K_d f^{bcd}{}_a \G^{IJK} \e, \label{d3}
\eea
and $\d'$ are the additional contributions to the supersymmetry 
transformations due to the flux
\bea
\d' X^I_a &=& 0,  \label{dp1} \\
\d' \At_\m{}^b{}_a &=& 0, \label{dp2}\\
\d' \Psi_a &=& \omega \G^{IJKL}\G^M \e X_a^M \tilde{G}_{IJKL}. \label{dp3}
\eea
Here 
$\Psi$ and $\e$ are eleven dimensional spinors satisfying the projector conditions
\be \label{c0}
\G_{012} \Psi =- \Psi, 
\ee
\be \label{c0p}
\G_{012} \e= \e.
\ee 
We shall examine the closure of the superalgebra after showing that the Lagrangian is invariant under the supersymmetry transformations $\d$, we have the variation
\bea
\d \cL = &&(i\o +2\b)\Tr(\Psib\G^\m\G^{PQRS}\G^I\e, D_\m X^I)\Gt_{PQRS} \nn\\
&&+\frac{i\o}{2}\Tr(\Psib\G^{IJ}\G^{PQRS}\G^K\e , [X^I, X^J, X^K])\Gt_{PQRS} \nn \\
&&-\frac{2\b}{6}\Tr(\Psib\G^{PQRS}\G^{IJK}\e\ , [X^I, X^J, X^K])\Gt_{PQRS} \nn \\
&&+4ic\Tr(\Psib\G^I\e, [X^J, X^K, X^L])\Gt_{IJKL} +im^2\d_{IJ}\Tr(\Psib\G^I\e,X^J)\nn\\
&&+2\b\o\Tr(\Psib\G^{IJKL}\G^{PQRS}\G^T\e,X^T)\Gt_{IJKL}\Gt_{PQRS}.
\eea
We may simplify this by setting $\b=-\frac{i\o}{2}$ so that the term linear in a derivative is zero. Then by substituting this value for $\b$ and using identities in the Appendix, we obtain
\bea
\d \cL = &&\frac{2i\o}{3}\Tr(\Psib\G^{IJKMNOP}\e, [X^I, X^J, X^K])\Gt_{MNOP} \nn\\
&&+(4ic-16i\o)\Tr(\Psib \G^I\e, [X^J, X^K, X^L])\Gt_{IJKL} \nn\\
&&+im^2\d_{IJ}\Tr(\Psib\G^I\e, X^J) \nn\\
&& -i\o^2\Tr(\Psib\G^{IJKL}\G^{PQRS}\G^T\e,X^T)\Gt_{IJKL}\Gt_{PQRS}.
\eea
Using Hodge duality of the gamma matrices \eq{hodge2}, also adopting Feynman slash notation for clarity, yields
\bea
\d \cL = && \frac{96i\o}{6}\left(-1 +\frac{c}{4\o} -\star\right)\Tr(\Psib\G^I\e, [X^J, X^K,X^L])\Gt_{IJKL} \nn\\
&&+i\Tr(\Psib(m^2 -\o^2{\not}{\Gt}{\not}{\Gt})\G^I\e, X^I).
\eea
So we see that supersymmetry requires  
the coefficients $\o$ and $\b$ to be  determined by the 
flux term 
\be \label{c2} 
\o = \frac{c}{8}, \quad \b = -i \frac{c}{16},
\ee
via the supersymmetric equations of motion:
\bea
\label{seom1}
0 &=&\left(-1 +\frac{c}{4\o} -\star\right)\Gt_{IJKL},\\
\label{seom2}
0 &=&(m^2 -\o^2{\not}{\Gt}{\not}{\Gt})\G^I\e.
\eea
The value of $\o$ is determined by considering the eigenvalue of $\star$ where $\Gt$ is the eigenvector. The eigenvalues of $\star\star =1$ are $\pm 1$, therefore the eigenvalue $-1 +\frac{c}{4\o}$ must be equal to $\pm1, \pm i$. The imaginary eigenvalues can be ruled out due to physical reasons, so we have two choices $\pm1$. The eigenvalue $+1$ gives $\o=c/8$ and $\Gt = \star \Gt$, the eigenvalue $-1$ gives $\o=0$ and an anti self-dual flux.

Moreover, since the flux $\Gt_{IJKL}$
has to be  self-dual for a non-trivial extension to the BLG theory, it implies
\be \label{G1}
\G^{012} \Gts = \Gts.
\ee
The flux also needs to satisfy the condition
\be\label{G2}
\Gts \Gts =  \frac{32m^2}{c^2} (1+ \G^{3456789(10)} ),
\ee
which implies immediately that
\be \label{c1}
G_{MN[IJ}G_{KL]}{}^{MN} =0
\quad \mbox{and}
\quad
m^2 = \frac{c^2}{32 \cdot 4!} G^2.
\ee
An analysis of the back reaction of the flux gives $c=2$.

The self-duality condition is solved by $\Gts$ of the form
$ \Gts = d (1+ \G^{012}) R$, 
where $d$ is a constant coefficient and $R$ is a sum of products of four
transverse $\G^I$'s, $I= 3, 4, \cdots, 10$. The condition \eq{G2} then implies 
that
\be \label{gen-G}
\Gts = 2 \mu \frac{1+ \G^{012}}{2} R, \quad R^2 =1,
\ee
for $\mu = \pm 2 m$, $m\geq 0$.
A simple solution is 
\be
R = \G^{3456}. 
\ee
This corresponds to  the flux
\be
\Gt =  \mu \;(dx^3\wedge dx^4\wedge dx^5\wedge dx^6 + dx^7\wedge dx^8\wedge dx^9\wedge dx^{10}),
\label{sf1}
\ee
and the Lagrangian \eq{LR1} reproduces precisely the deformed Bagger-Lambert 
Lagrangian of \cite{gomis} and \cite{hll}. 

\subsubsection{Closure of the Superalgebra}
Under the supersymmetry transformations \eq{d1}-\eq{dp3}, we can compute the closure relations of the gauge fields $\At_\m{}^b{}_a$, scalars $X^I$ and fermions $\Psi$. The closure relation of gauge field remains the same as in \eq{gtransA}, namely
\bea
[\d_1,\d_2]\At_\m{}^b{}_a = v^\n \Ft_{\m\n}{}^b{}_a + D_\m \Lt^b{}_a
\eea
where $v_\m$ and $\Lt^b{}_a$ are defined in \eq{translation},\eq{lambda}. 

For the scalars $X^I$, there is an additional term due to the modification to the supersymmetry transformation of the fermion $\d'\Psi$, namely $2i\o\eb_2\G^{PQRSIJ}\e_1X^J\Gt_{PQRS}$. We can use the Hodge duality \eq{hodge2} of a $\G^{(6)}$ term to give a $\G^{(2)}$ term:
\be
[\d_1,\d_2]X^I_a = v^\m D_\m X^I_a +\Lt^b{}_a X^I_b + iR^I{}_JX^J_a,
\ee
where $R_{IJ} = 48\o \eb_2 \G^{OP}\e_1 \Gt_{OPIJ}$ is the closure of terms under the SO(8) R-symmetry. Note this term in the closure relation is imaginary, it is consistent with the idea that R-symmetry is given by the matrices that leave the superalgebra $\{Q_\a^I,Q_\b^J\}$ invariant.

The closure of the fermions $\Psi_a$ gives us the off-shell equation of motion
\be
E' = \G^\m D_\m \Psi_a +\frac{1}{2}\G_{IJ}X^I_c X^J_d \Psi_b f^{cdb}{}_a -\o \G^{PQRS}\Psi_a \Gt_{PQRS}.
\ee
So on-shell, we have the closure relation
\be
[\d_1,\d_2]\Psi_a = v^\m D_\m \Psi_a +\Lt^b{}_a \Psi_b + \frac{i}{4}R_{PQ}\G^{PQ}\Psi_a.
\ee
We see that the fermions also get a contribution from the SO(8) R-symmetry.

We shall now review the $\cN=6$ theory coupled to flux in terms of the 3-bracket construction \cite{LR}. 

\section{Closed $\cN=6$ M2-branes with Flux}

We now turn to the $\cN=6$ theory with mass and flux terms 
given by Lambert-Richmond \cite{LR} and carry out a similar analysis to the previous section. 
The $S_C$ action for the $\cN=6$ theory is given by
\bea
S_C = &&\frac{1}{3!}\ve^{\m\n\l}\int \mathrm{d}^3 x\, \Big( N T_{M2} C_{\m\n\l} +\frac{3}{2}C_\m{}^A{}_B\Tr (D_\n \Zb_A, D_\l Z^B) \nn\\
&&+\frac{3}{2}C_\m{}_A{}^B\Tr (D_\n Z^A, D_\l \Zb_B) +\frac{3c}{2}C_{\m\n AB}{}^{CD}\Tr (D_\l \Zb_D, [Z^A, Z^B; \Zb_C]) \nn\\
&&+\frac{3c}{2}C_{\m\n}{}^{AB}{}_{CD}\Tr (D_\l Z^D, [\Zb_A, \Zb_B; Z^C])\Big).
\eea
The $\cN=6$ theory has a 3-algebra given by a matrix representation of the 3-bracket as stated in \eq{3bn6} and $A,B,C,D = 1,..,4$ are the $SU(4)$ R-symmetry indices. 
Repeating the argument as in the $\cN=8$ theory, we obtain the full Lagrangian of the flux 
deformed $\cN=6$ theory:
\be
\label{LRN6}
\cL = \cL_{\cN=6} + \cL_{flux} + \cL_{mass}.
\ee
where
\bea 
\cL_{\cN=6} &=& -\Tr( D^\m \bar{Z}_A ,D_\m Z^A) 
-i\Tr( \psib^A, \g^\m D_\m \psi_A )
-V +\cL_{\mathrm{CS}} \nn\\
&& -i\Tr (\psib^A,[\psi_A,Z^B;\bar{Z}_B])
+2i\Tr (\psib^A,[\psi_B,Z^B;\bar{Z}_A]) \nn\\
&& +\frac{i}{2}\,\ve_{ABCD}\Tr(\psib^A,[Z^C,Z^D;\psi^B]) 
-\frac{i}{2}\,\ve^{ABCD}\Tr(\bar{Z}_D,[\psib_A,\psi_B;\bar{Z}_C]), 
\label{L6-BL}\\
\cL_{\mathrm{CS}} &=& \frac{k}{4 \pi} \e^{\m\n\l} \left(
A_\m \del_\n A_\l + \frac{2i}{3} A_\m A_\n A_\l
- 
\Ah_\m \del_\n \Ah_\l - \frac{2i}{3} \Ah_\m \Ah_\n \Ah_\l
\right),\label{L6-CS} \\
\cL_{flux} &=& 
\frac{c}{4} \Tr(\bar{Z}_D,[Z^A,Z^B;\bar{Z}_C]) \Gt_{AB}{}^{CD} , 
\label{L6-flux} \\
\cL_{mass} &=& -m^2 \Tr(\bar{Z}_A,Z^A) 
+ \b\Tr(\psib^A, \psi_F)\Gt_{AE}{}^{EF}. \label{L6-mass}
\eea
Here $V$ is defined in \eq{n6BLpot} and we also define 
\be
\Gt_{AB}{}^{CD} = \frac{1}{4}\ve_{ABEF}\ve^{CDGH}G^{EF}{}_{GH}.
\ee
The supersymmetry transformations of the original $\cN=6$ theory are given by
\bea
\d_0 Z^A &=& i \eb^{AB}\psi_{B}, \\
\d_0 A_\m &=&  \frac{2\pi}{k}\left[Z^B\psib^A\g_\m\e_{AB} 
+ \e^{AB}\g_\m\psi_A\bar{Z}_B\right], \\
\d_0 \psi_{A} &=& \g^\m\e_{AB}D_\m Z^B +N_A, 
\eea
and their conjugates, where 
\be
N_A = \frac{2\pi}{k}\left[-\e_{AB}\left(Z^C\bar{Z}_CZ^B 
-Z^B\bar{Z}_CZ^C \right) +2\e_{CD}Z^c\bar{Z}_AZ^D\right]
\ee
and $\d'$ is the  additional contribution to the supersymmetry 
transformations due to the flux terms
\bea
\d' Z^A &=& 0,  \label{6dp1} \\
\d' A_\m &=& 0, \quad \d' \Ah_\m =0, \label{6dp2}\\
\d' \psi_{A} &=& \omega \e_{DF} Z^F \Gt_{AE}{}^{ED}. \label{6dp3}
\eea
The supersymmetry transformation parameter satisfies the reality condition
\be
\e_{FP} = \frac{1}{2} \ve_{IJFP} \e^{IJ}
\ee
and is in the $\mathbf{6}$ representation of SU(4), recall that raising and lowering SU(4) indices acts as complex conjugation.
For the action to be supersymmetric, the flux needs to take the form 
\be
\label{n6flux}
\Gt_{AB}{}^{CD} = \frac{1}{2}\delta^C_B\Gt_{AE}{}^{ED}
 -\frac{1}{2}\delta^C_A\Gt_{BE}{}^{ED} 
-\frac{1}{2}\delta^D_B\Gt_{AE}{}^{EC} + \frac{1}{2}\delta^D_A\Gt_{BE}{}^{EC},
\ee
where the matrix $\Gt_{AE}{}^{EB}$ is traceless
 $\Gt_{AE}{}^{EA} =0$ 
and squares proportional to the identity
\be
\Gt_{AE}{}^{EB} \Gt_{BF}{}^{FC} = \frac{m^2}{\o^2} \d^C_A.
\ee
Supersymmetry also relates the coefficients $\o, \b, m$ to the flux term:  
\be
\o = \frac{c}{4}, \quad \b = -i \frac{c}{4}, 
\quad m^2 = \frac{c^2}{32 \cdot 4!} G^2,
\ee
where $G^2 = 6 G_{AB}{}^{CD} G^{AB}{}_{CD}$. As before, one finds
$c=2$ by a  backreaction analysis~\cite{LR}.

Taking the flux  
\be \label{Gform}
\Gt_{AE}{}^{ED} = 
\begin{pmatrix}
  \ \m & \ 0 & 0 & 0 \\
  0 & \m & 0 & 0 \\
  0 & 0 & -\m & 0 \\
  0 & 0 & 0 & -\m
\end{pmatrix},
\ee
for $\mu = \pm 2m$, $m\geq 0$,  
one obtains immediately the deformed theory in \cite{gomis2,hll2} as in the $\cN=8$ case.

\chapter{Open M2-branes with Flux and Modified Basu-Harvey Equation}

\section{Boundary Condition for the BLG Theory Coupled to Flux}

\subsection{Flux modified supersymmetric boundary 
condition}

We now want to consider the open case of the 
flux modified BLG theory and derive the boundary condition.
In the previous chapter we ignored all boundary terms in the derivation of the supersymmetric actions coupled to a flux and mass terms, this was due to the closed boundary conditions of the M2-branes. In this chapter these boundary terms have to be kept carefully. An analysis of open M2-branes without the flux was carried out in \cite{BermanBH}.

Such boundary contributions arise from the fermion and scalar
kinetic terms in the Lagrangian $\cL_{\cN=8}$ as these are the terms which are first and second order in derivatives respectively and so can be written as total derivatives. When considering the boundary equations of motion we are only considering the supersymmetric variations of the action \eq{LR1}, this is because we wish to study BPS configurations of branes in M-theory and general variations of the action typically break all supersymmetry. So we restrict to the supersymmetric cases which breaks some translation invariance and then impose further boundary conditions and projectors to satisfy this breaking of Poincare symmetry. Any boundary conditions from the BLG theory which are usually dropped are satisfied by such further boundary conditions, such as the Dirichlet boundary conditions we will impose. Varying the Lagrangian \eq{LR1} with respect to the supersymmetry transformations \eq{d1}-\eq{dp3} we obtain
\be
\d \cL = \frac{i}{2}\del_{\mu}\Tr(\Psib\G^\m,\d \Psi) 
- \del_\m \Tr(\d X^I, D^\m X^I  ) + \mbox{bulk terms}, 
\ee
where the `bulk terms' denote non-total derivative terms and are precisely
equal to zero when the conditions 
\eq{c2}--\eq{c1} are satisfied. 
Let us choose a boundary $\s_2 =0$ on the M2-branes worldvolume so then the variation of the action yields 
\be
\d \int d^3 \s \cL = \frac{i}{2}\int d^2 \s\left(
\Tr(\Psib\G^{2},\d\Psi) 
-2\Tr(D_{2}X^I, \Psib\G^I\e) \right),
\ee
we then obtain the supersymmetric boundary condition
\be
\label{beom}
 0=D_{\a}X^I\Psib\G^2\G^\a \G^I\e 
-\frac{1}{6}[X^I,X^J,X^K]\Psib\G^2\G^{IJK}\e 
+\frac{1}{4} X^M\Psib\G^2\Gts\G^M\e  -D_2X^I\Psib\G^I\e, 
\ee
where $\a = 0,1$ and the trace $\Tr$ is omitted for clarity. 
This is the most general supersymmetric 
boundary condition one may have for a system of open M2-branes in 
the flux background \cite{LR}. This is a boundary equation of motion which so happens to satisfy the bulk equation of motion of Basu-Harvey (with fluxes), the boundary condition $\s_2=0$ can be taken at other constant values to obtain slices of the boundary equation along the bulk equation. Due to the different number of $\G$-matrices
in each term, the equation \eq{beom} generically only has a trivial solution due to the linear independence of the $\G$-matrices.  
We may obtain non-trivial solutions when 
we impose further conditions on the matter fields $X^I, \Psi$ and projector conditions on the supersymmetry parameter $\e$. Different conditions on these fields and parameters yield different configurations of multiple M2-branes ending on other M-theory objects.

As previously mentioned, the study of the boundary conditions without flux was carried out 
in \cite{BermanBH}, the type of solutions to the boundary condition
are determined by the number of scalars obeying a Dirichlet boundary condition
(or more precisely being set equal to zero). We consider the flux configuration \eq{sf1} and perform a similar analysis to \cite{BermanBH} for the boundary condition \eq{beom} and find how much supersymmetry is broken in the different configurations of open M2-branes with flux ending on different M-theoretic objects.

\subsection{Half Dirichlet: A flux modified Basu-Harvey equation}
The half Dirichlet case corresponds to an ansatz for half of the transverse scalars to be set to zero. We shall consider the case where we set
\be
\label{hd1}
X^{3,4,5,6} =0,
\ee
which corresponds to a breaking of the R-symmetry from $SO(8)$ to $SO(4)$. Breaking the supersymmetry in the directions of $X^{3,4,5,6}$ is imposed by the projector
\be
\label{hd2}
\G^{01789(10)}\e = \e,
\ee
this is obtained from $\G^{012}\e = \e$ and $\G^{0123456789(10)} =1$. It then follows that
\be
\label{hd3}
\G^{ijk} \e =\ve^{ijkl}\G^2\G^l\e, \quad i,j,k,l = 7,8,9,10;
\ee
this term is useful for reducing the number of gamma matrices in the 3-bracket term in the boundary equation of motion \eq{beom}. Another useful relation to reduce the $\Gts$ term in \eq{beom} is given by
\be
\label{hd4}
\G^{2}\Gts\e = 2 \mu  \, \e,
\ee
where we used the form of the flux \eq{sf1}.

Using the above relations \eq{hd1}-\eq{hd4}, the boundary condition \eq{beom} is reduced to boundary equation of motion
\be
\label{preind}
 0 =D_{\a}X^i\Psib\G^2\G^\a\G^i\e 
-\frac{1}{6}\e^{ijkl}[X^i,X^j,X^k]\Psib\G^l\e 
+\frac{\m}{2} X^i\Psib\G^i\e 
- D_2X^i\Psib\G^i\e.
\ee
The first term in \eq{preind} is identically zero when we impose the projector condition on the fermion
\be
\label{ferm5}
\G^{01789(10)}\Psi = -\Psi.
\ee
This reduction in the degrees of freedom is consistent with the 1/2 BPS nature 
of the projector \eq{hd2} on the supersymmetry parameter $\e$. 
After eliminating the first term, we obtain the boundary equation of motion by factorising the fermionic terms;
\be
\label{r2}
 D_2X^i = -\frac{1}{6}\varepsilon^{ijkl}[X^j,X^k,X^l] +  \frac{\m}{2} X^i
\ee
for $ i,j,k,l = 7,8,9,10$.
This is the Basu-Harvey equation in the presence of a specific flux \eq{sf1}. Adding mass terms to the Basu-Harvey equation was first considered in \cite{BL3}To summarise, the supersymmetric boundary conditions are given by
\bea
\label{sbc}
&&\G^{01789(10)}\Psi = -\Psi, \nn\\
&&D_2X^i = -\frac{1}{6}\varepsilon^{ijkl}[X^j,X^k,X^l] +  \frac{\m}{2} X^i, \nn\\
&&X^{3,4,5,6} =0, \nn\\
&&D_\a X^i =0, \quad \quad \a=0,1.
\eea

We must now check that the boundary conditions \eq{sbc} are indeed supersymmetric, i.e. we must check that they are invariant under the supersymmetry transformations $\d$ with supersymmetry parameter $\e$ obeying $\G^{01789(10)}\e =\e$. Indeed, it is simple to show that
\be
\d X^{i'} =0, \quad i' =3,4,5,6 
\ee
and 
\be
(1+ \G^{01789(10)})\; \d \Psi =0
\ee
using the conditions \eq{hd2} and \eq{ferm5}. 
For the Basu-Harvey equation \eq{r2}, invariance of supersymmetry requires that we impose the  fermionic boundary equation
\be \label{r3}
D_2 \Psi = -\frac{1}{2}\Gamma^2 \Gamma_{ij}[\Psi,X^i,X^j] +\frac{\m}{2}\Psi =0.
\ee
Indeed this equation is invariant under the supersymmetry transformations $\d$.
The last equation in \eq{sbc} must also be invariant under supersymmetry, this requires a new boundary condition
\be
\label{fbeom}
D_\a \Psi =0, \quad \a =0,1.
\ee
So we include \eq{fbeom} to the set of boundary conditions \eq{sbc}.

In \eq{hd2} we made a choice to preserve half the supersymmetry by the choice of the projector's sign, we can instead choose to preserve the other half; this results in
\be
\G^{01789(10)}\e = - \e.
\ee
The same analysis can then be repeated to derive the boundary conditions, thus the Basu-Harvey equation is
\be \label{r2p}
D_2X^i = \frac{1}{6}\varepsilon^{ijkl}[X^j,X^k,X^l] -  \frac{\m}{2} X^i.
\ee
Applying the supersymmetry transformations on this new Basu-Harvey equation \eq{r2p} yields
\be
D_2 \Psi = \frac{1}{2}\Gamma^2 \Gamma_{ij}[\Psi,X^i,X^j] -\frac{\m}{2}\Psi =0.
\ee
This can be interpreted as a system of M2-branes stretched between a single M5-brane at $\s_2=0$.

\subsubsection{No Dirichlet: The M2-M9 system}
We may now consider the case of keeping all eight scalar fields which corresponds to having no Dirichlet boundary conditions. The object in M-theory which would be described by this would be the M9-brane \cite{berg}, so in our case we will examine the supersymmetric boundary conditions of M2-branes with flux ending on an M9-brane.

The M9-brane projector condition is defined as 
\be
\G^{013456789(10)}\e = \e,
\ee
which implies
\be
\G^2\e = \e
\ee
as a result of using $\G^{0123456789(10)}=1$. We also have the M2-brane projector condition $\G^{012}\e=\e$, so we actually have:
\be
\G^2\e = \e = \G^{01}\e.
\ee
The condition corresponding projector condition
\be
\G^{013456789(10)}\Psi = \Psi
\ee
is applied on the fermion. 

We proceed to find the boundary equation of motion by considering the boundary condition \eq{beom} and using the above conditions on $\e$ and $\Psi$. It is clear that the first term in \eq{beom} is zero by inserting the projectors. The penultimate term is interesting as it is a linear combination of two products of three and five gamma matrices. Due to the independence of the gamma matrices as a basis of the Clifford algebra, we must impose that the coefficient of the $\G^{(5)}$ term is zero, namely $X^I=0$. Thus the term with the 3-bracket must also be zero. We have
\be
\label{M9-1}
D_2 X^I = 0
\ee
and
\be
\label{M9-2}
X^I=0.
\ee
So we see that there are no non-trivial solutions for a system of M2-branes with a non-zero flux ending on an M9-brane.

Turning off the flux in \eq{beom} results in the problematic term described above to have a combination of three and five gamma products to be zero. Then we only have the three bracket term, which is proportional to $\G^{(3)}$ and the derivative term proportional to $\G^I$. The first term in the boundary condition vanishes for the same reason as in the massive case. Again, independence of the gamma matrices yields
\be
\label{BermanM9}
D_2 X^I = 0, \quad [X^I,X^J,X^K] =0.
\ee
This has been interpreted as
an M9-brane occupying the directions
013456789(10) where the M2-branes end on \cite{BermanBH}.

In the presence of flux, the system of M2-branes cannot end on an M9-brane supersymmetrically as we only obtain a trivial solution for the scalar fields.
This is a result of our open M2-branes with flux analysis, it would be interesting to motivate this from another area of string theory. 
One can consider the projectors of the M2 and M9-branes with flux and show that they are incompatible to prove that they have indeed only the trivial solution.
To carry out this analysis, one needs 
to first construct the supergravity solution of an M9-brane with a constant flux 
and then determine the preserved supersymmetry as performed in \cite{berg} for the case without flux.

Another way to approach the problem is to compactify M-theory to \IIA theory on the $x^{10}$ direction. The M2-M9 system then reduces down to a D2-D8 system. The D8-brane is endowed with a worldvolume NS-NS $B$-field in the 78, 79 or 89 directions as a result of the reduction of the RR 4-form flux of M-theory and the projector condition is given by
\be
e^{-a/2} \G^{013456789(10)}e^{a/2} \e =\e, 
\ee
where $a = \frac{1}{2}Y_{IJ}\G^{IJ} \G^{(10)}$
and $Y$ is a nonlinear function of $B$ whose explicit form can be found in
\cite{d11}. Note that only the 78, 79 or 89 components are non-zero in our case.
It is then clear that the supersymmetry preserved by the D8-brane
is incompatible with $\G^{012} \e =\e$ of the D2-brane.
Therefore the D2-D8 system and the M2-M9 system are not supersymmetric.

\subsubsection{All Dirichlet: M-wave}
In M-theory there is an object called the M-wave, this is the uplift of the D0-brane to M-theory and is the (1+1)-dimensional gravitational wave analogue.
For a system of M2-branes ending on an M-wave, we set all the eight scalars to zero at the boundary. As a result, all the modifications due to flux vanish and 
the terms proportional to $X^I$ vanish. We are left with the boundary condition
\be
D_2 X^I \Psib \G^I \e=0.
\ee
This can be solved immediately if one imposes the projection conditions
\be
(1-\G^2) \e =0, \quad (1+ \G^2) \Psi =0.
\ee
The solution has been interpreted as an M-wave where the M2-branes end on \cite{BermanBH}.

\section{Boundary Condition for the ABJM Theory Coupled to Flux}

We turn to the $\cN=6$ theory of $N$ open M2-branes probing the orbifold 
$\mathbb{C}^4/\mathbb{Z}_k$ and repeat the analysis of the $\cN=8$ case to find the various boundary equations of motion. To proceed we note that the boundary contributions will come from a supersymmetric variation of the scalar and fermionic kinetic terms
\be
\d \cL = -2\del_\m \Tr(\d\bar{Z}_A,D^\m Z^A) 
- i\del_\m \Tr (\psib^A,\g^\m \d\psi_A) + \mbox{bulk terms},
\ee
Choosing the boundary condition $\s_2=0$ gives the 
boundary equation of motion
\bea
\label{bvn6}
0=&-&2i\Tr \Big(\psib^B \e_{AB},D^2 Z^A\big) 
- i\Tr\Big[\psib^A\g^2, \g^\m \e_{AB}D_\m Z^B \nn\\ 
&+& \frac{2\pi}{k}\Big(-\e_{AB}(Z^C \bar{Z}_C Z^B  - Z^B \bar{Z}_C Z^C )   
+ 2\e_{CD}Z^C\bar{Z}_A Z^D\Big) + \omega \e_{DF} Z^F \Gt_{AE}{}^{ED} \Big].
\quad \qquad
\eea
The nice form of the flux \eq{Gform} can be written simply as 
\be
\Gt_{AE}{}^{ED} = \m \d_A^D\eta_A,
\ee
where $\eta_A$ is the sign defined as 
\be
  \eta_A = 
  \begin{cases}
   +1 & \text{if $A=1,2$} \\
   -1 & \text{if $A=3,4$}
  \end{cases} .
\ee
Substituting the form of the flux in \eq{bvn6}, we obtain the boundary equation of motion
\bea \label{bcc1}
0= \psib^A&\Big[&2D^2 Z^B -\g^2 \g^\m D_\m Z^B 
+ \g^2 \frac{2\pi}{k} \left(Z^C \bar{Z}_C Z^B  
- Z^B \bar{Z}_C Z^C \right)  \nn \\ &&- \frac{\m}{2} \, \eta_A\g^2 Z^B \Big] 
\e_{AB}
-\frac{4\pi}{k}\psib^F\g^2 \e_{AB} Z^A \bar{Z}_F Z^B,
\eea
where the trace products are understood. 
This is the most general 
supersymmetric
boundary equation of motion for 
open M2-branes in the $\cN=6$ theory with our specific flux configuration.

To analyse the boundary condition, we
introduce the following notation; $A=(a,i)$ where we split the $SU(4)$ indices and 
denote $Z^A=(X^a,Y^i),\ \psi_A = (\chi_a,\xi_i)$, where $a=1,2$  corresponds 
to the spacetime directions $3456$ and $i=1,2$ corresponds to the spacetime directions $789(10)$. 
The antisymmetric supersymmetry parameter $\e_{AB}$ is in the $\textbf{6}$ 
representation of $SU(4)$, it decomposes as \cite{BermanBH}
\be
\e_{AB} = 
\begin{pmatrix}
  \ \varepsilon_{ab}\e & \e_{ai} \\
  -\e_{ai} & \ \varepsilon_{ij}\tilde{\e}
\end{pmatrix}.
\ee
Recasting the boundary equations of motion \eq{bcc1} in terms of the new fields yields:
\bea\label{1of4}
0= \chib^a\varepsilon_{ab}\Big[&&2D^2 X^b -\g^2 \g^\m D_\m X^b 
+ \g^2 \frac{2\pi}{k} \left(Z^C \bar{Z}_C X^b  - X^b \bar{Z}_C Z^C \right) 
- \frac{\m}{2} \g^2 X^b \Big] \e \nn  \\
&&-\frac{4\pi}{k}\e_{ab}\psib^F\g^2\e X^a \bar{Z}_F X^b, 
\eea
\bea \label{2of4}
0= \chib^a\Big[&&2D^2 Y^i -\g^2 \g^\m D_\m Y^i 
+ \g^2 \frac{2\pi}{k} \left(Z^C \bar{Z}_C Y^i  - Y^i \bar{Z}_C Z^C \right) 
- \frac{\m}{2} \g^2 Y^i \Big] \e_{ai} \nn \\
&&-\frac{4\pi}{k}\psib^F\g^2\e_{ai} X^a \bar{Z}_F Y^i, 
\eea
\bea \label{3of4}
0=\xib^i\Big[&&2D^2 X^a -\g^2 \g^\m D_\m X^a 
+ \g^2 \frac{2\pi}{k} \left(Z^C \bar{Z}_C X^a  - X^a \bar{Z}_C Z^C \right) 
+ \frac{\m}{2} \g^2 X^a \Big] \e_{ai} \nn \\
&&-\frac{4\pi}{k}\psib^F\g^2\e_{ai} Y^i \bar{Z}_F X^a,
\eea
\bea
\label{4of4}
0= \xib^i\varepsilon_{ij}\Big[&&2D^2 Y^j -\g^2 \g^\m D_\m Y^j 
+ \g^2 \frac{2\pi}{k} \left(Z^C \bar{Z}_C Y^j  - Y^j \bar{Z}_C Z^C \right) 
+ \frac{\m}{2} \g^2 Y^j \Big] \tilde\e \nn \\
&&-\frac{4\pi}{k}\e_{ij}\psib^F\g^2\tilde\e Y^i \bar{Z}_F Y^j.
\eea
These four equations are independent and we will have to impose further conditions on the fields in order to obtain the boundary equations of motion we expect.

\subsection{Flux modified Basu-Harvey equation}

As with the $\cN=8$ case we can consider the half Dirichlet case which amounts to setting half the scalar fields to zero, so we take $Y^i=0$. This condition reduces the R-symmetry from $SU(4)$ to $SU(2)$.
First we consider equations \eq{1of4} and \eq{3of4}, it turns out that  
the second term  ``$\g^\m D_\m X^a$''  in these equations
vanishes\footnote{We will come back to this later} for $\mu =0,1$. Assuming that this is true,  the boundary equations of motion \eq{1of4} and \eq{3of4} become
\be
\label{equiv1}
0= \chib^a\varepsilon_{ab}\Big[D^2 X^b 
+ \g^2 \frac{2\pi}{k} \left(X^c \bar{X}_c X^b  - X^b \bar{X}_c X^c \right) 
- \frac{\m}{2}\g^2 X^b \Big] \e 
-\frac{4\pi}{k}\ve_{cd}\chib^a \g^2 \e X^c \bar{X}_a X^d
\ee
and
\bea
\label{equiv2}
0=\xib^i\Big[D^2 X^a + \g^2 \frac{2\pi}{k} \left(X^c \bar{X}_c X^a  
- X^a \bar{X}_c X^c \right) + \frac{\m}{2}\, \g^2 X^a \Big] \e_{ai}.
\eea
The two equations are not compatible with each other in general, however
it is possible to impose suitable supersymmetry projection 
conditions on the spinors $\e$ and $\e_{ai}$ so that these two equations become
equivalent for the bosonic parts. The needed projector conditions are
\bea 
&&(1+\g^2)\e = 0 = (1+\g^2)\tilde\e \label{proj1}\\
&&(1-\g^2)\e_{ai} = 0, \label{proj2}
\eea
or equivalently
\bea 
&&(1-\g^2)\e = 0 = (1-\g^2)\tilde\e, \label{proj1p}\\
&&(1+\g^2)\e_{ai} = 0. \label{proj2p}
\eea
As a result, \eq{equiv1} and \eq{equiv2} are identical since we can use the simple identity
\be
X^c \bar{X}_c X^b  - X^b \bar{X}_c X^c = 
\varepsilon^{ba}\varepsilon^{cd}X^c\bar{X}_a X^d.
\ee
The modified Basu-Harvey equation with a mass term is then obtained
\be \label{abjm-BH}
D_2 X^a \pm \frac{2\pi}{k}\left(X^c \bar{X}_c X^a  
- X^a \bar{X}_c X^c\right) \pm \frac{\m}{2} X^a=0,
\ee
where the $+$ sign corresponds to the choice \eq{proj1}, \eq{proj2} and
the $-$ sign corresponds to the choice \eq{proj1p}, \eq{proj2p}.
We may also write the Basu-Harvey equation in terms of the Hermitian 3-bracket
\be
\label{n6BH}
D_2 X^a \pm \left[X^c, X^a; \bar{X}_c\right] \pm \frac{\m}{2} X^a=0.
\ee
This is the mass deformed Basu-Harvey equation for the flux modified $\cN=6$ theory. The complete set of supersymmetric boundary conditions are given by
\bea
\label{n6bcs}
&&D_2 X^a \pm \left[X^c, X^a; \bar{X}_c\right] \pm \frac{\m}{2} X^a=0 \nn\\
&&D_\a X^a =0 \nn\\
&&Y^i =0.
\eea
In the following we consider the choice of projectors \eq{proj1}, \eq{proj2} which give $+$ signs in the Basu-Harvey equation
\be \label{abjm-BHp}
D_2 X^a + \frac{2\pi}{k}\left(X^c \bar{X}_c X^a  
- X^a \bar{X}_c X^c\right) + \frac{\m}{2} X^a=0.
\ee 
The analysis for the other choice of projector conditions is exactly the same.
 
We now wish to check that the boundary conditions \eq{n6bcs} are indeed supersymmetric. We have
$\d Y^i = i \e^{ia} \chi_a + i \ve^{ij} \tilde \e \xi_j$, the boundary 
condition $Y^i =0$ is indeed supersymmetric invariant after imposing \eq{proj1} and \eq{proj2}, if 
\bea
(1+\g^2)\chi_a &=&0, \label{proj3}\\
(1-\g^2)\xi_i &=&0. \label{proj4}
\eea
The conditions \eq{2of4} and \eq{4of4}, which read
\be
\chib^a \e_{ai} D_2 Y^i =0,
\ee
and
\be
\xib^i \ve_{ij} \tilde\e D_2 Y^j =0,
\ee
after the boundary condition $Y^i=0$ are now satisfied immediately as a result of the projection conditions
\eq{proj1}, \eq{proj2}, \eq{proj3}, \eq{proj4}.
It is also easy to see that these projection conditions 
are supersymmetric invariant.
Moreover, supersymmetry on \eq{abjm-BH} requires the fermionic boundary 
equations \\
\be
D_2 \chi_c - [\chi_c,X^a;\bar{X}_a] + 2[\chi_d,X^d;\bar{X}_c]
-\frac{\m}{2} \chi_c =0,
\ee
\be
D_2 \xi_j + [\xi_j,X^a;\bar{X}_a] - \ve_{jk} \ve_{ab} [X^a,X^b;\xi_k]
-\frac{\m}{2} \xi_j =0.
\ee

As for the above assumption of the vanishing of the
terms of the form  ``$\g^\m D_\m X^b$''  in the equations
\eq{1of4} and \eq{3of4}, one can see that it follows immediately from
the projection conditions \eq{proj1} and \eq{proj3}, and respectively
\eq{proj2} and \eq{proj4}. 

The Basu-Harvey equation \eq{abjm-BHp} can be readily solved by employing 
the ansatz
\be
X^a(s) = f(s) R^a,
\ee
where $s=x_2$ and $R^a$ are $N\times N$ matrices satisfying the relation
\be \label{Q}
R^c R^\dag_c R^a -R^a R^\dag_c R^c = -R^a,
\ee
then we obtain
\be \label{f-dep}
f' - \frac{2\pi}{k} f^3 + \frac{\mu}{2} f=0.
\ee
The equation \eq{Q} has been solved in \cite{gomis2} and 
the irreducible solution is
\be
(R^1)_{mn} = \d_{m,n} \sqrt{m-1}, \quad
(R^2)_{mn} = \d_{m-1,n} \sqrt{N-m+1}, \quad m,n =1, \cdots, N.
\ee
A direct sum of such blocks is also a solution. 
The equation \eq{f-dep} is the same equation as in the $\cN=8$
theory. This is how the M2-M5 intersection is represented in the $\cN=6$  theory.

\subsection{M9 and M-Wave Solutions}

Finally, let us comment briefly on the no Dirichlet and all Dirichlet cases. 
For the no Dirichlet case, we find only 
the trivial solution $X^a=Y^i=0$ as in the $\cN=8$ theory.
As for the  all Dirichlet case,
since  the flux modifications all go away when
all the scalars are set to zero at the boundary,  the boundary conditions 
\eq{1of4}-\eq{4of4} reduce to exactly the same form as 
in flux-less case and one gets M2-branes ending on an M-wave \cite{BermanBH}.

\section{Summary}
In this Chapter we generalised the boundary analysis of \cite{BermanBH} to include fluxes as constructed in \cite{LR}. In general the boundary condition \eq{beom} has a trivial solution, this is due to the $\G$-matrices being linearly independent. To obtain non-trivial solutions, we had to impose conditions on the transverse scalar fields which required us to demand further supersymmetry projector conditions; preserving supersymmetry in the surviving directions. This lead to the Basu-Harvey equation with a mass deformation which describes the profile of multiple M2-branes ending on an M5-brane. The system is $1/2$-BPS due to having to impose the projector \eq{hd2}, \eq{ferm5}. The boundary analysis was carried out for both BLG and ABJM theories. A generalisation for the M-theory Myers terms was constructed in \cite{as} and it would be interesting to compare the analysis of this with the results obtained here.

It was also possible to consider the mass deformed M2-branes ending on an M9-brane, but it turns out that only the trivial solution is possible here. However in \cite{BermanBH}, we can see that there is a solution \eq{BermanM9}. So with the flux, it is not possible to find a supersymmetric solution.

\chapter{Lorentzian 3-Algebras and a Reduction to D2-Branes}

The BLG theory \cite{BL1,BL2,BL3} was constructed in terms of 
the Lie 3-algebra $\cA_4$, which turned out to be the unique indecomposable 3-algebra solution to the generators that satisfied the fundamental identity \eq{fi} for a Euclidean inner product \cite{papa1,gaunt}. 
The motivation for using 
$\cA_4$  was to construct the  
Basu-Harvey equation \cite{BH} which is a
generalisation of the Nahm equation for describing M2-branes ending on an abelian M5-brane.

In order to find a new 3-algebra which is not given by $\cA_4$ we must relax some assumptions, if we consider a Lorentzian inner product instead of a Euclidean one this can indeed give us a new 3-algebra. It was shown in \cite{gomis-lor,v-lor,Ho1,gomis3,schwarz,mukhi} that the Lorentzian BLG theory reduces to an exact $\cN=8$ SYM theory of multiple D2-branes, as opposed to the novel Higgs mechanism reduction for the original BLG theory \cite{MP} which gives higher order corrections.
 
In this chapter we review the Lorentzian BLG theory in the first section and derive the D2-brane action but will keep the boundary terms in the action. The literature \cite{gomis-lor,v-lor,Ho1,gomis3,schwarz,mukhi} is collated and made consistent, we will adopt the notation of \cite{Ho1}. In the second section we will apply the Lorentzian 3-algebra to the flux modified BLG theory and analyse its reduction, carefully keeping the boundary terms for the action. In the last section we will derive the supersymmetric boundary condition on open D2-branes with flux and find the mass deformed Nahm equation.

\section{Lorentzian 3-algebras and D-branes}
The main concept behind the Lorentzian 3-algebra is to introduce two extra generators to a standard Lie algebra $\cG$ and then to impose a 3-bracket which will give us {\it Lie algebra} structure constants, let us review the approaches of \cite{Ho1,schwarz} in particular.

Consider a Lie algebra $\cG$ of a compact gauge group $G$ with generators $T^\ri,\ri=1,...,\dim \cG$ such that
\be
[T^\ri,T^\rj]=f^{\ri\rj}{}_\rk T^\rk,
\ee
equipped with a Killing form $h^{\ri\rj}=\d^{\ri\rj}$. Then the Lorentzian 3-algebra is defined by a set of generators $T^a=\{T^+, T^-, T^\ri \}$, with a 3-bracket specified by
\bea
\left[T^a,T^b,T^c\right] &=& f^{abc}{}_d T^d,  \quad a=+,-,\ri,\\
\label{L3alg1}
\left[T^{-},T^a,T^b\right] &=& 0, \\
\label{L3alg2} \left[T^+,T^\ri,T^\rj\right] &=& f^{\ri \rj}{}_\rk T^\rk, \\
\left[T^\ri,T^\rj,T^\rk\right] &=& f^{\ri \rj \rk}T^{-}.
\eea
This algebra satisfies the fundamental identity \eq{fi}. We still impose the relation \eq{3fund} and this means that the invariant metric on this Lorentzian 3-algebra is
\bea
\label{lmetric}
 \Tr (T^{-},T^{+}) &=& -1, \nn \\
 \Tr(T^{\ri},T^{\rj}) &=& \d^{\ri \rj},
\eea
with all other products vanishing. 
One may expand all the fields of the BLG theory with respect to the generators $T^a$ in the following way
\be
X^I = X^I_a T^a = X^I_{-}T^{-} + X^I_+ T^+ + \Xh^I, 
\ee
where $\Xh^I = X^I_\ri T^\ri$ are the modes corresponding to the Lie algebra $\cG$, similar expressions can be obtained for the fermions $\Psi$ and gauge field $\At_\m$.
By applying all the above rules one obtains the action for a Lorentzian BLG theory \cite{gomis3},
\bea
\label{lorentz}
\cL_{Lorentz} = &&-\frac{1}{2}\Tr(\Dh_\m \Xh^I - B_\m X_+^I)^2 
+\del^\m X^I_+(\del_\m X_-^I-\Tr(B_\m, \Xh^I))
+\frac{1}{2}\varepsilon^{\m\n\l}\Tr(B_\l F_{\m\n})\nn\\
&&+\frac{i}{2}\Tr(\hat{\Psib}\Gamma^\m,(\Dh_\m\Psih -B_\m\Psi_+)) 
-\frac{i}{2}{\Psib}_+\Gamma^\m(\del_\m\Psi_- -\Tr(B_\m,\Psih)) 
-\frac{i}{2}\Psib_-\G^\m\del_\m\Psi_+ 
\nn\\
&&+\frac{i}{2}\Tr(\hat{\Psib}\G^{IJ}X_+^I[\Xh^J,\Psih]) 
+\frac{i}{4}\Tr(\hat{\Psib}\G^{IJ}[\Xh^I,\Xh^J]\Psi_+) 
-\frac{i}{4}\Tr(\hat{\Psib}_+\G^{IJ}[\Xh^I,\Xh^J]\Psih)\nn\\
&&-\frac{1}{12}\Tr\left(X_+^I[\Xh^J,\Xh^K]+X_+^J[\Xh^K,\Xh^I] 
+ X_+^K[\Xh^I,\Xh^J]\right)^2,
\eea
where $I=3,\dots,10$.
The gauge field $A_\m$ is for the compact gauge 
group $G$. The gauge field $B_\m$ is defined by
$B_\m = A_{\m \ri\rj}f^{\ri\rj}{}_\rk T^\rk$ and the theory is invariant under
an extra non-compact gauge symmetry associated with $B_\m$:
\bea \label{sgs}
\d B_\m &=& D_\m \z, \quad \d \Xh^I = \z X_+^I, 
\quad \d X_-^I = \Tr(\z,\Xh^I), \nn\\
\d \Psih &=& \z\Psi_+, \quad \d\Psi_- = \Tr(\z,\Psih).
\eea
The fields $A_\m, B_\m, \Xh^I, \Psih$ transform in the adjoint of the gauge group and all the other fields are singlets. The supersymmetry transformations decompose under the new generators as
\bea
\label{d2susys1}
\d_0 X_{-}^I &=&i\eb\Gamma^I\Psi_{-},\\ 
\d_0 X_+^I &=& i\eb\Gamma^I\Psi_+, \\
\d_0\Xh^I &=& i\eb\Gamma^I\Psih, \\
\d_0 \Ah_\m &=& \frac{i}{2}\eb\G_\m\G^I(X_+^I\Psih - \Xh^I\Psi_+), \\
\d_0 B_\m &=& i\eb\G_\m\G^I[\Xh^I,\Psih], \\
\d_0\Psi_{-} &=& (\del_\m X_{-}^I - B_\m X^I)\G^\m\G^I\e - \frac{1}{3} \Xh^I\Xh^J\Xh^K \G^{IJK}\e, \nn \\ 
\d_0\Psi_+ &=& \del_\m X_+^I \G^\m\G^I\e, \\
\label{d2susys2}
\d_0\Psih &=& \Dh_\m \Xh^I \G^\m\G^I\e 
-\frac{1}{2}X_+^I[\Xh^J,\Xh^K]\G^{IJK}\e.
\eea

A special feature of the Lagrangian \eq{lorentz} 
is that the fields $X_-^I, \Psi_-$  
appear linearly. For convenience let us collect the terms containing 
$X_-^I, \Psi_-$, they are
\bea
\cL_{gh} = \partial_\m X_+^I\partial^\m X_{-}^I 
-i\Psib_{+}\G^\m \partial_\m\Psi_-.
\eea
We have called these ghost terms since $\cL_{gh}$ has an 
indefinite metric and is hence non-unitary. To proceed, we may integrate out $X^I_-, \Psi_-$ and obtain the equations of motion:
\bea
\label{eom-a}\del^2X_+^I &=&0, \\
\label{eom-b}\G^\m\del_\m\Psi_+ &=&0.
\eea
A solution to \eq{eom-a} and \eq{eom-b} which preserves gauge symmetry 
and supersymmetry is given by 
\bea
\label{vev1}
X_+^I &=& v_0\d_{10}^I, \\
\label{vev2} \Psi_+ &=& 0,
\eea
where $v_0$ is some real constant. If we substitute \eq{vev1}, \eq{vev2} and integrate out the $B_\m$ field in the Lagrangian \eq{lorentz}, we obtain  
\bea
\label{n8sym}
\cL = &&-\frac{1}{2}(\Dh_\m \Xh^I)^2 
+ \frac{i}{2}\Tr(\hat{\Psib}, \G^\m D_\m \Psih) 
- \frac{1}{4v_0^2} \Tr(F_{\m\n},F^{\m\n}) 
-\frac{v_0^2}{4}\Tr([\Xh^I,\Xh^J],[\Xh^I,\Xh^J])\nn\\
&&+ \frac{iv_0}{2}\Tr(\hat{\Psib}\G^{(10)I}[\Xh^I,\Psih]) 
+\del_\l\left(\frac{\varepsilon^{\m\n\l}\Fh_{\m\n}\Xh^{10}}{2v_0}\right),
\eea
where we 
have kept the boundary term for later discussions on flux modified D2-brane theories. 
For a closed theory, the Lagrangian \eq{n8sym} is the maximally supersymmetric 
$\cN=8$ SYM theory in 2+1 dimensions describing the low energy D2-brane theory. 
The only dependence of $\Xh^{10}$ is in the boundary term for the open theory, this means we must impose a boundary condition in order to decouple the field from the theory to obtain the correct field content and $SO(7)$ invariance. Furthermore, supersymmetric invariance of the boundary theory gives non-trivial boundary conditions, namely the Nahm equation, so we will study these in the subsequent sections. In particular more projector conditions must be imposed on $\e$ and so this indicates the presence of a brane at the boundary, we will examine the D2-D4 1/2 BPS configuration.

\section{Multiple D2-branes in a background flux}
In this section we will extend the results of \cite{LR} to the case of Lorentzian 3-algebras and in doing so we will obtain the mass-deformed D2-brane theory. It is not difficult to see that by applying the rules of the previous section in \eq{L-flux} and \eq{L-mass} we obtain the flux and mass terms for the model
\bea
\label{Lf}
\cL_{flux} &=& 2\Gt_{IJKL} \Tr( X^I,[X^J,X^K,X^L]) \nn \\
&=& -8\Gt_{IJKL} X_+^I\Xh^L [\Xh^J,\Xh^K], \\
\label{Lm}
\cL_{mass} &=& -\frac{1}{2}m^2 \Tr( X^I, X^I ) 
-\frac{i}{8}\Tr( \Psib,\G^{IJKL}\Psi) \Gt_{IJKL} \nn \\
&=& -\frac{1}{2}m^2 \Xh^I \Xh^I +m^2 X_+^I X_{-}^I \nn\\
 &&-\frac{i}{8}\hat{\Psib}\G^{IJKL}\Psih\Gt_{IJKL} 
+\frac{i}{4}\Psib_{+}\G^{IJKL}\Psi_-\Gt_{IJKL}.
\eea
As an aside, we note it is easy to check that the gauge symmetry \eq{sgs}
extends to the flux and mass Lagrangians \eq{Lf}, \eq{Lm}. We will take
$\mu =2m$ in the following analysis as chosen in \cite{LR}.

The equations \eq{eom-a} and \eq{eom-b} are modified due to the presence of of new terms in \eq{Lf} and \eq{Lm} linear in $T^-$ modes, the resulting changes are
\bea
&&\partial^2 X_+^I -m^2 X_+^I = 0, \label{eom-X} \\
&&i\Gamma^\m\partial_\m\Psi_+ -\frac{i}{4}\G^{IJKL}\Psi_+\Gt_{IJKL} = 0.
\label{eom-Psi}
\eea
These are the Klein-Gordon and Dirac equations respectively. 
We will next show that one is able to 
pick solutions to $X_+^I$ and $\Psi_+$ which preserves gauge invariance 
and supersymmetry.

The supersymmetry transformations remains the same for the bosons 
as in \eq{d2susys1} but the flux modifies the supersymmetry 
transformations for the
fermions
\bea
\label{susy-1f}
&& \d\Psi_{-} = (\del_\m X_{-}^I 
-B_\m X^I)\G^\m\G^I\e 
- \frac{1}{3} \Xh^I\Xh^J\Xh^K  \G^{IJK}\e 
+ \frac{1}{4}\Gts\G^M\e X_{-}^M,\qquad  \\ 
\label{susy+mode}
&& \d\Psi_+ = \del_\m X_+^I \G^\m\G^I\e 
+ \frac{1}{4}\Gts\G^M\e X_+^M, \\
\label{susyf}
&& \d\Psih = \Dh_\m \Xh^I \G^\m\G^I\e 
-\frac{1}{2}X_+^I[\Xh^J,\Xh^K]\G^{IJK}\e
+ \frac{1}{4}\Gts\G^M\e \Xh^M.
\eea
Using a $(-++)$ metric for spacetime, the simplest non-trivial solution to \eq{eom-X} and \eq{eom-Psi} is
\bea
\label{psi+}
\Psi_+ &=& 0, \\
\label{coupling}
  X_+^I &=& 
  \begin{cases}
   v_0e^{m\s_1}\d_{10}^I,\ \ \  \text{or}\\
   v_0e^{imt}\d_{10}^I ,
  \end{cases}
\eea
where $v_0 \in \mathbb{R}$ and we pick the real solution in \eq{coupling} for physical reasons. 

We will now consider the first solution with the $\s_1$ dependence and for convenience we
will denote this solution with $v=v_0e^{m\s_1}$ below.
It is easy to see that the solution \eq{psi+} is supersymmetrically
invariant. In fact for the flux \eq{gen-G}, we have
$\delta \Psi_+ = m v \G^{2(10)} ( 1- \G^2 R') \e$,
where $R'$ is defined by $R \G^{(10)} = \G^{(10)} R'$. Therefore the
configuration \eq{psi+} is supersymmetrically invariant for $\e$ satisfying 
\be \label{GRp}
(1-\G^{2} R')\e = 0.
\ee
Since the projectors 
\eq{c0p}, \eq{GRp} commute, 
8 supersymmetries are preserved and so this is a 1/2-BPS configuration.

Now that we have examined what solutions to $X_+^I, \Psi_+$ we can have, we substitute the solution \eq{psi+} and the $\s_1$ dependent solution in \eq{coupling} and integrate
out the $B_\mu$ field. The Lagrangian that consists of \eq{n8sym}, \eq{Lf} and \eq{Lm} then reads
\bea
\label{md2}
\cL = &&-\frac{1}{2}(\Dh_\m \Xh^A)^2 
- \frac{1}{4v^2} \Tr(F_{\m\n},F^{\m\n}) 
-\frac{v^2}{4}\Tr([\Xh^A,\Xh^B],[\Xh^A,\Xh^B]) \nn\\
&& + \frac{i}{2}\Tr(\hat{\Psib}, \G^\m D_\m \Psih) 
+ \frac{iv}{2}\Tr(\hat{\Psib},\G^{(10)A}[\Xh^A,\Psih]) \nn\\
&& - 8v \Gt_{(10)ABC}\Tr(\Xh^A,[\Xh^B,\Xh^C]) 
-\frac{i}{2}\Tr(\hat{\Psib},\G^{(10)ABC}\Psi)\Gt_{(10)ABC}
\nn\\
&& -\frac{1}{2}m^2\Tr(\Xh^A,\Xh^A) 
-\frac{i}{8}\Tr(\hat{\Psib},\G^{ABCD}\Psi)\Gt_{ABCD} \nn\\
&& +\del_\l\left(\frac{\varepsilon^{\m\n\l}}{2v}\Tr(\Fh_{\m\n},\Xh^{10})\right) 
-\frac{m}{2}\del_1\Tr(\Xh^{10},\Xh^{10}),
\eea
where the indices $A,B,C,D  =3, \cdots, 9$ are the $SO(7)$ invariant indices of the R-symmetry for the D2-brane theory. 
The boundary terms in \eq{md2} are the total derivatives, these terms vanish for a closed boundary condition. They are non-trivial for the open boundary case and we shall discuss them in some more detail later. The RR coupling $\Gt$ splits into a 4-form and 3-form under the integral in the action, this is a feature that will be discussed later.

Since the Lambert-Richmond action is supersymmetric with 16 supercharges and the 
solution \eq{coupling} is 1/2-BPS, by construction our action 
\eq{md2} is supersymmetric and preserves 8 supersymmetries:
\bea
\d \Xh^A &=& i \eb \G^A \Psih,\\
\d\Psih &=& \Dh_\m \Xh^A \G^\m\G^A\e 
-\frac{1}{2v} \e_{\m\n\l}\Fh^{\n\l} \G^\m \G^{10} \e
-\frac{v}{2}[\Xh^A,\Xh^B]\G^{AB(10)}\e
+ \frac{1}{4}\Gts\G^A\e \Xh^A,\quad \;\;\; \\
\d \Ah_\m &=& \frac{i v}{2} \eb \G_\m \G^{(10)} \Psih.
\eea
These are found by substituting the solutions \eq{psi+}, \eq{coupling} into the supersymmetry transformations \eq{d2susys1}-\eq{d2susys2} and the contributions due to the flux in \eq{susy-1f}-\eq{susyf}. Note that here the variable $v$ depends on the mass $m$ and so the modifications to the supersymmetry transformations of SYM in $2+1$ dimensions is non-trivial but still preserves some supersymmetry.

The Lagrangian \eq{md2} can be understood as the worldvolume theory of  
D2-branes with a space(time) dependent
coupling $g_{\rm YM}= v $ and coupled to NS-NS and R-R fluxes.
In 10 dimensions, the flux 
$\Gt_{ABCD}$ is identified with the R-R 4-form flux of 
the 3-form potential $C_3$ 
and $\Gt_{(10)ABC}$  is identified with the NS-NS 3-form flux of 
the 2-form potential $B_2$ of the D2-brane theory. 
The term in \eq{md2} proportional to $\Gt_{(10)ABC}$ 
can be traced back 
as the low energy limit of the Myers' action \cite{myers}, 
together with its superpartner. 
The terms proportional to $m^2$ and $\Gt_{ABCD}$ are typical of 
couplings to the R-R fields. Supersymmetric Yang-Mills theories with a spacetime
dependent coupling were originally constructed in \cite{bak,dzf} 
and are known as Janus field theories. 

The massive D2-brane theory with a boundary contains terms with the field $\Xh^{10}$, these arise from the contributions from the total derivatives. The presence of this field means that there is an additional symmetry in the theory that we do not want if we are really describing D2-branes with an $SO(7)$ R-symmetry. These decouple in the closed case trivially but for the open case we need to examine this boundary condition further to check that this does indeed decouple in a way that will preserve the supersymmetry of the theory without breaking more or all of it, this shall be discussed next.

\section{Multiple D2-branes ending on a D4}

The supersymmetric boundary conditions for the flux
modified Lorentzian Bagger-Lambert theory are now derived and we will discuss the issues discussed at the end of the previous section. Since the 
field $X^I_-$ has been integrated out, the boundary condition \eq{r2} cannot be 
applied immediately and one needs to derive the boundary condition 
from the reduced action \eq{md2} directly.

There are two possible boundary conditions one can write down for \eq{md2}, these correspond to choosing $\s_1=0$ and $\s_2=0$ respectively. The reason why these two choices are unique is because rotational invariance is broken due to the $\s_1$ dependence of the coupling $v$ and so the two are no longer equivalent. The two cases which decouple the $\Xh^{10}$ are
\bea
\Tr(2\Fh_{02}\Xh^{10} +m v (\Xh^{10})^2) =0, \quad &&\mbox{boundary at $\s_1=0$}, 
\label{fbc1}
\\
\Tr(\Fh_{01} \Xh^{10}) =0, \quad && \mbox{boundary at $\s_2=0$}. \label{fbc2}
\eea
It so turns out that \eq{fbc1} breaks all supersymmetry, we will choose the boundary condition \eq{fbc2} for now and justify it in what follows.

To find the boundary equation of motion we repeat the procedure of the mass deformed BLG theory, so we perform a supersymmetric variation of the Lagrangian given by
\be
\d \cL = \frac{i}{2}\partial_\m\Tr(\hat{\Psib},
\Gamma^\m\d\Psih) - \partial_\m\Tr(\d \Xh^A,
 \Dh^\m \Xh^A) +\mbox{bulk terms}.
\ee
The boundary condition $\s_2 =0$ under the integral of the action gives 
\be \label{dL}
\d \int\mathrm{d}^3 \s \cL = \frac{i}{2} 
\int\mathrm{d}^2 \s\left( \Tr(\hat{\Psib},\Gamma^2\d\Psih)
-2 \Tr(\d X^A, \Dh_2 \Xh^A)\right),
\ee
so we obtain the boundary equation of motion
\bea \label{bc-lorf}
&&\Dh_\m \Xh^A\hat{\Psib}\Gamma^2\Gamma^\m\Gamma^A\e 
 -\frac{v}{2}[\Xh^A,\Xh^B]\hat{\Psib}\Gamma^2\Gamma^{(10)AB}\e 
 -\frac{1}{2v}\e_{\m\n\l}\Fh^{\n\l} \hat{\Psib} \G^2 \G^{\m (10)} \e \nn\\
 &&+\frac{1}{4}\hat{\Psib}\Gamma^2 \Gts \Gamma^A\e \Xh^A 
-2\Dh_2\Xh^A\hat{\Psib}\Gamma^A\e = 0, 
\eea
where, $\m=0,1,2$ and the $\Tr$ products are understood.

Now that we have a boundary equation of motion we can consider a BPS configuration of D2-branes ending on another brane, we will study the 1/2-BPS equation of multiple D2-branes ending on a D4-brane. In general a system of two intersecting D-branes is supersymmetric if  
the relative transverse space has dimension in multiples of 4, so we will consider a setup for a 
solution to the boundary condition \eq{bc-lorf} with 
\be \label{xzero}
\Xh^{3,4,5,6} =0.
\ee 
This corresponds to a half Dirichlet boundary condition for a D4-brane with its worldvolume lying in the 01789-directions, as a result the R-symmetry is broken from $SO(7)$ to $SO(3)$.
We also impose the condition
\begin{equation}
\label{d4}
 \G^{01789(10)}\e = \e.
\end{equation}
as well as \eq{c0p}, so we are left with 4 supercharges in our theory.
Note that this method of projecting out the supersymmetry is not the same as the D4-brane projector in the $\kappa$-symmetric formulation of D-branes as in \cite{d11,d22}, the effect of a background flux is already encoded in terms of 
the flux modified M2-branes description and so the D4-brane supersymmetry is
simply given by \eq{d4} in the Lorentzian M2-branes model.

From the above, we can derive the following relations which will be useful in simplifying the boundary equation of motion \eq{bc-lorf}:
\bea
\label{d2reln1}
&&\G^{(10)ij}\e = - \varepsilon^{ijk}\G^2\G^k\e, \\
\label{d2reln2}&&\G^2\Gts \e = 4m\e,
\eea
where we have 
used the indices $i, j, k =7,8,9$. Note that here we have also used the flux \eq{sf1} for the same reasons as that for the projector condition, here we also take $\m = 2m$.
We also impose the projector on the fermion
\be \label{psi-d4}
\G^{01789(10)} \Psi = -\Psi.
\ee
It follows that $\delta \Ah_\a=0$ for $\a =0,1$ and hence
$\Fh_{01}$ is supersymmetric invariant. 
Therefore one can impose the supersymmetric 
boundary condition
\be
\Fh_{01} =0,
\ee
which also implies \eq{fbc2}.

\begin{figure}[htb]
\centering
\includegraphics[width=1\textwidth]{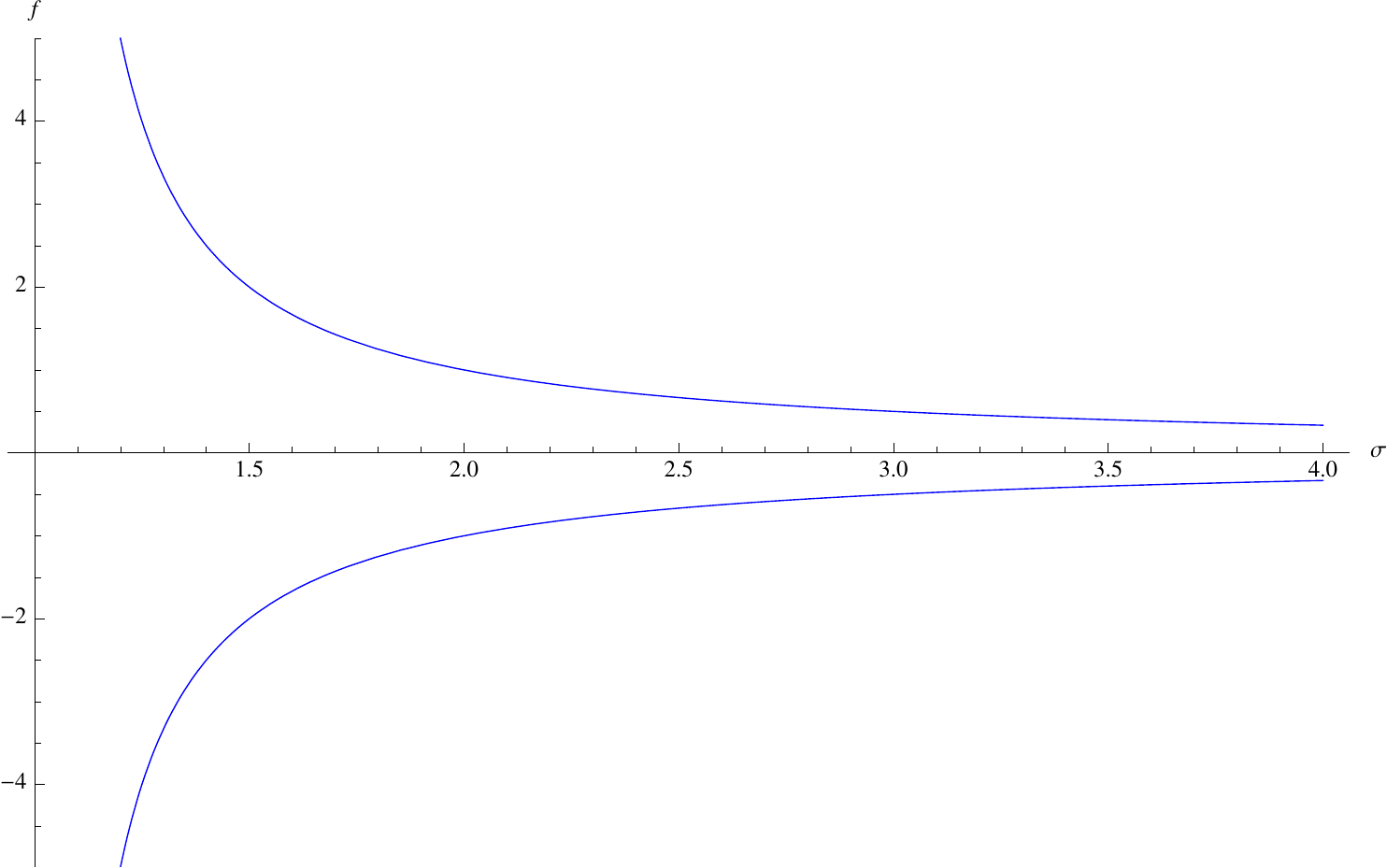}
\caption{The profile of multiple coincident D2-branes blowing up into an Abelian D4-brane via a fuzzy $S^2$ in the presence of a small mass term $m$ and $v_0 = s_0 = 1$.}
\label{d2profile}
\end{figure}

The boundary equation of motion \eq{bc-lorf} can now be simplified by using the relations \eq{d2reln1},\eq{d2reln2} to 
\bea \label{nahm-d4}
\Dh_2\Xh^i = \frac{1}{2}v \varepsilon^{ijk}[\Xh^j,\Xh^k] +m\Xh^i.
\eea
This Nahm equation differs from \eq{nahm} in a few ways but still describes the profile of the 
D4-brane where the D2-branes end. Note that the commutator term is also mass dependent as $v$ depends on $m$ and $\s_1$. Also we have the additional mass term proportional to $\Xh^i$. To see how these changes effect the ODE that we obtain from \eq{nahm-d4}, we propose the following ansatz for the fuzzy funnel solution: 
\be \label{x-ansatz}
\Xh^i(\s_2)= f(\s_2) T^i,
\ee
where $T^i$ obey the $SU(2)$ algebra\footnote{We use a different convention here to \eq{nahm} and so this takes into account the factor of $2i$ in the structure constant given here.} $[T^i,T^j] = \e^{ijk} T^k$ and $f(\s_2)$ obeys the ODE
\be
f'= vf^2 +mf.
\ee
A solution to this ODE is given by
\be
\label{d2soln}
f = \frac{m}{c e^{-m\s_2} - v},
\ee
where $c$ is a constant. We can expand the solution for a small mass $m$ as 
\be 
f = v_0^{-1}/(s_0-\s_2),
\ee
where $s_0$ is a constant, this is precisely the expected profile in the 
absence of flux and so we find an agreement in this certain small $m$ limit as in Figure~\ref{d2profile}. 

The solution \eq{d2soln} describes a fuzzy sphere, given by
\be
\sum_{i=7}^{9}(\Xh^i)^2 = R^2,
\ee
whose  radius $R =C f(\s_2)$ depends on the Casimir $C$ of the 
representation as well as $f(\s_2)$.
Since $v$ actually depends on $\s_1$ the fuzzy funnel has an $S^2$ 
cross section whose radius depends on both
$\s_1$ and $\s_2$, this is a new feature of the flux we consider.

The boundary conditions \eq{xzero}, \eq{psi-d4} and \eq{nahm-d4} are indeed invariant under supersymmetry and this can be checked with the projectors and supersymmetry transformations. 
Before we mentioned that we could have two possible boundary conditions, one at $\s_1 =0$ and one at $\s_2=0$. We now turn to the case $\s_1=0$ and discuss its implications. 
We can repeat the analysis above exactly in the same way with
only a straightforward substitution of the index 2 to 1 
in the equations \eq{dL}-\eq{psi-d4}.
However, it is easy to convince oneself that 
there is no way to impose a supersymmetric boundary condition  
such that \eq{fbc1} holds. This is due to the
fact that $\Xh^{10}$ has a non-trivial supersymmetry variation. Therefore we 
conclude that with the solution $v =v_0 e^{m \s_1}$, 
the flux modified Lorentzian Bagger-Lambert theory is 1/2-BPS if there is a boundary 
at $\s_2 =0$. On the other hand, if the
boundary is at $\s_1 =0$, all supersymmetries are broken.

\section{Summary}

In this Chapter, we explored the applications of the Lorentzian 3-algebra to our boundary analysis. The motivation behind this was to explore what could be studied about 3-algebras when we move away from the Lie 3-algebra $\cA_4$. The difference with the Lorentzian 3-algebra and the novel Higgs mechanism is that the Lorentzian reduction gives no higher order terms to the $\cN=8$ SYM theory in $(2+1)$-dimensions.

We then collated the literature on the Lorentzian 3-algebra into a consistent construction of the $\cN=8$ SYM in $(2+1)$ dimensions, here we kept the boundary terms after integrating out the decoupled ghost terms. A boundary condition must be imposed to obtain the correct $SO(7)$ R-symmetry of the multiple D2-brane theory.

The flux terms of the Lambert-Richmond model were added to the action in the Lorentzian 3-algebra; after some modifications to the ghost terms we were able to integrate them out, but this time the couplings were dependent on spacetime coordinates. Theories with couplings which are spacetime dependent are known as Janus field theories. Again all boundary terms were kept for consistency. We note that the 4-form $\Gt_{IJKL}$ splits into a RR 4-form $\Gt_{ABCD}$ and a NS-NS $\Gt_{(10)ABC}$, these terms can be traced back as Myers' terms from the Chern-Simons action.

Finally the multiple D2-brane system was studied in the D2-D4 configuration. In a similar fashion to the M2-M5 system, we obtained the mass deformed Nahm equation by imposing the D4-brane projector and turning off four transverse scalar fields; reducing the $SO(7)$ to an $SO(3)$ symmetry. A new type of fuzzy funnel solution was obtained for the $1/2$-BPS system with boundary at $\s_2=0$. What would be interesting is to find the D2-D6 system by imposing a suitable projector, both for the massive and massless case.

\part{Quantum Nambu Geometry of M5-branes in a C-Field}

\chapter{Quantum Geometries in String Theory}
This chapter focuses on the concept of matrix models in string theory and their compactifications, also we will look at noncommutative geometry within these models. Finally we will consider an example of a higher dimensional analogue of noncommutative geometry in M-theory, particularly on M5-branes. This will allow us to have a better understanding of the motivations of the following two chapters.

\section{Matrix Models}

In this section we will review what a matrix model is and why they are useful in string theory,  for a review of the general field see \cite{taylorreview}.
A matrix model is a theory in which the fields are matrix valued. These matrices can be put into multidimensional actions, i.e. have space-time dependence, or be constant random matrices for a zero dimensional action.

In string theory, we are mainly interested in the $0+0$ and $0+1$ dimensional matrix models named the IKKT and BFSS models respectively \cite{ikkt, bfss}. Starting from the 10D Super Yang Mills action, reductions of this give the various D-$p$-branes of string theory for $0\leq p+1 \leq 10$. The choices $p=-1$ and $p=0$ give the IKKT and BFSS models respectively, we shall now study these models in more detail.

\subsection{The IKKT Matrix Model}
Let us consider the IKKT model from the choice $p=-1$. The action for the 10D Super Yang Mills theory is given by
\be
\label{10dsym}
S = \int d^{10}x \Big(-\frac{1}{4g_{YM}^2} \tr (F_{\m\n}F^{\m\n}) +\frac{i}{2g_{YM}^2}\tr (\Psib \g^\m D_\m \Psi)\Big),
\ee
where $\m=0,...,9$ and $g_{YM}$ is the coupling constant. Here we choose 
\be
F_{\m\n} = \del_\m A_\n - \del_\n A_\m +i[A_\m,A_\n],
\ee
where the gauge field $A_\m$ and fermion $\Psi$ are in the adjoint representation of $SU(N)$. 

If we now consider a dimensional reduction to a zero dimensional action, we obtain
\be
\label{IKKT}
S = -\frac{1}{4g^2} \tr [X_\m,X_\n]^2 -\frac{1}{2g^2}\tr \Psib \g^\m [X_\m, \Psi].
\ee
Here we use the notation where the gauge field $A_\m$ is usually split into $A_\m = (A_i, X_{(9-i)})$ where $X$ is now a transverse scalar. In the case of a complete reduction to a zero dimensional action we can simply replace the $A$'s with $X$'s. The fields $X$ and $\Psi$ are $N\times N$ constant matrices. The action \eq{IKKT} is invariant under the supersymmetry transformations
\bea
\d X_\m &=& \eb \g_\m \Psi \\
\d \Psi &=& \frac{i}{2}[X_\m,X_\n]\g^{\m\n}\e.
\eea
The gauge symmetry acting on the fields is given by
\be
X_\m \rightarrow U^{-1}X_\m U, \quad \quad U \in SU(N).
\ee
The theory is also invariant under shifts proportional to the identity
\be
X_\m \rightarrow X_\m +a_\m \id, \quad \quad a_\m\in \mathbb{R}^{1,9},
\ee
as any term proportional to the identity in the commutator in the action \eq{IKKT} is trivially zero. It is also invariant under a $SO(10)$ Euclideanised rotational symmetry generated by $\L$,
\be
X_\m \rightarrow \L_\m{}^\n X_\n.
\ee
So we have a matrix model which is Poincare invariant, we will not have this luxury in the BFSS model as we will be working in a light cone gauge.

The action \eq{IKKT} describes an instantonic object, namely a  D(-1)-brane. It is possible to compactify this along tori to give us all the D-branes in type \IIB string theory, so this matrix model is seen as describing all the objects in type \IIB string theory.

As we shall see next, the other reduction to a D0-brane will give us the even numbered D-$p$-branes of type \IIA string theory. The examples of compactifying these two matrix models to give higher dimensional D-branes will be covered later in this section.

\subsection{The BFSS Matrix Model}
The BFSS model is a reduction of the 10D SYM theory to one time direction, hence it describes a quantum mechanical action. In their paper, BFSS considered 11D M-theory compactified on a circle $S^1$ to give a 10D theory with large momenta in the compactified $x^{11}$ direction.

The resulting action is given by
\bea
\label{BFSSa}
S = \frac{1}{2g_s l_s}\int dt\, \tr \Big[D_t X^\m D_t X_\m +\frac{1}{2}[X_\m,X_\n]^2 +i\Psib D_t \Psi -i\Psib\g^\m[X_\m, \Psi]\Big],
\eea
where we have nine $X^\m$ which are $N\times N$ matrices and $\Psi$ is the 16 component spinor which is Grassmann valued in an $N\times N$ matrix. Note that we can gauge fix the action with $A_0=0$ and so it reduces to
\bea
\label{BFSSb}
S = \frac{1}{2g_s l_s}\int dt\, \tr \Big[\Xd^\m  \Xd_\m +\frac{1}{2}[X_\m,X_\n]^2 +i\Psib  \dot{\Psi} -i\Psib\g^\m[X_\m, \Psi]\Big],
\eea
this is known as the BFSS matrix model. The remarkable thing about the BFSS matrix model is that although it describes D0-branes, and by extension all type \IIA D-branes, the value $g_s l_s$ is the M-theory radius $R_{11}$. This allows us to rewrite the action from an M-theoretic point of view.

The type \IIA  theory contains D0, D4, D6 and 
D8-branes in addition to the D2-branes, so what can we say about these other branes in the model? If we compactify M-theory on a circle and go to the infinite
momentum frame, then only states with positive D0-brane charge are
left in the physical description as shown in \cite{bfss}.
But these happen to be all the D-branes in the type \IIA theory since the worldvolume Born-Infeld terms for each of the D-branes are of the form $F\wedge \cdots \wedge F$ with $p/2$ terms, hence the D$p$-branes ($p$ even) are charged under the RR $C_1$. 

The reason why it is not necessary to include the higher 
D$p$-branes ($p$ even) is because they can be constructed out of the D0-branes and so they 
are already included. This is so is because in flat space, 
the worldvolume action for such a system of D$p$-branes is 
given by 
\be
S_{YM}  = \int d^{p+1} \s \; [X^I,X^J]^2,
\ee
where $X^I = (X^\m, X^i)$, $\m =0, 1, \cdots, p$, $i =p+1, \cdots, 9$
and $X^\m = i D^\m$ and a background with non-trivial $F^{\wedge n}$ is assumed.
In this way one can see that all the higher
D$p$-brane actions can actually be constructed from the D0-branes and so it is
sufficient to include only the D0-branes in the description. This explanation was not originally made in \cite{bfss}, it is the interpretation that we make from the results in \cite{sehmbi2} where couplings to the RR potentials are included. It demonstrates the power of the BFSS theory as a model for describing branes.

\subsection{Compactifications}
From the IKKT and BFSS models, it is possible to compactify these theories on tori to obtain the higher dimensional D-branes of type \IIB and \IIA theory respectively. For a review of toroidal compactifications of matrix models, see \cite{taylorreview} for example. 

Typically these matrix models are compactified along tori using T-duality in the type \IIA/\IIB theory we are working in respectively, i.e. the BFSS/IKKT model respectively. Before we see how this is done first recall that if we start in type \IIA theory, then a compactification along a circle of radius $R$ relates the type \IIA theory to type \IIB theory on a circle of radius
\be
\Rh = \frac{\a'}{R}.
\ee
In perturbative string theory the T-duality exchanges winding modes with momenta for closed strings along the compactified direction. The Dirichlet and Neumann boundary conditions for open strings are exchanged via T-duality; this means that D$p$-branes compactified along a transverse direction become D$(p+1)$-branes, while longitudinal compactifications give D$(p-1)$-branes.

D$p$-brane actions are SYM theories with a $U(N)$ gauge symmetry obtained by a dimensional reduction from 10D SYM to $p+1$ dimensions. For example, we can reduce 10D SYM to $0+1$ dimensions to obtain \eq{BFSSb}, the BFSS matrix model as previously mentioned which describes D0-branes. So T-duality exchanges transverse scalars with covariant derivatives with a $U(N)$ valued gauge field $A_i$, i.e.
\be
X^i \to 2i\pi \a' D_i,
\ee 
where $D_i = \del_i -iA_i$. So the BFSS model on a $q$-torus $T^q$ is equivalent to a dimensional reduction of 10D SYM to a $q+1$ dimensional theory, namely D$q$-branes \cite{taylor}.

\section{Noncommutative Geometry in Matrix Models}

In this section we discuss the origins of noncommutative geometry and the basic underlying principles. We then show how we can use noncommutative gauge theories in string theory, particularly in matrix models.  This provides the foundation for the next chapter on the Quantum Nambu Geometry.

\subsection{Noncommutative Geometry}

In quantum theories noncommutativity is a very important concept, for example:  the Heisenberg commutation relation $[x,p]=i\hbar$ is key to writing down the Heisenberg uncertainty relation. There is no reason, a priori, that the spacetime coordinates $x^\m$ should commute. Theories where we write a relation such as
\be
\label{ncg1}
[x^\m, x^\n]=i\th^{\m\n},\quad \th^{\m\n} \text{\, some antisymmetric parameter},
\ee
are called noncommutative field theories. Here the $\th^{\m\n}$ has some dependence on $\hbar$. See \cite{ncg-review1, ncg-review2} for a review of the subject. These theories are constructed by taking a field theory Lagrangian and apply the usual perturbative quantisation techniques except we must also take \eq{ncg1} into account. Noncommutative field theories tend to be non-local due to \eq{ncg1} for length scales of order $\th$, for larger scales the theories are effectively local. Another reason for studying noncommutative field theories is because the fields now have an explicit cut-off at short distances and so these theories are believed to be finite. Yang-Mills theories with simple gauge groups have no dimensionless parameters to use for perturbative expansions so a noncommutative Yang-Mills gauge theory could provide a finite theory.

String theory, 
as a candidate for a theory of quantum gravity, provides an interesting 
setup to address some of these questions.  
One of our  motivations is to discover  
new types of quantum geometries in string theory and 
to study the physics on these quantum
spaces. 

\subsection{Noncommutative Gauge Theories}

There are a a few known examples where
a quantum geometry could emerge in string theory in terms of noncommutative models. 
One possibility is when we consider open string theory ending on a D-brane with a background NS-NS $B$-field on it and use the open string to probe the 
geometry of the D-brane. The resulting form of the noncommutative geometry
could be either of Moyal type \cite{sch1,dh,CH-non1,CH-non2,sw}, 
\be \label{ncg}
[X^\m,X^\n] = i \th^{\m\n},
\ee
or a fuzzy sphere \cite{sch2,sch3}
\be
\label{ncg2}
[X^i,X^j] = i \l \e^{ijk} X^k.
\ee
Noncommutative geometry 
arises in matrix models such as the IKKT and BFSS models \cite{bfss,ikkt} as a classical solution \cite{bss,bkgd-indep,makeenko,smith-2b}.
Myers effects \cite{myers} could introduce additional terms to the
matrix model and lead to new solutions 
\cite{iso-fuzzy}.
We note that 
all these quantised geometries are characterised by a
commutator and could be referred to as of Lie-algebra type.\footnote{For the fuzzy sphere it is indeed a Lie-algebra valued commutator, but for noncommutative geometry solutions the algebra is mapped to a central element.}
By considering small fluctuations around the solutions \eq{ncg},\eq{ncg2} we obtain a noncommutative gauge theory \cite{bkgd-indep}. 

In the paper \cite{bkgd-indep}, Seiberg considered the matrix model 
\be
\label{seibergaction}
S(X^I) = \int \mathrm{d}t\, (D_t X^I)^2 -g_{IK}g_{JL}[X^I,X^J][X^K,X^L],
\ee
which admits the equation of motion 
\be
\label{seibergeom}
D^2_t X^I +g_{JL}[X^J,[X^I,X^L]]=0.
\ee
This has a static non-trivial solution $X^I = x^i, i=1,...,p$ where $p$ is even, satisfying
\be
[x^i,x^j]=i\th^{ij}\id.
\ee
Around such a classical solution, we make a small perturbation
\bea
\label{pertsol}
\begin{cases} x^i = x^i +\th^{ij}A_j \equiv x^i +\At^i& i=1,...,p \\
x^{a+p} =\phi^a &a=1,...,9-p.
\end{cases}
\eea
Then $S(x +\At,\phi)=\tilde{S}(\At,\phi)$ gives us a noncommutative SYM theory\footnote{This can be related to a commutative DBI action via a Seiberg-Witten map \cite{sw}.} with a field strength for the gauge fields given by
\be
\Ft^{ij}=[\At^i,\At^j].
\ee

\subsection{Three Algebra Quantum Geometries}
Given the mathematical structure of a Lie 3-algebra, it is natural to ask if it is possible to construct a higher dimensional analog of noncommutative geometries and fuzzy spheres. Moreover, one may wonder if the physics of the fluctuations around such a solution may 
lead to new kinds of gauge theories. 
In the paper \cite{CS1} such a generalisation was made to the system of M2-branes ending on an M5-brane in a constant $C$-field.\footnote{
In \cite{CS1}, it was also
shown that the standard
noncommutative geometry \eq{ncg} of D-branes in 
a constant 2-form NS-NS $B$-field could be derived similarly by
considering the intersecting system of F1-strings and D3-brane in a
$B$-field.} There was a boundary analysis in which the Basu-Harvey equation was derived and then was modified non-trivially by the $C$-field.
The proposal for the quantum geometry of the M5-brane in the presence of a constant 3-form $C$-field takes the form
\be \label{3bkt-geom}
[X^\m,X^\n,X^\l] =i\th^{\m\n\l},
\ee
where $\th$ is a constant and the 3-bracket is given by a Lie 3-bracket satisfying the fundamental identity \eq{fi} and $\m,\n,\l=0,...,5$.

The Lie 3-bracket appears in the analysis because 
the geometry of the M5-brane was 
inferred from the boundary dynamics 
of the open M2-branes which end on it; the BLG model with boundary 
was used to describe the open M2-branes.
In a quantum theory, it is necessary to understand the relation \eq{3bkt-geom} as an operator 
relation acting on some states in the theory. However, the representation of the
Lie 3-algebra relation as transformations on vector spaces 
or maybe some kind of generalisation is still an open question \cite{sz1,sz2,cw1,cw2}. 
The main issue with constructing representations of \eq{3bkt-geom} is due to the insistence of the the fundamental identity, this leads to the question of what types of mathematical structures could be used instead of the Lie 3-algebra to avoid this issue of the fundamental identity being required.
Since the ABJM theory involves operators as the fields, one could naively think that the   analysis of \cite{CS1} with the open ABJM theory could be carried out
instead and we could obtain a 
similar relation to \eq{3bkt-geom} where $X^i$'s would be operators. 
However the fuzzy funnel solution of the ABJM theory with a $C$-field is classically a fuzzy two sphere \cite{ram}. So the interpretation of the M2-M5 brane system does not apply here necessarily. It is possible that this is describing the D2-D4 system and this could be a further direction to be explored.

\chapter{D1-Strings in Large RR 3-Form Flux, Quantum Nambu Geometry and M5-Branes in C-Field}

\section{A Proposal, The Quantum Nambu Geometry}

The fundamental identity \eq{fi} of the multiple M2-brane theory places a constraint on the 3-bracket used to describe the quantum geometry on the M5-brane theory in \eq{3bkt-geom}. A priori, there is no reason why a 3-bracket structure with a different origin than the M2-brane theory should obey the fundamental identity. For the case of M2-branes, it was required to write down the gauge symmetry and for the supersymmetry algebra to close.

In this chapter we propose a new type of quantum geometry based on the completely anti-symmetric quantum Nambu bracket
\be
\label{n-3bkt}
[A,B,C] :=ABC +BCA +CAB -BAC-CBA-ACB,
\ee
where $A,B,C$ are three operators in an algebra with the usual operator product. We will see that this algebra is natural when considering D1-strings of type \IIB string theory in a large RR 3-form flux background.

The quantum Nambu bracket \eq{n-3bkt} does not obey the fundamental identity, this can easily be checked, see \cite{zachos} for a discussion of this
as well as some other algebraic properties of the 3-bracket \eq{n-3bkt}. 
The relaxation of the fundamental identity allows us to consider a gauge group with large rank, this was one of the key constraints of the BLG theory and it will not affect the quantum Nambu bracket. In the absence of the fundamental identity the most natural 3-bracket to
consider would be the one given by \eq{n-3bkt} which is the most natural higher
order  generalisation of the  commutator. 
The 3-bracket \eq{n-3bkt} was
originally introduced by Nambu \cite{nambu} as a possible
candidate of the quantisation of the classical Nambu bracket 
\be
 \{f,g,h\}
: = \e^{ijk} \del_i f \del_j g \del_k h.
\ee
The geometry \eq{3bkt-geom} for the quantum Nambu bracket \eq{n-3bkt} is called the {\it quantum Nambu geometry}.

In the previous chapter we gave an overview of noncommutative geometry in string theory, the key equation is given by \eq{ncg}, which describes the breakdown of commutativity for the scalar fields. A 2-form field
strength can be written as
\be
F^{\m\n} = -i [X^\m,X^\n]
\ee 
when the fluctuation over the noncommutative geometry background is taken 
into account. This is what was mentioned in the previous chapter regarding expanding around a solution to the noncommutative geometry.
We now turn to the quantum Nambu geometry, we propose a straightforward uplift from the noncommutative case to be given by
\be
[X^\m,X^\n,X^\l]=i\th^{\m\n\l},
\ee
where the bracket is the quantum Nambu bracket \eq{n-3bkt}.
It is natural to interpret the quantum Nambu bracket of the target space coordinate fields $X^\m$ as a 3-form field strength
\be \label{H-id}
H^{\m\n\l} = -i [X^\m,X^\n, X^\l].
\ee
This provides a suggestive connection to M-theory, in particular the 3-form field strength $H_{\m\n\l}$ on an M5-brane.
To check this idea, we have to look for a way to connect the D1-strings model to the non-abelian theory of multiple D4-branes\footnote{Where the 3-form field strength would be the
Hodge dual of a 2-form field strength.} and hence multiple M5-branes through the duality presented in the first chapter, namely \cite{d1,d2}.

The D4-brane is a BPS object in type \IIA string theory. We will show that starting with D1-strings of type \IIB string theory we may obtain matrix models for not only \IIB string theory, but for M-theory and \IIA theory. Compactifying these matrix models gives the various branes within each theory. The matrix models we will find in our analysis are the large RR flux dominated terms, this is achieved in a certain double scaling limit of the theories. These matrix models admit the quantum Nambu geometry as a solution.

The quantum Nambu geometry can be expanded by some small fluctuation to give a Lagrangian in terms of a non-abelian 1-form potential. This potential will be of the same form as the dimensionally reduced version of the PST action \cite{pst1,pst2,pst3,pst4}, hence describing the D4-branes. Note that this is a non-abelian description of the D4-branes whereas the PST action is given in terms of an abelian gauge symmetry. Physically we interpret this as the D1-strings expanding out over the quantum Nambu geometry into D4-branes (in a certain limit).

Using the recent proposal for the duality between M5-branes on $S^1$ and 5D super Yang-Mills \cite{d1,d2}; we generalise in a very natural way how to promote the D4-brane theory into the M5-brane theory. The interpretation here is that the M5-branes are in a constant self-dual $C$-field. Quite remarkably, the requirement of self-duality of the $H_{\m\n\l}$ of M5-brane theory is automatically satisfied in our model.

\section{Matrix model of D1-strings in a limit of a large RR 3-form $F_3$}

In this section we begin our construction by considering D1-strings in a constant RR 3-form flux background which is the natural coupling associated with the 2-form RR gauge field $C_2$. We find that there is a low energy - large flux double scaling limit in which the action of the D1-strings are dominated by the RR terms. We then show that the resulting D1-branes matrix model has the quantum Nambu geometry as a classical solution.

\subsection{A Type \IIB supergravity background}

An exact \IIB supergravity background with a constant RR 3-form flux was considered in \cite{CH-nambu}, it was constructed by turning on a constant RR 3-form flux  
in the $AdS_5$ sector of the $AdS_5 \times S^5$ background.
The background has a spacetime which is a direct product of the $AdS_5$ and $S^5$ contributions;
\be
\cM = \cM_5 \times \cM_5'
\ee
respectively and has a dilaton, axion $\chi \equiv C_0$,
RR potentials $C_2$ and $C_4$ specified by:
\bea
e^{-\Phi} &=& \chi /(2 \sqrt{2}) = \mbox{constant}, \\
F_3 &=&
\begin{cases}
f \e_{ijk}, & i,j,k =1,2,3, \\
0, & {\rm otherwise},
\end{cases} \label{F-ans}\\
F_5 &=& \begin{cases} 
c \ve_5 & \mbox{on $\cM_5$}, \\
c \ve'_5 & \mbox{on $\cM_5'$}, \\
0 & \mbox{otherwise}.
\end{cases} 
\eea
We note that $F_3 = dC_2$ and $F_5 = dC_4$ are the fluxes associated to the RR potentials.
Here $f$ and $c$ are real constants,  
$\ve_5$ and $\ve_5'$ are the volume forms on $\cM_5$ with $\mu =0,1,2,3,4$ 
and $\cM_5'$ with $\mu = 5,6,7,8,9$ 
\be
\ve_{\m_0 \cdots \m_4} = \sqrt{-\det G_5} \; \e_{\m_0 \cdots \m_4}, \qquad
\ve'_{\m_5 \cdots \m_9} = \sqrt{-\det G'_5} \; \e_{\m_5 \cdots \m_9}, 
\ee 
and  $\e_{\m_0 \cdots \m_4}$, $\e_{\m_5 \cdots \m_9}$  
are the Levi-Civita symbols with convention:\\ $\e^{01234} = - \e_{01234} =1, \e^{56789} =\e_{56789} =1$. 

It was shown in \cite{CH-nambu} that it is possible to construct a consistent background if the fluxes of the RR potentials $C_2$ and $C_4$ are chosen to be
\be
f^2 = \frac{2}{3} c^2.
\ee
The metric of the background now takes the form $\RR^3 \times AdS_2 \times S^5$:
\be \label{metric}
ds^2 = \sum_{i=2}^4 (dX^i)^2 + R^2 (\frac{-dt^2 +dU^2}{U^2})
+ R'{}^2d \Omega_5^2,
\ee
where 
\be
R^2 = 2e^{-2\Phi}/f^2, \qquad  
R'{}^2 = 80 e^{-2\Phi}/f^2,
\ee
and 
$d \Omega_5^2 = \hat{G}_{i'j'} dX^{i'} d X^{j'}$ 
is the metric for an $S^5$. The radius $R$ is the $AdS$ radius while the radius $R'$ is that of the $S^5$.
The key point here is that we have warped the spacetime such that there is a flat $\RR^3$ where the $F_3$ lives, as we will see later this is the term which will dominate in the double scaling limit. 

As previously mentioned, we will be considering a double scaling limit of the system in which the RR potential $C_2$ becomes the dominant term in the action for the matrix model. So let us write explicitly the 2-form in a specific coordinate basis 
\be \label{C-ans}
C_2 = f \e_{ijk} X^i dX^j dX^k, 
\qquad i,j,k =1,2,3.
\ee
The convention we shall use is the following 
\be
F_{\m\n\l} = \frac{1}{3}(\del_\m C_{\n\l} + \del_\n C_{\l\m} + \del_\l C_{\m\n}).
\ee
The expressions for $C_4$ is more complicated, we will see that these terms do not survive in the double scaling limit. 
Later we will consider a large $f$ limit for a system of 
D1-strings in this background. 
For our purpose, it is enough to 
note that
$C_{\m_1 \cdots \m_5}$ with $\mu_i = 0,1, \cdots, 4$ is proportional to $1/f$ and 
$C_{\m_1 \cdots \m_5}$ with $\mu_i = 5, 6, \cdots, 9$ is proportional to  $c R^5 \sim 1/f^4$.

\subsection{Matrix model of D1-strings in limit of large $F_3$}

Let us consider a system of $N$ parallel D1-branes in the background \eq{metric},
the worldvolume action for the D1-branes is given by the Non-abelian
Born-Infeld action plus the Chern-Simons term of the Myers type given by 
\cite{myers}
\be
\label{myerscs}
S_{CS} = \m_1\int \Tr P(e^{i \l \ri_\Phi \ri_\Phi} 
\sum_n C_n )e^{\l F}.
\ee
Here $\mu_1 = 1/(g_s 2 \pi \a' )$, $\l= 2 \pi \a'$ and $X^I=2 \pi \a' \Phi^I$, the product $\ri_\Phi$ is the interior product.
We have set the NS-NS $B_2$ field to zero. With the axion $\chi$, RR fields
$C_2$ and $C_4$ turned on, the Chern-Simons term reads
\bea \label{S-CS}
S_{CS} &=& \m_1 \int  \Tr \left[  \l F \chi +
P\,C_2 + i \l^2 F \ri_\Phi \ri_\Phi C_2 + 
i \l P\,  \ri_\Phi \ri_\Phi C_4 - \frac{\l^3}{2} F \ri_\Phi^4 C_4 
\right] \\
&:=& S_{\chi} + S_{C_2} + S_{C_4}, \label{S-Cs}
\eea
where $S_\chi, S_{C_2}, S_{C_4}$ denote 
the terms in $S_{CS}$ that depend on the RR-potentials $\chi, C_2$ and $C_4$
respectively.
Substituting the explicit expression for $C_2$ \eq{C-ans} into the term $S_{C_2}$ above, we obtain
\bea \label{S-C2}
S_{C_2} /\m_1 &= & f \int  d^2 \s \Tr (
\frac{1}{2} \e_{ijk} X^i D_\a X^j D_\b X^k \e^{\a\b}) 
+ 
 f\int  d^2 \s \Tr (
i  F X^i X^j X^k \e_{ijk})\nn\\
&\equiv&  
f\int d^2\s\, (L_1 + L_2),
\eea
here $\e^{01} = -\e_{01}=1$, 
$F=F_{01}$. From now on we will use $F$ to refer to either 
the curvature two-form or the component $F_{01}$.
It should be clear from the context which is which. 
Naively, if we take a large RR flux $F_3$ limit, then the D1-branes action 
is dominated by $S_{CS}$. But we can be more precise than this, we will show that there is a certain double scaling limit where the dynamics of the system of D1-branes is 
dominated by the $C_2$ coupling term $S_{C_2}$. 
To do this, we need to include the Non-abelian Born-Infeld action,
examine the large $f$ limit of the 
equations of motion and keep the parts of the action that contribute
in the limit.

Typically non-abelian D-branes in curved space times are poorly understood, this is due to several reasons which we shall touch upon in the following. This poses a potential problem for the usual Born-Infeld term for the $N$ parallel D1-branes. The Yang-Mills action is obtained from the Born-Infeld action in flat space, this term is fine in general but higher flat space terms have ordering ambiguities in the worldvolume gauge field strength $F^n$ \cite{NBI}, \cite{HT}. For a general curved spacetime the metric becomes a function of the coordinates, i.e. the scalar fields, $G_{IJ}(X)$. The action for the D1-branes is then given by \cite{curved-d1,curved-d2}; 
\bea \label{S-X}
S_{X}/\m_1 := \int d^2 \s \sqrt{-\det G} \Big (&&
 G_{IJ}(X) D_\a X^I D_\b X^J G^{\a\b} \nn\\
&&+ \frac{1}{\a'} G_{IJ}(X) G_{KL}(X)[X^I,X^K] [X^J,X^L]
\Big),
\eea
where $I,J =2,...,9$.
But this is highly ambiguous again for the non-abelian theory as the metric has an ordering problem in terms of scalar fields $X^I$. The problems highlighted above will not survive the particular limit we are taking of a large $F_3$ with a small $\a'$, so we can happily ignore these.

Let us assume that in the small $\a'$ limit,
the system of D1-branes is described by
an action of the form \eq{S-X} together with 
the Chern-Simon coupling \eq{S-CS}. Note that for our metric, 
the ambiguity of the action $S_X$ 
is concentrated entirely in the $S^5$ part. 
The full D1-strings action is given by
\be
S_{D1}:= S_X+ S_{CS} + S_{YM},
\ee 
where the Yang-Mills term is
\be
S_{YM}/\m_1 =  \a'{}^2 
\int \sqrt{-\det{G_{\a\b}}} \; F_{\a\b}F_{\a'\b'} G^{\a\a'} G^{\b\b'}
\ee
and the metric is 
\bea
G_{\a\b} &=& R^2 \s^{-2} \eta_{\a\b}, \qquad\quad\qquad \;\; \a,\b =0,1,\\
G_{ij} &=& \d_{ij} , \qquad\qquad\qquad\qquad\quad i, j =2,3,4, \\
G_{i'j'} &=& R'{}^2 \times \hat{G}_{i'j'}(X^{k'}),
\qquad\quad\qquad i',j' =5,6,7,8,9,
\label{G3}
\eea
with $\hat{G}_{i'j'}$ being the metric for a unit 5-sphere and $\s$ is the physical gauge coordinate. All other metric entries are zero.

We now comment on a few consequences of our choice of set up for the system of D1-strings in the background \eq{metric}:

\begin{enumerate}
\item The scalars $X^i$ and $X^{i'}$ decouple from each other 
in the action $S_{D1}$ as the metric is in a block diagonal form and does not mix primed and unprimed coordinates.
\item  
The contributions to the equations of motion 
of $X^i$ and $X^{i'}$ from the various pieces of the actions 
\eq{S-CS} and \eq{S-X} are given by: 
\be
\begin{tabular}{|c|cccc|}
\hline
\mbox{} & $S_X$  & $S_{C_2} $ 
& $\ri_\Phi^2 C_4$ &  $\ri_\Phi^4 C_4$ \\
\hline
\mbox{EOM of $X^i$:} & $ O(1/\a') $ &  $O(\frac{f}{\a'}) $ 
&  $O(\frac{1}{f\a'{}^2}) $ &  $0 $ \\
\mbox{EOM of $X^{i'}$:} & $ O(\frac{1}{f^2 \a'}) $ &  $0 $ 
&  $O(\frac{1}{f^4 \a'{}^2})  $ &  $O(\frac{1}{f^4 \a'{}^2}) $\\
\hline 
\end{tabular}
\ee 
\item The equation of motion of $X^{i'}$ can be solved
with $X^{i'} =0$.
\end{enumerate}
These remarks are independent of the 
ambiguity of the form of the metric $G_{i'j'}$ 
in  the action \eq{S-X} as we are only considering the powers of the couplings to $f$ and $\a'$. 

Since the equations of motion imply $X^{i'} =0$, we may drop these terms from the action and focus on the sector 
with only the scalars $X^i$ and the gauge field activated. Finally we come to the double scaling limit that has been mentioned previously. The action
$S_{C_2}$ is of order $O(f/\a')$ and the relevant piece of the Myers term $\ri_\Phi^2 C_4$ in \eq{S-CS} is of order $O(1/f\a'^2)$. 
Therefore 
if we take a double scaling limit $\e\to 0$:
\bea \label{scaling}
\a' &\sim& \e, \nn \\
f &\sim& \e^{-a}, \quad a>0,
\eea
such that $a>1/2$, then $S_{C_2}$ dominates. Moreover, 
$S_{YM}$ can be ignored compared to $S_{C_2}$ if $a<2$. All in all, in the 
double scaling limit \eq{scaling} with $1/2 <a<2$, the low energy action of
$N$ D1-branes in a large $F_3$ background is given by
\be
\lim_{\e \rightarrow 0} S_{D1} = S_{C_2}.
\ee

\subsection{Quantum Nambu Geometry as a classical solution}

We can now find the equations of motion to the action $S_{C_2}$, these are
\bea
&&\ve^{\a\b}\ve_{ijk}[X^j,D_\b X^kX^i] + \ve^{\a\b}\ve_{ijk}
[D_\b, X^iX^jX^k] =0, \\
&&\frac{3}{2}\ve_{ijk}D_\a X^j D_\b X^k \ve^{\a\b} +\ve_{ijk}[F;X^j,X^k]' =0,
\eea
where $[A;B,C]' := [B,C]A + A[B,C] +BAC-CAB$ is antisymmetric only 
in the exchange of $B$ and $C$. This bracket arises since 
$\Tr [A,B,C]D = \Tr [D;B,C]' A$, in analogy to the relation
$\Tr D[A,B] = \Tr [D,A] B$ which is useful in ordinary Yang-Mills theory.\footnote{Note that the 3-bracket which satisfies the fundamental identity has a more straightforward generalisation \eq{3fund}, up to a sign, whereas the case for the quantum Nambu bracket does not.}

The first equation of motion is solved with any covariantly constant configuration
\be \label{soln1}
D_\a X^i=0.
\ee
The second equation of motion becomes
$ \ve_{ijk}[F;X^j,X^k]'=0$
and is solved by 
\be \label{soln2}
F=0.
\ee
Recall that $F$ is the worldvolume field strength of the non-abelian gauge field.
The standard noncommutative geometry
\be
[X^i,X^j] = i \th^{ij}
\ee
is allowed as the equation of motion is anti-symmetric in the two $X^i$ entries, but there is also a new solution, the quantum Nambu geometry,
\be \label{xxx}
[X^i,X^j,X^k] = i \th \e^{ijk},
\ee
where $\th$ is a constant and the 3-bracket is given by \eq{n-3bkt}. This relation is also known as the Nambu-Heisenberg commutation relation.
We note that
the solution \eq{xxx} is not allowed in the standard matrix models 
\cite{bfss,ikkt}
where  no external $F_3$ is turned on.

The 3-bracket \eq{n-3bkt} was
originally introduced by Nambu \cite{nambu} as a possible
way to write down the quantisation of the classical Nambu bracket 
\be 
\{f,g,h\}
: = \e^{ijk} \del_i f \del_j g \del_k h.
\ee 
This quantisation was thought of as generalising Hamiltonian
mechanics to the form
\be \label{N-mech}
\frac{d f}{d t} = \{ H_1, H_2, f \},
\ee
which involves two ``Hamiltonians'' $H_1, H_2$. This would require a generalised Jacobi identity, the fundamental identity, but was not considered here.
In fact one can easily check that 
the fundamental identity is not satisfied for \eq{n-3bkt}.
The concept of 
fundamental identity was introduced almost 20 years later by
Takhtajan \cite{tak} 
as a natural condition for his definition of a Nambu-Poisson manifold
which allows him to formulate the Nambu mechanics in an invariant
geometric form similar to that of Hamiltonian mechanics.
For example, the fundamental identity implies that the time evolution
preserves the Nambu bracket. However for this purpose, 
a weaker form of the fundamental identity, where two of the elements
are fixed in \eq{fund}: $\a=H_1, \b=H_2 $, is sufficient. 
What we have shown above is that a quantised geometry characterised by the
Nambu bracket \eq{n-3bkt} is allowed as a solution in string theory and we will
refer to the quantised geometry \eq{xxx} as quantum Nambu geometry. An analysis of the quantum Nambu geometry will be carried out in the next chapter, two explicit infinite dimensional representations will be discussed.

\section{Matrix Theories in Large RR  Flux Background}
 
In the previous chapter we examined the expansion around a noncommutative geometry to find a new gauge theory, namely noncommutative Yang-Mills, we would like to extend this concept to that of the quantum Nambu geometry in the most straightforward and natural way. In noncommutative geometry, the expansion around the solution $[x^\m,x^\n]=i\th^{\m\n}$ gives a natural noncommutative 2-form field strength
\be
F^{\m\n} = -i [X^\m,X^\n].
\ee 
In the case of the quantum Nambu geometry, it is natural to suggest that the quantum
Nambu bracket of the target space coordinate fields $X^\m$ 
is a 3-form field strength
\be 
H^{\m\n\l} = -i [X^\m,X^\n, X^\l].
\ee
We now need to look for objects in string theory where such a non-abelian field strength would live and then check if these theories are consistent with the quantum Nambu geometry.

Within string theory the place where we can find a non-abelian 3-form field strength would be in the theory of multiple D4-branes, here the 3-form field strength would be 
the Hodge dual to a 2-form field strength on its five-dimensional worldvolume. If we consider M-theory, the theory of multiple M5-branes has a self-dual 3-form field strength of a 2-form tensor gauge field living on the worldvolume of the branes. The D4-brane is an object in type \IIA string theory whereas the D1-brane matrix model we have is in type \IIB string theory. So we must find a connection between these theories in the large flux limit. As such, we construct the type \IIB, M-theory and type \IIA matrix models in the large flux \eq{S-C2} from which all their respective branes can be found by compactifications.

\subsection{Type \IIB Matrix Theory}

The \IIB matrix model 
can be obtained by a large $N$ reduction of the 
D1-string action. Let us first denote the covariant derivative 
\be
D^\a = \del^\a -i A^\a, \quad \a=0,1
\ee
as 
\be
 i D^\a  = X^\a,
\ee
and rewrite $L_1$, $L_2$ in terms of the $X$'s as
\be\label{L1}
L_1 = - \frac{1}{2} \Tr X^i [X^\a,X^j] [X^\b,X^k] \e_{\a\b} \e_{ijk},  
\ee
\be\label{L2}
L_2 =-\Tr[X^0,X^1][X^2,X^3,X^4].
\ee
It is simple to show that the two terms can be combined together to give
\be
L_1+ L_2 =\frac{1}{40} \Tr[X^a,X^b][X^c,X^d,X^e]\e_{abcde}.
\ee 
It is quite remarkable that the D1-branes' Chern-Simons coupling to a
constant RR $F_3$ flux can be written in such a simple form. 
The action for $N$ D1-branes in a large $F_3$ double scaling limit can then
be written as
\be
\label{D1}
S_{D1} = \frac{\m_1 f}{40}\int  
d^2\s\,\Tr[X^a,X^b][X^c,X^d,X^e]\e_{abcde}
= \frac{3 \m_1 f}{10 }\int  d^2\s\,\Tr X^a X^b X^c X^d X^e\e_{abcde},
\ee
where $a,b,c,d,e=0,1,2,3,4.$ 
The large $N$ reduction\footnote{For a review of how to perform a large $N$ reduction for matrix models, see \cite{taylorreview}} gives the D-instantonic action, 
up to an unimportant overall numerical constant,
\be
\label{D-1}
S_{IIB} = 
\frac{f}{g_s l_s^2} \Tr X^a X^b X^c X^d X^e\e_{abcde}, \quad
a,b,c,d,e=0,1,2,3,4. 
\ee
This gives the matrix model description for the \IIB string theory
in the limit of a large constant RR 3-form flux, and in the
sector with  $X^{a'} =0$, $a' = 5,6,7,8,9$.
In this limit, the Myers term dominates 
over the standard Yang-Mills term in the IKKT matrix model \cite{ikkt} and so this matrix model can be thought of the large constant RR 3-form flux version of the IKKT model.

\subsection{Matrix Model of M-theory}
Now we shall T-dualise the type \IIB theory to type \IIA, compactify the system to describe D0-branes and then take the M-theory limit.
The type \IIB background \eq{metric} is invariant under the Killing
vector $\del/\del x^i, i =2,3,4$ along the $\RR^3$ directions. We can then compactify along the 
$x^2$ direction, for example, on a circle of radius $R_2$ and T-dualise the system. The corresponding theory is a type \IIA background and has
\bea
\mbox{metric}:&& S^1 \times \RR^2 \times AdS_2 \times S^5, 
\label{IIA-bkgd-1}\\
\mbox{constant RR field strength}: && \nn\\
F_{ij} &=& F_{2ij}, \quad i,j =3,4, \nn \\
F_{abcd} &=& F_{2abcd}, \quad a,b,c,d = 0,1,3,4, \label{IIA-bkgd-2}\\
F_{2a'b'c'd'e'} &=& F_{a'b'c'd'e'}, \quad a',b',c',d',e'= 5,6,7,8,9, \nn\\
\mbox{constant dilaton:} && 
e^{\phi'} = e^\phi \frac{\sqrt{\a'}}{R_2}. \label{IIA-bkgd-3}
\eea

The T-duality we have performed turns the D1-branes into D2-branes, we must now check what happens to the flux. In the double scaling
limit \eq{scaling}, the D2-branes action is given by 
the T-dual of the D1-branes action \eq{D1} by applying  the usual T-duality rule \cite{bfss, taylor} to the D1-branes action;
\bea
X^2 = i R_2 D_2, \label{Td1} \\
\Tr \to \int \frac{l_s d \s_2}{R_2} \Tr. \label{Td2}
\eea 
The resulting action for the D2-branes is given by 
\be \label{D2}
S_{D2} = \frac{f}{g_s l_s } \int d^3 \s \Tr X^a X^b X^c X^d X^e\e_{abcde},
\ee
where $a,b,c,d,e =0,1,2,3,4$ 
and we have ignored an  unimportant overall numerical constant again.
Note that since the Chern-Simons coupling is topological,  
the $R_2$ dependence gets cancelled in \eq{D2}. 
We can also obtain \eq{D2} directly from the 
Chern-Simons coupling of the D2-branes in the \IIA RR flux background \eq{IIA-bkgd-2}, so the solution for the D2-brane action \eq{D2} can be checked for consistency. The Chern-Simons action is given by
\be
S_{CS} = \frac{1}{g_s l_s^3} \left[
\int P(C_1)\l F  + \int P(C_3) + 
\int P(i \l \ri_\Phi \ri_\Phi C_5)   
\right].
\ee
The $C_3$ and $C_5$ terms of type \IIA theory have their origin from the RR
5-form of \IIB theory under T-duality and so they can be ignored in the 
double scaling limit \eq{scaling}. The $C_1$ term then
reproduces precisely \eq{D2}.

The BFSS model proposed that M-theory in a flat space background, while in the infinite momentum frame, is given by the large $N$ quantum mechanics of D0-branes \cite{bfss} as described in the previous chapter. So if we reduce to an action describing D0-branes, we can first obtain a matrix model for M-theory in the flux background and then, after some tricks, obtain the type \IIA matrix model in the 2-form flux background.

For us, we would like to derive the quantum mechanical 
description of M-theory in a curved background that corresponds to
\eq{IIA-bkgd-1} - \eq{IIA-bkgd-3} uplifted to 11 dimensions.
The eleven dimensional background reads
\bea
\mbox{metric}:&& ds^2 = e^{-\frac{2}{3} \phi} ds^2_{IIA} +
 e^{\frac{4}{3} \phi}  (dx^{11} - dx^i C_i)^2,
\label{M-bkgd-1}\\
\mbox{3-form potential}: && C^{(3)} = \frac{1}{6} C_{abc} dx^a dx^b dx^c, 
\label{M-bkgd-2}
\eea
where $\phi$ is the dilaton in \IIA theory, $C_i$ and $C_{abc}$ are the RR
1-form potential and RR 3-form potential which appear 
in \eq{IIA-bkgd-2}.

In the following, we denote the compactified $x^{11}$ radius by $R_{11}$. 
In general, with a suitable worldvolume flux turned on, 
the higher D$p$-branes ($p$ even) of type \IIA theory carry D0-brane charges as discussed in the previous chapter and so in principle 
should be kept in the infinite momentum frame. However as in the flat case,
it is sufficient to select a
subset of degrees of freedom in such a way that 
all the other degrees of freedom as well as their dynamics could be
recovered. 
Now what is different for our background  is that there is a set of
non-vanishing RR gauge potentials which lead to explicit
Chern-Simons terms in the action of the D$p$-branes.

We now examine this more rigorously and then obtain the M-theory matrix model.
As previously mentioned, the double scaling limit \eq{scaling} eliminates the terms $C_3$ and $C_5$ (which originate from the $F_5$ of the \IIB theory), 
we can also ignore the Yang-Mills term in this limit and concentrate on the Chern-Simons coupling of $C_1$. 
Moreover in the sector where the fields in the sphere directions are set to zero: $X^{a'} =0$, $a'
= 5,6,7,8,9$, the Chern-Simons couplings for D4, D6 and D8-branes are
zero. So we are only left with the D0-branes and the D2-branes. For the D0-branes, we obtain the action, up to an unimportant overall numerical constant,
\be\label{D0}
S_{D0} = \frac{1}{g_s l_s} \int P(C^{(1)}) 
=  \frac{f}{g_s l_s} \int dt \e_{ij} X^i D_t X^j , \quad
i,j =3,4,
\ee 
where
\be
C_1 = f \e_{ij} X^i dX^j.
\ee
The $f$ is the same and the $C_1$ and $C_2$ are related under T-duality.
Now  the action \eq{D2} is equivalent to it's
dimensional reduction
\be \label{red}
\frac{f}{g_s l_s} \int  dt \,\Tr X^a X^b X^c X^d X^e\e_{abcde},
\ee
since one can always recover $S_{D2}$ by compactifying $X^1, X^2$ 
and then decompactify using the rules \eq{Td1}, \eq{Td2}. 
Since the action \eq{D0} can be
considered as a special case of 
\eq{red}
in a background $[X^1,X^2] = 1$, 
we propose that in the large flux limit and in the
sector with  $X^{a'} =0$, $a' = 5,6,7,8,9$, M-theory in our curved background 
\eq{M-bkgd-1}, \eq{M-bkgd-2} 
is described by the quantum mechanical action 
\be
\label{M}
S_M 
= -\frac{if}{g_s l_s } \int dt \,\Tr D_t X^b X^c X^d X^e\e_{bcde},\quad
b,c,d,e=1,2,3,4.
\ee
Here we have substituted $X^0 = -i D_t$ and we have 
ignored an  unimportant overall numerical constant.

\subsection{Type \IIA Matrix String Theory}

Now that we have the Matrix model \eq{M} for M-theory, it is simple to apply the procedure of DVV in \cite{dvv} and derive
the corresponding type \IIA matrix model. This involves 
rewriting \eq{M} in terms of the M-theory eleventh dimensional radius
\be
R_{11} = g_s l_s,
\ee
then compactify the scalar $X^2$ on a circle of radius $R_2$, then finally perform an 11-2 flip which
exchanges the role of the $11^{th}$ and the $2^{nd}$
direction of the torus $T^2$ where our M-theory is now compactified on. Keeping all the powers of the constants in the theory is important here.
The process described above is simply applying the `rules'
\be
R_2 = g_s l_s, 
\ee
and 
\be
R_{11} =N,
\ee
where a normalisation of lightcone momentum $p_+ =1$ is adopted, see the appendix of 
\cite{dvv}. We then obtain the matrix model for the type \IIA theory
\be
\label{IIA}
S_{IIA}
= \frac{f}{N}
\int  d^2 \s \,\Tr X^a X^b X^c X^d X^e\e_{abcde}, \quad
a,b,c,d,e=0,1,2,3,4, 
\ee
where 
\be
X^\a = i D^\a, \quad X^i = \mbox{scalars}, \qquad \a= 0,1, \quad i = 2,3,4
\ee
and we have ignored an  unimportant overall numerical constant.

We note that the D1-strings action \eq{D1} 
and the \IIA Matrix string action \eq{IIA} are indeed the same up to a constant
coefficient. This is similar to what was found in \cite{dvv,ver,bonora} 
where the same 
2-dimensional supersymmetric Yang-Mils theory could have different string
interpretations depending on how one associates its parameters with the string 
theories. We remark that the type \IIB matrix model is given by an instantonic action, the M-theory action is given by a quantum mechanical action and the type \IIA action is given by a matrix string action.

\section{Multiple D4-Branes and M5-Branes}
Starting with Matrix string theory action for the type \IIA theory in a large flux, we wish to construct the model of D4-branes in such a limit. Then we will make use of the duality between M5-branes and D4-branes to obtain a theory of M5-branes, at least the non-abelian 3-form field strength sector, in a large constant $C$-field.

\subsection{D4-branes in large RR 2-form flux}

Since the Matrix string theory \eq{IIA} takes the same form as the
original D1-strings action \eq{D1}, it admits the classical solution:
\be \label{soln-23}
[X^\a, X^\b] =0, \quad [X^\a, X^i ] =0, \quad \a=0,1,\; i =2,3,4.
\ee
As before, the commutation relations of $X^i$ among themselves are 
not constrained at the classical level. Let us now consider the solution  
$X^i_{cl} =x^i$ of quantum Nambu geometry
\be \label{d4-xxx}
[x^2,x^3,x^4] = i\th
\ee
and 
consider a fluctuation around it in a similar fashion to the standard noncommutative gauge theory case. 
In the large $N$ limit, our matrix model is built up out of large $N$ matrices $x^i$.
The representation chosen of the quantum Nambu geometry determines whether the basis spans the whole $N\times N$ matrices. We will assume the $x^i$ of the quantum Nambu geometry do not span the whole set of $N \times N$ matrices in general. Then every $N\times N$ matrix can be expressed as a $K\times K$ matrix whose entries are functions of
$x^i$ \cite{bkgd-indep}.
The expansion of the dynamical variables around the classical solution
can be parameterised as
\be
X^i = x^i \id_{K \times K} + A^i(\s,x^j).
\ee
The action \eq{IIA} for the type \IIA model expands over the quantum Nambu geometry as a five-dimensional integral
\be \label{S5}
S_5 =  \frac{f}{N} \int_{\S_5} \tr X^a X^b X^c X^d X^e\e_{abcde}
\ee
where   $\int_{\S_5} = \int d^2 \s \int_x$  and $\int_x$ is an integral on
the quantum Nambu geometry which can be constructed from a representation of
the geometry. In the large $N$ limit, the trace over large $N$ 
matrices decomposes as usual as $\Tr = \int_x \tr$.  

We would like to interpret this as $K$ parallel D1-branes expanding over the quantum Nambu geometry to give $K$ parallel D4-branes.
To do this, let us 
introduce a three-form $H$-field whose components are defined
by
\bea 
H^{abc} &= & - i [X^a,X^b,X^c], \label{H-id-1} \\
H^{de 5} &= & - i  [X^d, X^e], \qquad a,b,c,d,e = 0,1,2,3,4.
\label{H-id-2}
\eea
We remark that a similar
identification has also been proposed in
\cite{CS1} in the analysis of the M5-brane geometry in a large $C$-field but with a Lie 3-algebra valued 3-bracket and not our quantum Nambu geometry.
The action \eq{S5} simplifies to
\be \label{SHH}
S_5 =  \int_{\S_5} \tr H^{abc} H^{de5} \; \e_{abcde},
\ee
where have ignored the unimportant overall constant here.

In order to see the connection of \eq{SHH} with D4-branes, we consider the abelian case. This would mean a dimensional reduction on $x^5$ of the PST action \eq{pst1}. The 
first term $\e_{abcde} H^{abc} H^{de5}$  in the dimensionally
reduced, gauge fixed, PST action \eq{pst1} is precisely equal to \eq{SHH}.
This reduction is quite remarkable and reaffirms the conjecture between M5-branes and 5D Super Yang-Mills theory. It may at first seem that it is inconsistent to find the absence of the second term $H^{*ab5} H^*_{ab5}$ in the action \eq{SHH}, it is actually identified with the
D4-branes' Yang-Mills Lagrangian $\sqrt{-g} F_{ab}^2$ by performing a 
Hodge dualisation after the reduction to five dimensions. 
But we have already shown above that the Yang-Mills term does not contribute in the double scaling limit, therefore the  $H^{*ab5} H^*_{ab5}  $ 
term is absent from \eq{SHH}.
A dimensionally reduced M5-brane is simply a D4-brane, 
so this means \eq{SHH}, for the abelian case, 
does describe a D4-brane in the large RR flux background. 
We would like to use this evidence to propose that, for the non-abelian case, the action \eq{SHH} describes the non-abelian 3-form sector of multiple D4-branes theory (where $X^{a'} =0$) in a large RR 2-form flux background.

The relations \eq{H-id-1} and \eq{H-id-2} give us the relation to the PST action \eq{pst1} from the action \eq{SHH}, we have confidence that the reduction matching consistency condition is sufficient to show that the non-abelian D4-brane theory is indeed given by \eq{SHH}.
We emphasise that the reason that it is possible to write \eq{S5} in 
terms of the $H$'s is entirely due to the fact that the
D1-branes' Chern-Simons action could be 
combined nicely into the remarkable form  \eq{D1},  which
is true only for our constant RR-flux in the \IIB background.

\subsection{Multiple M5-branes using a 1-form gauge field} 

There has been a recent proposal \cite{d1,d2} that
the instantons on multiple D4-branes could be 
identified with the Kaluza-Klein modes associated with the compactification of non-abelian M5-branes on a circle. By including all the KK modes, it was proposed that 
the low energy 5D SYM theory of D4-branes is 
a well-defined quantum theory and is actually the theory of multiple
M5-branes compactified on a circle. We would like a way to include these KK modes in our description of D4-branes in the background flux \eq{SHH}, there is a very natural generalisation of this action to indeed describe the M5-branes.
It comes from the identification \eq{H-id-2}, we can think of the commutator as the following quantum Nambu bracket
\be
H^{ab5} = -i[X^a, X^b]=-i[X^a, X^b, \id],
\ee
this allows us to make the generalisation to promote the identity $\id$ to a dynamical field $X^5$. Thus we have
\be \label{H-id-3}
H^{de5} =  -i [X^d,X^e,X^5]
\ee
 with 
\be \label{X5-1}
X^5 = \id 
\ee
for the D4-brane theory.

We propose that the scalar field $X^5$ is along the compactified 
$X^5$ direction transverse to the D4-branes, then one can understand
the relation \eq{H-id-2} and \eq{X5-1} as saying only the zero mode of
the M5-branes has been included, i.e. a dimensional reduction to D4-branes.   
So it is suggestive to include the higher KK modes 
by promoting $X^5 =\id$ to a general field as in \eq{H-id-3}.
We may now write the 3-form field strength in terms of all six scalar fields as
\be \label{H-id-4}
H^{ \m\n\l} = -i [X^\m, X^\n,X^\l].
\ee
Note that this way of writing the 3-form field strength is different to the conventional way of writing the field strength in terms of the non-abelian 2-form potential $B$.
In the conventional description, the non-abelian 2-form potential 
$B$ is written as $H =dB + \cdots$ where the $\cdots$ term denotes terms necessarily for
the non-abelianisation. For the abelian case, this field strength is well defined in the literature, however for the non-abelian case it is not so simple as we do not fully understand how to write a tensor-gauge connection\footnote{See \cite{sati,gerbe1} for some recent proposals using gerbes. }. So what we are proposing here is that there is a dual description of 
the non-abelian 3-form field strength in terms of the 1-form variables 
$X=X_\m d\s^\m$; 
so the $B$-field and the $X$-fields are related, although one 
can expect the relation to be very complicated. This has not yet been achieved.

To justify our proposal, one needs to show that $H^{\m\n\l}$  satisfies the correct equation of motion \eq{sdeqn} and describes three
on-shell degrees of freedom. We will now propose an action for the tensor-gauge sector of the M5-brane theory on a quantum Nambu geometry, this is a generalisation of \eq{pst} but has an additional term which is allowed in the general case. It reads
\be \label{SM5-six}
S_{M5} = -\frac{1}{4} \int_{\S_6} \tr
\left(
 \frac{1}{6} \e_{abcde}H^{abc} H^{de5} + \sqrt{-g}{\big(} 
c_2 H^{abc} H_{abc}
+ c_3  H^{ab5} H_{ab5}
 {\big)} 
\right),
\ee
where $\S_6 = \S_5 \times S^1$ is the worldvolume of the M5-branes and
\eq{H-id-4} is the definition of the 3-form field strength, here we will consider a constant metric.
The action \eq{SM5-six} is the most general quadratic 
action that can be constructed out of the components 
$H^{abc}$ and $H^{ab5}$ and which is compatible with the $SO(1,4)$ Lorentz
symmetry. For a non-abelian generalisation of the gauged PST Lagrangian \eq{pst1}, 
it is expected that $c_2=1$ and $c_3 =0$. A priori there is no reason to
expect that our action will be exactly the same as the PST action and so we will consider arbitrary coefficients and find solutions for them.

Our goal is to construct an action for the non-abelian 3-form field 
strength living on a system of M5-branes.
Generally, one can turn
on a constant $C$-field on the worldvolume of the M5-branes. 
How could one incorporate a $C$-field in 
\eq{SM5-six}? It is useful to recall a similar 
story for the case of 
D-branes where it is well known that a constant NSNS $B$-field can be naturally 
included as a classical solution (which corresponds to a noncommutative geometry) 
of  matrix models \cite{bss,bkgd-indep,makeenko,smith-2b}. The 
remarkable feature
of this construction is that 
the different backgrounds that correspond to different $B$-fields arise as 
different classical solutions of the same degrees of freedom of the
underlying matrix model. 
Therefore let us follow the same route and  consider a reduction of
the matrix model  to a point. As a result, we obtain the matrix model 
\be \label{S0}
S_0 =   \frac{1}{4} \Tr \left(
c_1 \e_{abcde} X^a X^b X^c X^d X^e X^5 
+ c_2 [X^a,X^b,X^c]^2
+ c_3   [X^a,X^b,X^5]^2
\right),
\ee
where
\be
\label{cc}
c_1 = 2
\ee 
and the parameters $c_2, c_3$ are to be determined.
We would like to have an equation of motion which is precisely the self-duality equation for $H$ from our action. Quite remarkably this can be 
achieved with a particular choice of the parameters. 
 
There are two equations of motion for $S_0$, varying with respect to $X^5$ yields
\be \label{eom1}
c_1 \e_{abcde} X^a X^b X^c X^d X^e +2 c_3 [[X_a,X_b,X_5], X^a, X^b]'=0, 
\ee
and with respect to $X^a$ yields
\bea \label{eom2}
&& c_1  \e_{abcde} 
\left( X^b X^c X^d X^e X^5 + X^b X^c X^d X^5 X^e +X^b X^c X^5 X^d X^e 
   \right. \nn\\
&&\qquad \qquad \left. +X^b X^5 X^c X^d X^e+ X^5 X^b X^c X^d X^e \right) \nn\\
& +& 6 c_2 [[X_a,X_b,X_c],X^b,X^c]' 
+ 4 c_3 [[X_a,X_b,X_5],X^b,X^5]' 
=0.
\eea
The first equation of motion \eq{eom1} can be rewritten as
\be
(\frac{c_1}{6} -c_3) \e_{abcde} X^a X^b H^{cde} 
+ 2c_3[H_{ab5} +\frac{1}{6} \e_{abcde}H^{cde}, X^a,X^b]' =0.
\ee
Since we want to interpret $H^{\m\n\l}$ of \eq{H-id-4}  as the gauge covariant
non-abelian field strength on M5-branes, $H^{\m\n\l}$ must satisfy a 
Bianchi identity. The most natural gauge covariant version would be
\be \label{bianchi}
[X^{[\m}, H^{\n\l\rho]}]=0.
\ee
We use a convention of $[X^{[a}, H^{bcd]}] = [X^a, H^{bcd}] -  [X^b, H^{cda}]+  
[X^c, H^{dab}] - [X^d, H^{abc}]$.
Let us assume this condition holds, particularly 
\be \label{bi1}
[X^{[a}, H^{bcd]}] =0,
\ee
then
we see that the self-duality condition
\be \label{sd1}
H_{ab5} = -\frac{1}{6} \e_{abcde} H^{cde}
\ee
solves \eq{eom1}. 

We now look at the second equation of motion \eq{eom2}. Using
the conditions \eq{bi1} and the 
self-duality condition \eq{sd1}, one can show that
the LHS of \eq{eom2} can be written as
\be \label{eom2'}
(\frac{c_1}{2} + 18c_2)\{H_{ade}, X^d X^e \} 
+\{\frac{1}{4} E^{bcd5} , X^e\} \e_{abcde}
+ (\frac{c_1''}{4} - \frac{2 c_3}{3}) (X^b X^5 H^{cde} - H^{cde} X^5 X^b) ,
\ee
where
\be
E^{bcd5}:= c_1'[H^{bcd},X^5] - 12 c_2 [H^{5[bc}, X^{d]}] 
\ee
and the constants $c_1', c_1''$ satisfy
\be \label{c-1}
c_1' + c_1'' = c_1.
\ee
To get this, we have 
split the term proportional to $c_1$ 
of \eq{eom2} into two terms (with coefficients $c_1'$ and $c_1''$) and
used the $c_1'$ term to combine with the $c_2$ term and the $c_1''$ term to 
combine with the $c_3$ term to arrive at \eq{eom2'}. We note that the 
term $E^{bcd5}$ is of the form of the Bianchi identity
\be\label{bi2}
[X^{[5},H^{bcd]}] = 0
\ee
if $c_1' = 4 c_2$.
Therefore the equation of motion \eq{eom2} is satisfied if the coefficients are 
such that
\be \label{c-2}
c_1' = 4 c_2, \quad c_1'' = -40 c_2, \quad c_3 = -15 c_2
\ee
and the condition \eq{bi2} is satisfied. But the Bianchi identity \eq{bi2} is natural from \eq{bianchi}, so we see that our proposal is justified through the equations of motion (with self-duality).
 
So far we have obtained that the equations of motion \eq{eom1}, \eq{eom2} are satisfied if
the self-duality condition \eq{sd1} and the condition \eq{bianchi}
are satisfied and if the coefficients $c_i$ are given by
\be \label{c-values}
c_2=  -\frac{1}{18} (c_1/2), \quad c_3 = \frac{5}{6} (c_1/2).
\ee 
It is quite remarkable that a set of parameters can be found in a consistent way so that the 
self-duality condition of the 3-form field strength $H$ emerges from a matrix model where the 3-form is a product of 1-forms. This is not guaranteed a priori and provides evidence that the matrix model \eq{S0} describes the tensor-gauge sector of M5-branes with a self-dual 3-form field strength. 

We obtained the Bianchi identity and the 
self-duality condition as a solution of
the reduced matrix model description. 
However we need to establish
that it is the only non-trivial solution to give strength to the proposal.
We recall that in the PST action \eq{pst1},
one does not get the self-duality condition \eq{sdeom}
as the equation of motion immediately.
To do this, one needs to make
crucial use of  the symmetry \eq{T2} which acts on the $B$-field.
For our case, it is possible that there is a counterpart of 
the symmetry \eq{T2} which acts on the $X$'s; and this symmetry is needed 
to derive the  self-duality equation, hence the Bianchi identity. 
It is important to understand whether such a symmetry really exists 
in our model, and if so, how it acts.

The (2,0) supermultiplet \cite{tensor, tensor2} demands that the field strength be self-dual. This gives the on-shell degrees of freedom, namely three. If we were to supersymmetrise our theory by adding in the fermions and transverse scalar fields, self-duality would be an automatic feature of the theory. The issue here is that we are looking at just the tensor-gauge sector of the M5-brane theory and so we cannot argue that self-duality is automatically satisfied, hence also the Bianchi identity.

To obtain a six dimensional field theory for the worldvolume of the M5-branes, we need to
consider classical solutions to the equations of motion and
expand them by a fluctuation around the solution to build the six dimensional theory in a similar fashion to the Yang-Mills case \eq{pertsol}.
We will consider the following solution $X^\m =x^\m$ such that
\be \label{6dnambu}
[x^\m,x^\n,x^\l] = i \th^{\m\n\l} \id,
\ee
where $\th^{\m\n\l}$ are arbitrary constants.
Clearly, the Bianchi identity \eq{bianchi} is satisfied as the solution for 
\be
\label{htheta}
H^{\m\n\l} = \th^{\m\n\l}\id
\ee
is proportional to the identity and so the commutator in the identity vanishes.
Moreover the self-duality condition \eq{sdeqn} is satisfied if the parameter
$\th^{\m\n\l}$ is self-dual due to \eq{htheta}.
Thus we obtain a six dimensional quantum Nambu geometry 
parameterised by self-dual parameter $\th^{\m\n\l}$. 

The fluctuation around the solution \eq{6dnambu} can be written as
\be \label{flu}
X^\m = x^\m \id_{K\times K}+ A^\m(x),
\ee
where $A^\m$ are $K\times K$ matrices whose components are valued on the worldvolume of the quantum Nambu geometry of the M5-branes. The large $N$ trace becomes
\be
\Tr = \int_x \tr,
\ee
where $\int_x$ is determined from the representations of the quantum
Nambu geometry \eq{6dnambu} and our proposal for a theory of $K$ M5-branes
(or more precisely,  $K$ non-abelian 3-form) is
\be \label{SM55}
S_{M5, \th} = -\frac{1}{4} \int_x \tr
\left(
 \frac{1}{6} \e_{abcde}H^{abc} H^{de5} + {\big(} 
\a\; \frac{-1}{3} H^{abc} H_{abc}
+ (1-\a) H^{ab5} H_{ab5}
 {\big)} \sqrt{-g}
\right),
\ee
with $\a =1/6$ which is precisely the solutions \eq{c-values}.  We note that with 
the self-duality condition, 
the second and the third term in \eq{SM55} can be summed together 
and is equal to
$H^{ab5} H_{ab5}$ for any value of $\a$; 
and therefore 
the action \eq{SM55} has in fact precisely the same
form (including the coefficients) as the 
non-abelian generalisation of \eq{pst1}.
However  only for
$\a =1/6$ can one identify a Bianchi identity \eq{bianchi} and the self-duality condition \eq{sdeqn}.

We have shown that the worldvolume of the non-abelian M5-brane theory has a quantum Nambu geometry. Now we consider the origin of this quantum Nambu geometry as a quantised spacetime.
In the case of D-branes, the presence of a noncommutative worldvolume on a brane is generally due to a background gauge potential being turned on in its worldvolume i.e. an NS-NS $B$-field.
The self-duality of the quantisation parameter $\th^{\m\n\l}$ suggests to
identify it with the self-dual 3-form $C$-field on the worldvolume of 
the M5-branes. We can perform a dimensional reduction along $X^5$ to show that this is consistent with the D4-brane theory by putting $X^5 =\id$. The relation \eq{6dnambu} reads
\be
[X^a,X^b, \id] = [X^a,X^b] = i \th^{ab5}. 
\ee
This is the noncommutative geometry over D4-branes with a $B$-field\footnote{Recall we are considering the linearised limit of the DBI action which contains the $B$-field.}  whose
components are $B_{ab} = \th_{ab 5}$.
Since the $B$-field is related to the 11-dimensional $C$-field as 
$B_{ab} =C_{ab5}$, it is correct to identify 
$\th^{\m\n\l}$ with the constant $C$-field $C^{\m\n\l}$. 
All in all, we conclude that  the geometry \eq{6dnambu} is the result of having a self-dual 3-form $C$-field 
\be \label{C-th}
C_{\m\n\l} = \th_{\m\n\l}
\ee
turned on in the worldvolume of the M5-branes. So the action \eq{SM55} describes the 3-form field strength of multiple M5-branes in a self-dual constant $C$-field.
 
In the conventional description of the 3-form field strength $H_{\m\n\l}$ of the 2-form tensor-gauge field $B_{\m\n}$ the definition is given by $H = \mathrm{D} B$, where ${\rm D}$ is the tensor-gauge covariant derivative which is not known for the non-abelian case. The correct number of degrees of freedom is three which is reduced down from fifteen by the tensor-gauge and self-duality, we need to make sure that this is what we have in the 1-form formalism to ensure that we are describing the same object. What we have proposed is that there is a dual description to this in terms of non-abelian 1-forms $X$ such that the degrees of freedom match in both cases and that $H$ is self-dual and satisfies the Bianchi identity. 
Naively we may think that we have too many degrees of freedom in the 1-form formalism by simply counting the fields in \eq{H-id-4}, this gives six fields. But we also have an equation of motion which must be satisfied, namely the self-duality equation \eq{sdeqn}, which reduces the degrees of freedom by half, i.e. to three. So the two formalisms are equivalent except we do not choose to write the 3-form as a derivative of a 2-form, but rather as a product of 1-forms.

\section{Discussions}
In this Chapter, we demonstrated that a new novel structure exists both in String Theory and M-Theory known as the quantum Nambu geometry (QNG). This geometry is unique and is not of the same form of the Lie algebraic type noncommutative geometries. We found that the D1-strings in a low energy large flux double scaling limit gives rise to D4-branes as an expansion over the QNG. This leads to the promotion of the D4-brane action to that of the M5-brane, here we are only considering the gauge sector. The M5-brane is in a constant $C$-field which is also described by the QNG, hence the $C$-field is self-dual $\th^{\m\n\l} = C^{\m\n\l}$.

The construction of the 3-form field strength of the M5-brane was constructed as a product of 1-form gauge fields $X$. This is not the same as the conventional description of the field equations of the M5-brane, as these are written as a derivative of the tensor-gauge field $B_{\m\n}$. It is expected that the two descriptions are equivalent but the relation between the two could be very complicated. It is important to understand this for completeness. The construction in terms of the quantum Nambu bracket covers just the gauge sector, it does not give the transverse scalars or the fermions of the theory. The supersymmetrisation of this model is important in understanding the full multiple M5-brane theory.

The action \eq{SM55} is of the M5-brane in a constant $C$-field over a QNG. So let us discuss an analogy with D-brane physics. A D-brane in a NS-NS $B_{\m\n}$ field can be described in terms of a commutative DBI action {\it or} a noncommutative Yang-Mills action. This dual description is called the Seiberg-Witten map \cite{sw}, it relates the noncommutative Yang-Mills action to the DBI action with commutative coordinates and finds a relation between $\th$ and $B$. Our action \eq{SM55} is analogous to the noncommutative YM action as it is described over a fluctuation around the QNG similar to the noncommutative YM action over a noncommutative geometry. The equivalent full DBI action for the M5-brane in a $C$-field to a first order approximation has not been constructed. Some attempts of the case with $C=0$ have been constructed in \cite{chu,pm,sezgin}. 

The fluctuation analysis leads to the 3-form field strength $H_{\m\n\l}$ on the QNG, so we have a 3-bracket describing a 3-form field. For the noncommutative YM case we have the commutator with a noncommutative geometry fluctuating to give a 2-form $F_{\m\n}$. It seems natural for a quantum $N$-bracket to exist and to describe an $N$-form gauge field strength. It would be interesting to explore this further; although we only know of one higher tensor gauge field strength, it is possible that this would be useful for describing the Hodge duals of field strengths as in the D4-brane picture.

An emphasis must be placed on the properties of the quantum Nambu bracket, this is not a quantisation of the Nambu-Poisson bracket. The Nambu-Poisson bracket was introduced by Takhtajan and obeys the fundamental identity \eq{fi}. The quantisation of this bracket is a difficult problem. The quantum Nambu bracket does not obey the fundamental identity, it is the completely anti-symmetrised sum of three operators on the QNG.

\chapter{Representations of the Quantum Nambu Geometry}
In this chapter we will analyse the mathematical properties 
of the quantum Nambu geometry. Infinite dimensional representations of the quantum Nambu geometry are constructed. We will show that the infinite dimensional 
representations imply that the quantum Nambu geometry is different from the ordinary Lie algebra type geometry. 

\section{Finite Lie-algebraic Representations}
Before we begin to analyse the representations for the quantum Nambu geometry, it is useful to discuss finite representations of the relation \eq{xxx} which were constructed by Nambu where the generators of the algebra are in a Lie algebra. If we let $X^i = \a l_i$ for a constant $\a$ and $l_i$
are the generators of the $SU(2)$ algebra
\be
[l_i, l_j ] = -i \e_{ijk} l_k, 
\ee
then 
\be
[X^i,X^j,X^k] = i  \e^{ijk} \a^3 C_{\rm R}, 
\ee
where $C_{\rm R}$ is the quadratic Casmir 
for the representation ${\rm R}$ where $X^i$ is in. 
For $N\times N$ matrices, $C_N = (N^2-1)/4$ and so if we choose 
$\a^3 = \th/C_N$, 
then we can realise the relation \eq{xxx} with finite $N\times N$
matrices. Nambu has also constructed a representation of the relation \eq{xxx}
in terms of $SU(2)\times SU(2)$ representations. In these representations Nambu 
constructed, the quantum Nambu bracket is embedded in  
an underlying Lie algebra ($SU(2)$ or
$SU(2) \times SU(2)$ algebras) as a Casmir. As such, the relation \eq{xxx} is not 
fundamental but is a result of an underlying Lie algebraic structure.
So for finite $N$, the quantum Nambu bracket has a reducible representation in terms of just the usual Lie algebra $SU(2)$.
What we will show next is that in the large $N$ limit, there are new infinite dimensional 
representations of \eq{xxx} that are not representations of any Lie algebra. It is the existence of these representations that demonstrates the fundamental and novel nature of the 
Nambu-Heisenberg commutation relation \eq{xxx}.

\section{Infinite dimensional representations}

An infinite dimensional representation of \eq{xxx} has been 
constructed by Takhtajan \cite{tak}, however his representation is 
complex as the 
operators $X^i$ are not represented as Hermitian operators; as a result the quantum space is six dimensional.  In his paper, the Nambu bracket satisfies the fundamental identity, in our analysis we do not demand this. In this subsection, we give two examples of representations where the quantum space is three dimensional as opposed to six dimensional.

So let us begin by constructing representations of quantum Nambu geometry with three operators $X^1,X^2,X^3$\,:
\be \label{x123}
[X^1,X^2,X^3] = i \th, 
\ee
where $\th$ is real.

\subsection{A representation in terms of $Z,\Zb,X$}
In the case of representations of two Hermitian fields (as in the Heisenberg commutation relation for example), it is natural to construct a complex field in the usual manner of $Z = X_1 +iX_2;$ for the case of the quantum Nambu geometry we have 3 fields, so let us consider Hermitian $X^i$'s and introduce the complex fields
\be
Z:= X^1+ i X^2, \quad \Zb := X^1 -i X^2,
\ee
and $X=X^3$.
We can then rewrite the the bracket \eq{x123} in the form
\be \label{xzz}
[X, Z, \Zb] =2 \th.
\ee
We consider an ansatz for a representation
\bea
Z\ket{\o}&=&f_1(\o)\ket{\o+\b}+f_2(\o)\ket{\o-\b}, \label{repa}\\
\Zb\ket{\o}&=&f_2^*(\o+\b)\ket{\o+\b}+f_1^*(\o-\b)\ket{\o-\b},\label{repb}\\
X\ket{\o}&=&g(\o)\ket{\o}, \label{repc}
\eea
where the state $\ket{\o}$ is parameterised by a number $\o$ and 
$\b$ is a fixed ``step''.
It is clear the domain of $\o$ is one-dimensional. Without loss
of generality we can take $\b$ real and $\o \in \RR$.
The form of \eq{repb} is fixed by \eq{repa}, this is by requiring
$\Zb = Z^\dagger$.  Since $X$ is Hermitian, it follows that $g$ is real.
It would be natural to consider the representation \eq{repa}-\eq{repc} with $f_2=0$ or
$f_1=0$ if we think of the $Z,\Zb$ as raising and lowering operators naively. 
However this 
always gives a constraint of the form
$Z \Zb + \Zb Z = \cZ(X)$ for some function $\cZ$ and so describes at most a 
2-dimensional space. As a result, we are prompted to 
try the more general ansatz 
stated above. 

By calculating 
\be
[X,Z,\Zb]\ket{\o} = X[Z,\Zb]\ket{\o} +Z[\Zb,X]\ket{\o}+\Zb[X,Z]\ket{\o},
\ee
and evaluating each of the individual terms we obtain
\be
[X,Z,\Zb]\ket{\o}=I_2(\o)\ket{\o+2\b}+I_{-2}(\o)\ket{\o-2\b}+I_0(\o)\ket{\o},
\ee
where
\bea 
\label{raise} I_2(\o)&=& G(\o) K(\o), \\ 
I_{-2}(\o)&=& I_2(\o-2\b)^*, \\
I_0(\o) &=&  F(\o) \big( 2g(\o)-g(\o-\b)\big)  -F(\o+\b) \big( 2g(\o)-g(\o+\b)\big)
\eea
and
\bea
G(\o)&:=& g(\o+2\b)+g(\o)-g(\o+\b),\\
K(\o)&:=&f_2(\o+\b)^*f_1(\o+\b)-f_1(\o)f_2(\o+2\b)^*, \\
F(\o)&:=&|f_1(\o-\b)|^2-|f_2(\o)|^2.
\eea
We would like to find functions $g, f_1, f_2$ such that
\be
\label{raiselower0}
I_2= I_{-2} =0
\ee
and 
\be \label{i0}
I_0 = 2\th, 
\ee
this is due to the condition \eq{xzz}.
The condition \eq{raiselower0} can be solved by requiring 
$K(\o)=0$ or $G(\o) =0$, due to \eq{raise}.
The choice of $K=0$ implies that 
$[Z, \Zb] \ket{\o} = F(\o) \ket{\o}$ and so there is a relation of the form 
$[Z,\Zb] = \cZ(X)$ for some function $\cZ$. This means the relation \eq{xzz}
is not the fundamental relation of the representation but is reducible to a statement about commutators, we do not want this as we are looking for new representations of the quantum Nambu geometry as a new kind of geometry in string theory. So let us consider
\be \label{G0}
g(\o+2\b)+g(\o)-g(\o+\b)=0.
\ee 
It is easy to see that it
implies a  pseduo-periodic condition
\be \label{p-g}
g(\o+3\b)=-g(\o),
\ee
so
\be
g(\o+6\b)=g(\o),
\ee
and it follows that
\be
I_0(\o) =  F(\o) A(\o) -F(\o+\b) A(\o-\b),  
\ee
where
\be
A(\o) := g(\o)+g(\o+ \b),
\ee
and
$ A(\o+3\b) = -A(\o)$, $F(\o+3 \b) = - F(\o)$.
The condition \eq{G0} is solved by
\be \label{g-general}
g (\o) = \sin \a \o, \quad \cos \a \o, \qquad \mbox{where} \quad 
\a = \frac{\pi}{3 \b}(6 p \pm 1), \quad p \in \ZZ,
\ee
or generally a Fourier sum of these modes. For simplicity, let us 
construct a representation for the simple mode
\be
g (\o) = \cos \a \o, 
\ee
where $\a$ is as specified in \eq{g-general}. 
Consider the ansatz
\be
F(\o) = k \sin ( \a \o- \frac{\a\b}{2}) .
\ee
By substituting this ansatz into the relations, we find \eq{i0} is solved with
\be
k = - \frac{2 \th}{\sin{\a \b}  \cos\frac{\a \b}{2}}.
\ee
This provides a constraint on the two functions $f_1$ and $f_2$. For example,
a simple solution is
\be
|f_1(\o)|^2 = |f_2(\o)|^2 = k_0 - \frac{8\th}{3} \cos \a \o, 
\ee
where $k_0 > 8\th/3$ 
is  any constant such that the right hand side above is positive. Without loss 
of generality, we can take $\b=1$. The representation space is 
given by the 1-dimensional lattice
\be
\{ \ket{ \o + n}: n \in \ZZ  \},
\ee
and is of countably infinite dimension 
for each fixed $\o$.

\subsection{A representation with $Z_3$ symmetry}

We now demonstrate that there is another way to construct a representation of 
\eq{xxx} such that the quantum space it represents is 
3-dimensional. 
In this construction, we assume no reality condition on the fields $X^i$, thus far we have 6 degrees of freedom. 
Instead, let us introduce a unitary operator, 
\bea
U\ket{\o} = &&\ket{\r^2 \o}, \nn \\
U^\dagger \ket{\o} = &&\ket{\r\o},
\eea
and assuming 
\be
X^1\ket{\o} = (\o +a)\ket{\o +1},
\ee
one obtains 
\bea
U^\dagger X^1 U \ket{\o} = &&(\r^2\o +a)\ket{\o +\r}, \nn \\
U^\dagger{}^2 X^1 U^2 \ket{\o} = &&(\r\o +a)\ket{\o +\r^2},
\eea
where $a \in \mathbb{C}$ and $\r$ 
is a cubic root of unity ($\r^3=1$) which is not equal to 1.

Now if the fields $X^1$, $X^2$ and $X^3$ are unitarily related to each other by
\bea
X^2 = &&U^\dagger X^1 U, \nn\\
X^3 = &&U^\dagger X^2 U, 
\eea
then
\bea
\label{evals1}
X^1\ket{\o} && = (\o + a)\ket{\o +1}, \nn\\
X^2\ket{\o} && = \r^2(\o + a\r)\ket{\o +\r}, \nn\\
X^3\ket{\o} && = \r(\o + a\r^2)\ket{\o +\r^2}
\eea
and it easy to see that 
\be
[X^1,X^2,X^3]\ket{\o} = 3(a^2-a)(\r-\r^2)\ket{\o},
\ee
where $a \in \mathbb{C}$ and $\r-\r^2$ is pure imaginary. 
In this representation the fields $X^1,X^2,X^3$ are not Hermitian. 
They are however related through a unitary transformation,
$U = e^{i\Th}$, where $\Th$ is some Hermitian operator. 
So in this representation, we have 2 degrees of freedom from $X^1$ 
and one from $\Th$ giving us 3 real dimensions.

We note that in this representation, 
the operators $X^i$ can be constructed as pseudo-differential 
operators acting
on functions $\ipr{\o}{\psi}=\psi(\o).$ Let us start with $X^1$ and note that
$ \bra{\o~+~1}X^1=(\o+a)\bra{\o}$ and so
$ X^1\psi(\o)=\bra{\o}X^1\ket{\psi}=(\o+a-1)\psi(\o-1)$. 
Therefore, we obtain 
\be
X^1=(\o+a-1)e^{-\frac{\pa}{\pa \o}}.
\ee
Similarly
\be
X^2=(\r^2\o+a -1)e^{-\r\frac{\pa}{\pa \o}},
\ee
\be
X^3=(\r\o+a-1)e^{-\r^2 \frac{\pa}{\pa \o}},
\ee
and for the unitary operator
\be\label{eq:UnitaryPseudo}
U=\exp\left[\ln(\r)\o\frac{\pa}{\pa\o}\right].
\ee
The Hermitian conjugate of the unitary operator is
\be
U^\dagger=\exp\left[\ln(\r^2)\o\frac{\pa}{\pa\o}\right].
\ee

In this construction, the representation space is 
given by  the 2-dimensional lattice
\be
\{ \ket{ m+ n \rho }: m,n \in \ZZ  \}
\ee
and is of countably infinite dimension.

In conclusion, we have shown that 
there are at least two ways to represent \eq{xxx} as a three-dimensional 
quantum space: either a real representation, or having one complex field 
and introducing a unitary operator relating $X^2,X^3$. 
This is in contrast to the representation in \cite{tak} where all the fields 
are complex and not unitarily related.

\section{Discussions}

In this Chapter, we considered the representations of the QNG. Although the geometry is quantised, the action is classical. It would be interesting to construct a QFT on such a QNG. As with noncommutative geometry, it is interesting to construct the operators as an algebra of functions with a $*$-product. It is not immediately obvious how to construct this however, some attempts were made in \cite{CS1}.

In a similar spirit to noncommutative geometry where a fuzzy sphere solution can be analysed, the relation 
\be \label{xxx-x}
[X^i,X^j,X^k] = i \l \e^{ijkl} X^l,
\ee
can also be defined. It would be interesting to examine the physics of this and construct the representations where the bracket is given by the QNB. The QFT's one can construct from this is still an open question.

\chapter{Conclusions}
In this Chapter we will discuss some recent advancements in the theory of M2 and M5-brane theories, we will highlight interesting areas which can be explored as a result of the work in this thesis.

\subsubsection{The non-abelian (2,0) tensor supermultiplet}
We now discuss the results obtained on the non-abelian $(2,0)$-tensor multiplet coupled to an auxiliary vector as an attempt to describe the non-abelian M5-brane on the level of equations of motion. Since the equations of motion are linear in the 3-form field strength $H_{\m\n\l}$, we do not have the issues associated with self-duality. 

Very little is known about the non-abelian theory of $N$ M5-branes. On the level of supersymmetry transformations and equations of motion, there has been some recent progress \cite{tensor,tensor2} by introducing an auxiliary vector. 
We now consider these results on the worldvolume field theory on multiple M5-branes.
Originally, the free six dimensional (2,0) tensor supermultiplet was found in \cite{6dsusy,howe2}
\bea
\label{abelian20}
\d X^I &=& i\e\G^I\Psi \nn\\
\d \Psi &=& \G^\m\G^I\del_\m X^I\e +\frac{1}{2}\frac{1}{3!}\G^{\m\n\l}H_{\m\n\l}\e \nn\\
\d H_{\m\n\l} &=& 3i\eb\G_{[\m\n}\del_{\l]}\Psi,
\eea
where $\m=0,...,5$ and $I=6,...,10$. The convention for the self-dual field strength is $H_{\m\n\l}=3\del_{[\m}B_{\n\l]}$ and one could write the transformations \eq{abelian20} in terms of the 2-form $B$ also. The chirality of the supermultiplet implies that the projector conditions are given by
\be
\G_{012345}\e =\e, 
\ee
also for the fermion we have
\be
\G_{012345}\Psi=-\Psi.
\ee
The algebra closes on-shell, satisfying the equations of motion;
\bea
\del^2X^I &=&0 \\
\G^\m\del_\m\Psi&=&0 \\
\del_{[\m}H_{\n\l\r]}&=&0,
\eea
which are the usual free scalar and fermionic equations of motion along with the Bianchi identity for $H_{\m\n\l}$. 

This algebra was generalised to the non-abelian case in \cite{tensor}, the proposal relies on introducing a new vector field which turns out to be auxiliary. All fields are now promoted to live in a vector space (this turns out to be a 3-algebra), with the same form of covariant derivative as in the BLG case \eq{cderiv}. The theory of interacting multiple M5-branes should have (2,0) supersymmetry and an $SO(5)$ R-symmetry acting in the transverse directions to the brane. The system upon reduction should reduce to the D4-brane theory, this is described by 5D SYM. To this end, we know what the interactions for the M5-brane theory should look like if they are to reduce down in the correct way.

The key to writing the correct transformations for the interacting theory rely on using a 3-bracket  with structure constants $f^{abc}{}_d$;

\bea
\label{20mult}
\delta X^I_a &=& i \bar \epsilon \Gamma^I \Psi_a \nn\\
\delta \Psi_a &=& \Gamma^\mu \Gamma^I D_\mu X_a^I \epsilon + \frac{1}{ 3!}\frac{1}{2} \Gamma_{\mu\nu\lambda}H_a^{\mu\nu\lambda}\epsilon- \frac{1}{2}\Gamma_\lambda \Gamma^{IJ} C^\lambda_b X^I_c X^J_d {f^{cdb}}_a\epsilon \nn\\
\delta H_{\mu\nu\lambda\; a} &=& 3 i \bar \epsilon \Gamma_{[\mu\nu}D_{\lambda]} \Psi_a +  i\bar \epsilon \Gamma^I \Gamma_{\mu\nu\lambda\kappa}C^\kappa_{b} X^I_c \Psi_df^{cdb}{}_a \nn\\
\delta \At_\m {}^b{}_a &=& i \bar \epsilon \Gamma_{\mu\lambda} C^\lambda_c \Psi_d {f^{cdb}}_a\nn\\
\delta C^\mu_a  &=& 0.
\eea
The structure constants are fully anti-symmetric $f^{abcd}=f^{[abcd]}$ once we introduce the metric $h^{ab}=\Tr(T^a,T^b)$. The structure constant also satisfies the fundamental identity;
\be
f^{[abc}{}_e f^{d]ef}{}_g=0.
\ee
Again the algebra can be made to close on-shell subject to the equations of motion
\bea
D^2 X_a^I  &=&\frac{i}{2}\bar\Psi_cC^\nu_b\Gamma_\nu\Gamma^I \Psi_d f^{cdb}{}_a + C^\nu_b C_{\nu g} X^J_cX^J_eX^I_f f^{efg}{}_{d}f^{cdb}{}_a
 \cr
 D_{[\mu}H_{\nu\lambda\rho]\;a}  &=&-\frac{1}{4}\epsilon_{\mu\nu\lambda\rho\sigma\tau}C^\sigma_b X^I_cD^\tau X^I_df^{cdb}{}_a - \frac{i}{8}\epsilon_{\mu\nu\lambda\rho\sigma\tau}C^\sigma_b \bar\Psi_c\Gamma^\tau \Psi_df^{cdb}{}_a \cr
\Gamma^\mu D_\mu\Psi_a  &=& -X^I_cC^\nu_b\Gamma_\nu\Gamma^I\Psi_d f^{cdb}{}_a \cr
\tilde F_{\mu\nu}{}^b{}_a  &=& -C^\lambda_cH_{\mu\nu\lambda\; d}f^{cdb}{}_a\;,
\eea
and the conditions 
\bea
C^\rho_cD_\rho X^I_d f^{cdb}{}_a = 0 &\;,& D_\mu C^\nu_a  = 0 \cr
 C^\rho_cD_\rho \Psi_d f^{cdb}{}_a = 0&\;,& C^\mu_cC^\nu_df^{bcd}{}_a = 0\cr
C^\rho_cD_\rho H_{\mu\nu\lambda\;a} f^{cdb}{}_a = 0,&&
\eea
on $C^\m_a$. From these conditions, we see that the fields of the $(2,0)$ theory cannot be parallel to the direction of $C^\m_a$ and that the $C^\m_a$ itself must be constant. This means that the coupling of the $(2,0)$ supermultiplet to this vector seems to give us a five-dimensional theory as opposed to a six-dimensional one. But in \cite{tensor2}, the energy-momentum tensor was calculated and was found to carry momenta in all six directions, so the interpretation is that the theory does indeed describe a six-dimensional theory. The momenta in the $C^\m_a$ field however corresponds to the instantons of a compactified D4-brane theory. This relates back to the duality between M5-branes on $S^1$ and 5D SYM as discussed in Chapter 1. If a space-like vacuum expectation value for $C^\m_a$ is taken with a Lorentzian gauge group then the system reduces to that of 5D super Yang-Mills, this was discussed in \cite{tensor}. It is also possible to take a null reduction of $C^\m_a$ to obtain a system of instantons in one of the lightcone dimensions and the four space dimensions \cite{tensor2}.

It would be interesting to see if one could add matter fields to the action \eq{SM55} and construct the supersymmetric action. The 3-form in the action obtained from the quantum Nambu geometry does not obey the fundamental identity, so this poses extra difficulties when trying to close supersymmetry algebras and indeed showing that an action would be invariant under supersymmetry.

\subsubsection{Tensor-gauge symmetries and non-abelian actions}
On the M5-brane, a solitonic object exists known as the self-dual string \cite{sdstring}. It was proposed in \cite{CS2} that a certain analysis of the boundary dynamics on M2-branes in the ABJM theory gives the {\it multiple} self-dual strings action with a $U(N)\times U(N)$ gauge symmetry. This is given by a Wess-Zumino-Witten action with couplings to matter terms; the action arises by considering a boundary to the Chern-Simons ABJM theory and instead of imposing a boundary condition, extra degrees of freedom are added to cancel the gauge non-invariance of the boundary term. 

More recently in \cite{chu}, it was shown that the induced $U(N)\times U(N)$ gauge symmetry on the self-dual strings is in fact valid on their whole spacetime i.e. the M5-branes. A $G\times G$ gauge symmetry is then found and the explicit tensor-gauge symmetry is constructed with the use of two auxiliary Yang-Mills gauge fields. 
The construction starts with finding the tensor-gauge symmetries in any dimension and then building a free field action in arbitrary dimensions. The $G\times G$ {\it gauge} symmetries are given by
\be \label{GGA}
\begin{array}{lll}
\d_{\L} A_\mu &=& \del_\m \L+ [A_\m, \L],  \qquad\d_{\L} A'_\m= [A'_\m,\L],\\
\d'_{\L'} A_\mu &=&  [A_\m, \L'],  \qquad \d_{\L'} A'_\m= \del_\m \L'+[A'_\m,\L'],
\end{array}
\ee
where (un)primed fields are (left)right gauge fields in the direct product. This can be used to construct gauge transformations on $\cB_{\m\n}$ as 
\bea
\d_{\L} \cB_{\m\n} = [\cB_{\m\n}, \L], \\
\d_{\L'} \cB_{\m\n} = [\cB_{\m\n}, \L'],
\eea
where $\cB_{\m\n} : = B_{\m\n} - \frac{1}{2}(F_{\m\n} -F'_{\m\n})$. Finally, we can write down the {\it tensor-gauge} transformations given by $\L_\a$;
\bea
\d_{\L_\a} B_{\m\n} &=& \frac{1}{2} \left[
(D_\m + D'_\m) \L_\n - (D_\n + D'_\n) \L_\m 
\right] \label{t1}\\
&=& \left[
\del_\m + \frac{1}{2}(A_\m + A'_\m), \L_\n 
\right] - (\m \leftrightarrow \n), \nn\\
\d_{\L_\a} A_\m &=& \L_\m, \label{t2}\\
\d_{\L_\a} A'_\m &=& - \L_\m. \label{t3}
\eea
which implies
\be
\d_{\L_\a} \cB_{\m\n}=0.
\ee
The 3-form field strength of the tensor-gauge invariant $\cB_{\m\n}$ is given by
\be
\cH_{\m\n\l} \equiv [\del_\m+ \cA_\m\; , \cB_{\n\l}] 
+ \mbox{($\m,\n,\l$ cyclic)},
\ee
with modified Bianchi identity
\be
\cD_{[\m} \cH_{\n\l\r]} = \frac{3}{2} [\cF_{[\m\n}, \cB_{\l\r]}].
\ee
A self-duality condition may then be imposed with this tensor-gauge symmetry when applying  to M5-branes, this is a consequence of the $(2,0)$ supermultiplet equations of motion. In a similar fashion to only the non-perturbative M2-brane theory admitting maximal supersymmetry, Chu\footnote{Also \cite{sezgin} independently.} proposed that the full $(2,0)$ supersymmetry on the non-abelian M5-brane theory may not be seen. The proposal is that the M5-branes have a $(1,0)$ supersymmetry using this model.

Self-duality is a crucial ingredient in reducing the fifteen component tensor-gauge field $B_{\m\n}$ in the $(2,0)$ supermultiplet to three components. In our formalism presented in Chapter 6, we do not have a tensor-gauge symmetry but we do have the correct degrees of freedom with a gauge symmetry $G$. In a $G \times G$ gauge group, which contains tensor-gauge symmetry, the gauge symmetry can be fixed to a gauge group $G$; so this agrees and supports both descriptions of the 3-form and to some extent, the new type of Seiberg-Witten map.

In \cite{sezgin}, a non-abelian $(1,0)$ superconformal model was also constructed in six dimensions using a tensor hierarchy where the 3-form gauge potential is included with the tensor field and YM gauge fields. The $G\times G$ structure described above is a non-trivial solution to the constraints presented in this paper. However, the gauge group is restricted to $SO(5)$ or a nilpotent gauge group; the former restricts the rank (hence number of branes), while the second is difficult interpret physically.

A generalisation of the Perry-Schwarz action \cite{schw} was recently constructed in \cite{chuko}; this gives the non-abelian action of the 2-form $B$ in six dimensions and admits the self-duality equation. This action has a modified 6D Lorentz symmetry and admits the self-duality equation for the 3-form as an equation of motion. Upon a compactification along an $S^1$, the action gives 5D SYM plus corrections. So it has been argued that the gauge sector of the multiple M5-brane theory has been found. Since this action is describing the non-abelian 3-form part of the M5-brane theory, it may give a relation to the quantum Nambu geometry as a generalisation of the Seiberg-Witten map. This would be an exciting new relation of M-brane generalisations of the mapping.

\subsubsection{Higher gauge groups and M-branes}

The construction \cite{gerbe1} provides a way to describe M2-brane and M5-brane theories in terms of Lie 2-groups from their 3-algebras, these Lie 2-groups are higher gauge theories which arise from non-abelian gerbes. The motivation for this development comes from an attempt to write down a multiple self-dual string equation in spacetime coordinates instead of loop space \cite{gut-sds}. This would require a tensor-gauge connection and so this is where the non-abelian gerbe theory becomes useful.
The reformulation to Lie 2-groups gives a much more general framework for writing down M-brane models than 3-algebras allow and so it is an interesting avenue of research to explore. One possibility is that the $\cN=8$ BLG theory in such a framework will have a higher rank gauge group, thus allowing it to describe $N$ M2-branes.
The current work has only described these models with matter fields as sections of vector bundles. Once the tensor bundles have been established; it is hoped that a Basu-Harvey type equation will be obtained, describing the chiral tensor field $B_2$ on the M5-brane. Another interesting area of research would be to obtain the mass-deformed Basu-Harvey equation in this language and to find the qualitative differences to the results obtained in this thesis. The key to finding the dualisation between the 3-form $H^{\m\n\l}=-i[X^{\m},X^\n,X^\l]$ and the definition $H_3 = \rm{D} B_2$ in the QNG could very well rely on the structures proposed here.

\subsubsection{Free energy of M5-branes}

In a SYM gauge theory, the free energy from the gravitational dual of a stack of $N$ D-branes scales as $N^2$; this is well understood. In \cite{m2-entropy}, the authors used the full path integral on a sphere of the $\cN=6$ ABJM theory which localised into a matrix model. This allowed them to compute the free energy in the large 't Hooft limit to obtain the famous $N^{3/2}$ scaling for $N$ M2-branes. For the M5-brane theory, the scaling goes as $N^3$. In \cite{N3-3} and more recently in \cite{N3-1, N3-2,N3-4}, some progress has been made in understanding the $N^3$ scaling for the $(2,0)$-theory in a Coulomb phase. In \cite{N3-3} the scaling was indeed found for the Coulomb phase of the $(2,0)$-theory and more recently in \cite{N3-1}, the $(2,0)$ Coulomb branch was related to the 4D SYM theory. This allows one to compute the conformal anomaly for the SYM theory and then relate it back to the M5-brane theory. These works are still early ideas on solving the $N^3$ mystery, but are promising initial results.

\appendix

\chapter{Supersymmetry}

In this Appendix we state some identities which are useful in Chapters 1,2,3,4.

We begin with the eleven dimensional gamma matrix identities $\G^a$, the Clifford algebra is given by
\be
\{\G^a,\G^b\} = 2\eta^{ab}\id,
\ee
where $a,b=0,...,10$ and the spacetime signature is mostly positive. Anti-symmetrised gamma matrix products are defined as
\be
\G^{a_1...a_n} = \G^{[a_1}\G^{a_2}\dots\G^{a_n]}.
\ee

\section{BLG Identities}
For the BLG theory, we have the eleven dimensional gamma matrices decompose as $\G_a = (\G_\m,\G_I)$, where $\m=0,1,2$ and $I=3,...,10.$ We also have the following identities
\bea
\G &=& \G^{0123456789(10)} = -1,\\
\G^{012}\e&=&\e, \\
\G^{012}\Psi&=&-\Psi,
\eea
for the BLG theory. These can be used to derive the duality relations for the transverse directions
\bea
\label{hodge1}
\G^{I_1...I_k}\Psi &=& \frac{(-1)^{\frac{1}{2}(8-k-1)(8-k)}}{(8-k)!}\ve_{I_1...I_8}\G^{I_1...I_{(8-k)}}\Psi,\\
\label{hodge2}
\G^{I_1...I_k}\e &=& -\frac{(-1)^{\frac{1}{2}(8-k-1)(8-k)}}{(8-k)!}\ve_{I_1...I_8}\G^{I_1...I_{(8-k)}}\e.
\eea
The Fierz identity for the eleven dimensional theory which helps to close the supersymmetry transformations is given by
\begin{eqnarray}
\label{fierzblg}
 &&(\bar\epsilon_2\chi)\epsilon_1 -
(\bar\epsilon_1\chi)\epsilon_2 =\\
\nonumber&&-\frac{1}{16}\left(2(\bar\epsilon_2\Gamma_\mu\epsilon_1)\Gamma^\mu\chi
-(\bar\epsilon_2\Gamma_{IJ}\epsilon_1)\Gamma^{IJ}\chi
+\frac{1}{4!}(\bar\epsilon_2\Gamma_{\mu}\Gamma_{IJKL}\epsilon_1)\Gamma^\mu\Gamma^{IJKL}\chi
\right).
\end{eqnarray}
The following gamma matrix product identities are very useful
\begin{eqnarray}
\Gamma_M\Gamma^{IJ}\Gamma^M&=& 4\Gamma^{IJ},\\
  \Gamma_M\Gamma^{IJKL}\Gamma^M &=&0,\\
 \Gamma^{IJP}\Gamma^{KLMN}\Gamma_P&=&-\Gamma^I\Gamma^{KLMN}\Gamma^J +\Gamma^J\Gamma^{KLMN}\Gamma^I,\\
   \Gamma^I\Gamma^{KL}\Gamma^J - \Gamma^J\Gamma^{KL}\Gamma^I &=& 2\Gamma^{KL}\Gamma^{IJ}  - 2\Gamma^{KJ}\delta^{IL}
+ 2\Gamma^{KI}\delta^{JL}- 2\Gamma^{LI}\delta^{JK}\nn\\
&& + 2\Gamma^{LJ}\delta^{IK}- 4\delta^{KJ}\delta^{IL}+4\delta^{KI}\delta^{JL},   \\
   \Gamma^{IJM}\Gamma^{KL}\Gamma_M &=&2\Gamma^{KL}\Gamma^{IJ}  - 6\Gamma^{KJ}\delta^{IL}
+ 6\Gamma^{KI}\delta^{JL}- 6\Gamma^{LI}\delta^{JK}\nn\\
&& + 6\Gamma^{LJ}\delta^{IK}+ 4\delta^{KJ}\delta^{IL}-4\delta^{KI}\delta^{JL}.
\end{eqnarray}

\section{ABJM Identities}
The three dimensional gamma matrices are denoted $\g_\m$ with $\m=0,1,2.$ 
The Fierz transformation is
\begin{equation}
(\bar\lambda\chi)\psi_\a = -\frac{1}{2}(\bar\lambda\psi)\chi_\a
-\frac{1}{2} (\bar\lambda\gamma_\nu\psi)\gamma^\nu\chi_\a,
\end{equation}
it is useful in moving the spare index for the closure of the supersymmetry transformations. 
The following are also useful identities for the closure of the supersymmetry transformations:
\begin{eqnarray}
\nonumber
\frac{1}{2}\bar\epsilon^{CD}_1\gamma_\nu\epsilon_{2CD}\,\delta^A_B&=&\bar\epsilon^{AC}_1\gamma_\nu\epsilon_{2BC}
-\bar\epsilon^{AC}_2\gamma_\nu\epsilon_{1BC},\\
 \nonumber
2\bar\epsilon^{AC}_1\epsilon_{2BD}
-2\bar\epsilon^{AC}_2\epsilon_{1BD}
&=&\bar\epsilon^{CE}_1\epsilon_{2DE}\delta^A_B
-\bar\epsilon^{CE}_2\epsilon_{1DE}\delta^A_B\\
\nonumber&-&\bar\epsilon^{AE}_1\epsilon_{2DE}\delta^C_B+\bar\epsilon^{AE}_2\epsilon_{1DE}\delta^C_B\\
&+&
\bar\epsilon^{AE}_1\epsilon_{2BE}\delta^C_D-\bar\epsilon^{AE}_2\epsilon_{1BE}\delta^C_D\\
\nonumber &-&
\bar\epsilon^{CE}_1\epsilon_{2BE}\delta^A_D+\bar\epsilon^{CE}_2\epsilon_{1DE}\delta^A_D,\\
\nonumber \frac{1}{2}\varepsilon_{ABCD}
\bar\epsilon^{EF}_1\gamma_\mu\epsilon_{2EF}&=&\bar\epsilon_{1AB}\gamma_\mu\epsilon_{2CD}-\bar\epsilon_{2AB}\gamma_\mu\epsilon_{1CD}\\
&+&\bar\epsilon_{1AD}\gamma_\mu\epsilon_{2BC}-\bar\epsilon_{2AD}\gamma_\mu\epsilon_{1BC}\\
\nonumber&-&\bar\epsilon_{1BD}\gamma_\mu\epsilon_{2AC}+\bar\epsilon_{2BD}\gamma_\mu\epsilon_{1AC}.
\end{eqnarray}

\section{Superspace Conventions}
In three dimensions the Dirac matrices satisfy
\be
\g^\m\g^\n = \eta^{\m\n} + \e^{\m\n\r}\g^\r,
\ee
where we choose our basis of gamma matrices to be $(\g^\m)_\a{}^\b = (\s_1,\s_2,\s_3)$, where $\s_i$ are the usual Pauli matrices. Spacetime indices $\m,\n,\r=0,1,2,$ while spinor indices $\a,\b = 1,2.$ We note that when both indices are down the gamma matrices become symmetric
\be
(\g^\m)_{\a\b} = (\g^\m)_{\b\a},
\ee
where the metric on the Clifford algebra is given by $\e^{\a\b}$. We take $\e^{12}=-\e_{12} =1$.

Spinors are Weyl ordered and so we have
\bea
\th^\a\th^\b = -\frac{1}{2}\th\th\e^{\a\b},\quad \th_\a\th_\b = \frac{1}{2}\th\th\e_{\a\b}, \\
\th^\a\thb^\b = -\frac{1}{2}\th\thb\e^{\a\b},\quad \th_\a\thb_\b = \frac{1}{2}\th\thb\e_{\a\b}, \\
\thb^\a\thb^\b = -\frac{1}{2}\thbb\e^{\a\b},\quad \thb_\a\thb_\b = \frac{1}{2}\thbb\e_{\a\b}.
\eea

We note the useful relations
\bea
(\th\thb)(\th\thb) &=& -\frac{1}{2}\th\th \thbb, \\
(\th\thb)(\th\g^\m\thb)&=&0, \\
(\th\g^\m\thb)(\th\g^\n\thb)&=&\frac{1}{2}\eta^{\m\n}\th\th\thbb.
\eea
By choosing chiral spacetime coordinates to be
\be
y^\m = x^\m +i\th^\a\g^\m_{\a\b}\thb^\b,
\ee
where $x^\m$ are the spacetime coordinates, we can define commutation relations
\be
[y^\m,y^\n] = [y^\m,\th^\a] = [y^\m,\thb^\a]=0.
\ee
As a result of the above relations, one is able to derive the following in terms of the $x^\m$ variables
\be
[x^\m,\th^\a]=0 ,\quad [x^\m , x^\n] = 0.
\ee

We can define the supercovariant derivatives where spacetime derivatives $\del_\m$ are defined as $\del_\m \equiv \frac{\del}{\del y^\m}$;
\bea
D_\a &=& \del_\a +i\g^\m_{\a\b}\thb^\b\del_\m , \nn\\
\Db_\a &=& -\delb_\a -i\th^\b\g^\m_{\b\a}\del_\m,
\eea
and the supercharges as
\bea
Q_\a &=&\del_\a -i\g^\m_{\a\b}\thb^\b\del_\m , \nn\\
\Qb_\a &=& -\delb_\a +i\th^{\b}\g^\m_{\b\a}\del_\m.
\eea
The supercovariant derivatives and supercharges satisfy anti-commutator relations;
\bea
\{D_\a,D_\b\} &=& 0, \quad \{\Db_\a,\Db_\b\} = 0,\nn\\
\{\Db_\a,D_\b\} &=& -2i\g^\m_{\a\b}\del_\m, \nn\\
\{D_\a,Q_\b\} &=& \{\Db_\a,Q_\b\} = \{D_\a,\Qb_\b\} = \{\Db_\a,\Qb_\b\} = \{Q_\a,Q_\b\}=0, \nn \\
\{\Qb_\a,Q_\b\} &=& 2i\g^\m_{\a\b}\del_\m, \nn\\
\{\Qb_\a,\Qb_\b\} &=& 0.
\eea

We define the following integration measures
\bea
d^2\th &\equiv& -\frac{1}{4}d\th^\a d\th_\a, \\
d^2\thb &\equiv& -\frac{1}{4}d\thb^\a d\thb_\a, \\
d^4\th &\equiv& d^2\th \,d^2\thb,
\eea
so we have the normalisations
\bea
\int d^2\th\, \th\th = 1, \\
\int d^2\thb\, \thbb = 1, \\
\int d^2\th d^2\thb\, \th\th\thbb = 1.
\eea




\bibliographystyle{hieeetr}
\bibliography{bibliography}

\end{document}

%% file: format.tex
\usepackage{fancyhdr}
\usepackage{epsfig}
\usepackage{cite}
\usepackage{graphicx}
\usepackage{amsmath}
\usepackage{theorem}
\usepackage{amssymb}
\usepackage{latexsym}
\usepackage{epic}

\newcounter{ind}

{\theorembodyfont{\rmfamily}}
{\theorembodyfont{\rmfamily}}
{\theorembodyfont{\rmfamily}}
{\theorembodyfont{\rmfamily}}
{\theorembodyfont{\rmfamily}}
{\theorembodyfont{\rmfamily}}

\def\bv#1{\mbox{\boldmath$#1$}}



%% file: frontpage.tex
\pagenumbering{roman}

\setcounter{page}{1}

\newpage

\thispagestyle{empty}
\begin{center}
  \vspace*{1cm}
  {\Huge \bf Branes and Geometry in \\String and M-Theory}

  \vspace*{2cm}
  {\LARGE\bf Gurdeep Singh Sehmbi}

  \vfill

  {\Large A Thesis presented for the degree of\\
         [1mm] Doctor of Philosophy}
  \vspace*{0.9cm}
  
   \begin{center}
   \includegraphics{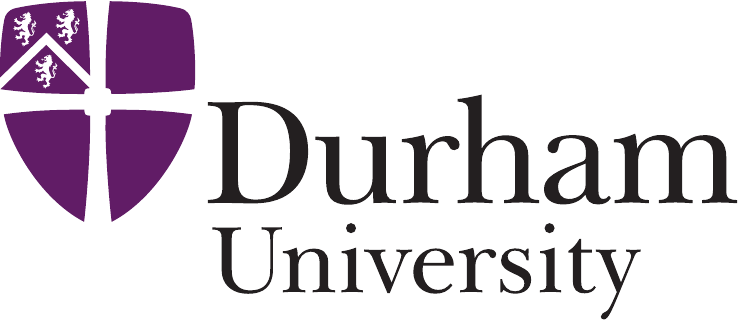}
   \end{center}

  {\large Centre for Particle Theory\\
          [-3mm] Department of Mathematical Sciences\\
          [-3mm] University of Durham\\
          [-3mm] England\\
	  [-3mm] g.s.sehmbi@durham.ac.uk\\
          [1mm]  June 2012}

\end{center}

\newpage
\thispagestyle{empty}
\begin{center}
 \vspace*{2cm}
  \textit{\LARGE {Dedicated to}}\\ 
 Mum and Dad
\end{center}

\newpage
\thispagestyle{empty}
\addcontentsline{toc}{chapter}{\numberline{}Abstract}
\begin{center}
  \textbf{\Large Branes and Geometry in String and M-Theory}

  \vspace*{1cm}
  \textbf{\large Gurdeep Singh Sehmbi}

  \vspace*{0.5cm}
  {\large Submitted for the degree of Doctor of Philosophy\\ June 2012}

  \vspace*{1cm}
  \textbf{\large Abstract}
\end{center}

This thesis is based on the two papers by the author \cite{sehmbi1,sehmbi2} and consists of two parts. In the first part we give an overview of the recent developments in the theory of multiple M2-branes and 3-algebras leading to multiple D2-brane theories. The inclusion of flux terms for the supersymmetric BLG and ABJM theories of closed M2-branes is discussed and then generalised to the case of open M2-branes. Here the boundary condition is derived and different BPS configurations are examined where we find a mass deformed Basu-Harvey equation for the M2-M5 system. The Lorentzian 3-algebra is then employed for obtaining a theory of D2-branes in a flux background, we then obtain the new fuzzy funnel solution of the system of D2-D4 branes in a flux. We then review matrix theories and their compactifications as well as noncommutative geometry and noncommutative gauge theories with a discussion on their generalisations to three dimensions to be used to describe the M-theory three form potential $C_3$. A new feature of string theory is then obtained called the quantum Nambu geometry $[X^\m,X^\n,X^\l]=i\th^{\m\n\l}$, this is another attempt to generalise noncommutative geometry to three dimensions but here we employ the Nambu bracket. It begins with considering the action for D1-strings in a RR flux background and show that there is a large flux double scaling limit where the action is dominated by a Chern-Simons-Myers coupling term. A classical solution to this is the quantised spacetime known as the quantum Nambu geometry (QNG). Matrix models for the type \IIB and type \IIA theories are constructed as well as the matrix model for M-theory. These are the large flux dominated terms of the full actions for these matrix models. The QNG gives rise to an expansion of D1-strings to D4-branes in the \IIA theory, and so we obtain an action for the large flux terms for this action which is verified by a dimensional reduction of the PST action describing M5-branes. Given the recent proposal of the multiple M5-brane theory on $S^1$ being described by 5D SYM and instantons, we make a generalisation of the D4-brane action to describe M5-branes. We are describing the 3-form self-dual field strength in a non-abelian generalisation of the PST action, the QNG parameter is identified with a constant $C_3$-field and is self-dual. The 3-form field strength is constructed from 1-form gauge fields.

\chapter*{Declaration}
\addcontentsline{toc}{chapter}{\numberline{}Declaration}
The work in this thesis is based on research carried out at the
Centre for Particle Theory, the Department of Mathematical Sciences, Durham, England.  No part of this thesis has been submitted elsewhere for any other degree or qualification and it all
my own work unless referenced to the contrary in the text. 

Chapters $1,2$ and 5 are reviews of published works. Chapters $3,4,6,7$ and 8 consists of original work published \cite{sehmbi1,sehmbi2} by the author in collaboration with my supervisor Prof. Chong-Sun Chu.

\vspace{2in}
\noindent \textbf{Copyright \copyright\; 2012 by Gurdeep Sehmbi}.\\
``The copyright of this thesis rests with the author.  No quotations
from it should be published without the author's prior written consent
and information derived from it should be acknowledged.''

\chapter*{Acknowledgements}
\addcontentsline{toc}{chapter}{\numberline{}Acknowledgements}
First and foremost I would like to thank my supervisor and mentor Chong-Sun Chu, his guidance and patience have been paramount to the work I have carried out throughout my PhD. I would also like to thank Douglas Smith for his guidance and helpful discussions on M-theory.

I would like to thank my parents, my sisters Kam and Sukhy, Matt and also my lovely niece Sofia for their unconditional love and support during my time in Durham. I would also like to thank my school boys Raj, Dhar, Taj, Parm, Kalsi and everyone else who has encouraged me to work hard and come home fast! A special thanks to the teachers that inspired and encouraged me during my school days; Mrs Bridge, Mr Nice and Mr Walton. Thanks to Atifah, David and Paul from KCL.

My office mates Dan(ny) and James have been great, from deep discussions on various aspects of mathematics and physics to discussing whatever comes up on the BBC news website. I would also like to thank everyone else in the department who have made my time there so enjoyable; Jonny, Dave, Andy, Jamie, Sam, Luke, Angharad, James B, Ruth, John, Simon, James A, Harry, Rafa, Pichet, Sheng-Lan and all the chaps from the 1st year. From the IPPP; Andrew, James and Katy. I would also like to thank the Old Elvet guys; Christian, Filip, Ilan, Mau.

Josephine Butler College has been a very large part of my life in Durham, especially the MCR. I would like to thank all my college friends who have given me such a good time while being in Durham. In particular I would like to thank `The Lad's Flat' in second year, Maccy, Kelly, Stich and Crafty. Also Alex, Amy, Chris, George, Henry, Hillman, Jack, Liam, Luke, Rich, Sam, Sarah, Tim.

Finally I would like to thank my good friend Daniel Fryer, you have always been there for me whether we were next door or 300 miles away!

\tableofcontents
\clearpage
